\documentclass[lettersize,journal]{IEEEtran}
\usepackage{amsmath,amsfonts}
\usepackage{algorithmic}
\usepackage{array}
\usepackage[figuresright]{rotating}
\usepackage[caption=false,font=footnotesize,labelfont=rm,textfont=rm,subrefformat=parens]{subfig}
\DeclareSubrefFormat{parens}{#1(#2)}
\usepackage{stfloats}
\usepackage{url}
\usepackage{upgreek}
\usepackage{cite}
\usepackage{makecell}
\usepackage{verbatim}
\usepackage{graphicx}
\def\BibTeX{{\rm B\kern-.05em{\sc i\kern-.025em b}\kern-.08em
		T\kern-.1667em\lower.7ex\hbox{E}\kern-.125emX}}
\usepackage{balance}
\usepackage{multirow}
\usepackage{url}
\usepackage{color}
\usepackage{nomencl}
\usepackage{float}
\usepackage{subfig}
\usepackage{booktabs}

\makenomenclature

\begin{document}
	\title{Intelligent Multi-Modal Sensing-Communication Integration: Synesthesia of Machines}
	\author{Xiang Cheng,~\IEEEmembership{Fellow,~IEEE,} Haotian Zhang,~\IEEEmembership{Graduate Student Member,~IEEE,} Jianan Zhang,~\IEEEmembership{Member,~IEEE,} Shijian Gao,~\IEEEmembership{Member,~IEEE,} Sijiang Li,~\IEEEmembership{Graduate Student Member,~IEEE,} Ziwei Huang,~\IEEEmembership{Graduate Student Member,~IEEE,} Lu Bai,~\IEEEmembership{Member,~IEEE,} Zonghui Yang,~\IEEEmembership{Graduate Student Member,~IEEE,} Xinhu Zheng,~\IEEEmembership{Member,~IEEE,} Liuqing Yang,~\IEEEmembership{Fellow,~IEEE}
		\thanks{
			Xiang Cheng, Haotian Zhang, Jianan Zhang, Sijiang Li, Ziwei Huang, and Zonghui Yang are with the State Key Laboratory of Advanced Optical Communication Systems and Networks, School of Electronics, Peking University, Beijing 100871, P. R. China (e-mail: xiangcheng@pku.edu.cn; haotianzhang@stu.pku.edu.cn; zhangjianan@pku.edu.cn; pkulsj@pku.edu.cn;  ziweihuang@pku.edu.cn; yzh22@stu.pku.edu.cn).
			
			Shijian Gao is with the Samsung Semiconductor, Samsung SoC Research and Development Lab, San Diego, CA 92121, USA (e-mail: gao00379@umn.edu).
			
			Lu Bai is with the Joint SDU-NTU Centre for Artificial Intelligence Research (C-FAIR), Shandong University, Jinan 250100, P. R. China (e-mail: lubai@sdu.edu.cn).
			
			Xinhu Zheng is with the Intelligent Transportation Thrust, The Hong Kong University of Science and Technology (Guangzhou), Guangzhou 510000, P. R. China (e-mail: xinhuzheng@ust.hk).
			
			Liuqing Yang is with the Internet of Things Thrust and Intelligent Transportation Thrust, The Hong Kong University of Science and Technology (Guangzhou), Guangzhou 510000, P. R. China, and the Department of Electronic and Computer Engineering, The Hong Kong University of Science and Technology, Hong Kong SAR, P. R. China (email: lqyang@ust.hk).
	}}
	
	
	\maketitle
	
	\begin{abstract}
		In the era of sixth-generation (6G) wireless communications, integrated sensing and communications (ISAC) is recognized as a promising solution to upgrade the physical system by endowing wireless communications with sensing capability. Existing ISAC is mainly oriented to static scenarios with radio-frequency (RF) sensors being the primary participants, thus lacking a comprehensive environment feature characterization and facing a severe performance bottleneck in dynamic environments. To date, extensive surveys on ISAC have been conducted but are limited to summarizing RF-based radar sensing. Currently, some research efforts have been devoted to exploring multi-modal sensing-communication integration but still lack a comprehensive review. To fill the gap, we embark on an initial endeavor with the goal of establishing a unified framework of intelligent multi-modal sensing-communication integration by generalizing the concept of ISAC and providing a comprehensive review under this framework. Inspired by the human synesthesia, the so-termed Synesthesia of Machines (SoM) gives the clearest cognition of such an intelligent integration and details its paradigm for the first time. We commence by justifying the necessity and potential of the new paradigm. Subsequently, we offer a rigorous definition of SoM and zoom into the detailed paradigm, which is summarized as three operational modes realizing the integration. To facilitate SoM research, we overview the prerequisite of SoM research, that is, mixed multi-modal (MMM) datasets, and introduce our work. Built upon the MMM datasets, we introduce the mapping relationships between multi-modal sensing and communications, and discuss how channel modeling can be customized to support the exploration of such relationships. Afterward, aiming at giving a comprehensive survey on the current research status of multi-modal sensing-communication integration, we cover the technological review on SoM-enhance-based and SoM-concert-based applications in transceiver design and environment sensing. To corroborate the rationality and superiority of SoM, we also present simulation results related to dual-function waveform and predictive beamforming design tailored for dynamic scenarios. Finally, we propose some open issues and potential directions to inspire future research efforts on SoM.

	\end{abstract}
	
	
	
	\section*{List of Acronyms}
	\begin{tabbing}
		\hspace*{80bp}\=article\quad \=ÄÕÂÀà\kill
		2D \> Two-Dimensional \\
		3D \> Three-Dimensional \\
		3GPP \> Third Generation Partnership Project \\
		4G\> Fourth-Generation\\
		5G\> Fifth-Generation\\
		6G \> Sixth-Generation\\
		AI \> Artificial Intelligence \\
		ANN   \> Artificial Neural Networks\\
		B5G   \> Beyond Fifth-Generation \\
		Bi-LSTM \> Bidirectional Long Short-Term Memory \\
            BS \> Base Station \\
		CIR \> Channel Impulse Response \\
		CNN \> Convolutional Neural Network \\
		COST \> European COoperation in the field of  \\
		\> Scientific and Technical research \\
		CRB \>  Cramér-Rao Bound \\
		CSI \> Channel State Information \\
		DFRC \>  Dual-Functional Radar and \\
		\> Communications \\
		DL \> Deep Learning \\
		DNN \> Deep Neural Network  \\
		DoA \> Direction of Arrival \\
		DoD \> Direction of Departure \\
		EKF \> Extended Kalman Filter \\
		FMCW \> Frequency Modulated Continuous Wave \\
		FoV \> Field of View \\
		GBDMs \> Geometry-Based Deterministic Models \\
		GBSMs \> Geometry-Based Stochastic Models \\
		GPS \> Global Positioning System \\
		IM-OFDM \> Index Modulation-OFDM \\
		IMU \> Inertial Measurement Unit \\
		IoTs \> Internet of Things \\
		ISAC \> Integrated Sensing and Communications \\
		JCRS \> Joint Communications and Radar Sensing \\
		JRC \> Joint Radar and Communications \\
		KF \> Kalman Filter \\
		LiDAR \> Light Detection and Ranging \\ 
		LoS \> Line-of-Sight \\
		METIS \> Mobile and wireless communications \\
		\> Enablers for the Twenty-twenty \\ 
		\> Information Society \\
		MIMO \> Multiple-Input Multiple-Output \\
		ML \> Machine Learning \\
		mMIMO \> Massive Multiple-Input Multiple-Output \\
		MMM \> Mixed Multi-Modal \\
		NGSMs \> Non-Geometry Stochastic Models \\
		NLoS \> Non-Line-of-Sight \\
		OFDM \> Orthogonal Frequency Division \\
		\> Multiplexing \\
		QoS \> Quality of Service \\
		Rad-Com \> Radar-Communications \\
		RF \> Radio-Frequency \\
		RGB-D \> Red-Green-Blue-Depth \\
        RL \> Reinforcement Learning \\
		RMSE \> Root Mean Square Error \\
		ROI \> Region of Interest \\
		RSU \> Road Side Unit \\
		SAGSIN \> Space-Air-Ground-Sea Integrated Network \\
		SNR \> Signal-to-Noise Ratio \\
		SoM \> Synesthesia of Machines \\
		THz \> Terahertz \\
		UAV \> Unmanned Aerial Vehicle \\
		V2I \> Vehicle-to-Infrastructure \\
		V2V \> Vehicle-to-Vehicle\\
		VCN \> Vehicular Communication Networks \\
		VR \> Visibility Region  \\
	\end{tabbing}
	
	\section{Introduction}
	\label{Background}

	\IEEEPARstart{T}{he} fifth-generation (5G) wireless communication networks have entered the stage of commercial deployment and are changing many aspects of life and industrial structures. Currently, the development and integration of advanced mobile communication techniques, artificial intelligence (AI), and big data techniques are propelling the evolution of 5G towards the development of beyond 5G (B5G) and sixth-generation (6G) wireless communication networks to satisfy the demands of future networks beyond 2030. Compared with 5G, the purposes of B5G/6G networks are not only to increase network capacity and data transmission rate but also to accomplish the ultimate goal of connecting everything \cite{ntt2020white}. There are increasing research interests for 6G \cite{zhang20196g, chowdhury20206g}. 6G is expected to provide a peak data rate of 1-10 Tbps and the users' experience data rates can reach Gbps. Moreover, the energy efficiency and spectrum efficiency of 6G are expected to be 10-100 times and 3-5 times higher than those of 5G. 6G will support a variety of potential new application scenarios, such as accurate indoor positioning and high-quality communication services on aircraft \cite{dang2020should}. With the in-depth integration of mobile communication techniques with new enabling technologies such as big data and AI, the B5G/6G networks also present some new development trends. Among them, integrated sensing and communications (ISAC) has emerged as a key candidate technology in future wireless systems.
	
    {\color{black}ISAC is anticipated to significantly enhance spectral and energy efficiencies while reducing both hardware and signaling costs since it aims to integrate radio-frequency (RF) based radar sensing and communications into one system \cite{liu2022integrated}.
    Such an integration is feasible due to the following reasons. Firstly, radar and communications exhibit strong similarities and integration feasibility in hardware structure and system components \cite{Paul2017}. Secondly, they are increasingly similar in antenna structure, such as phased array radar and multiple-input multiple-output (MIMO). Thirdly, the bandwidth of communications is becoming wider and even closer to that of radar.} The earliest ISAC-related research emerged decades ago. Radar, one of the first sensors, received a lot of attention in ISAC-related research in the early stage. Since the 1990s, RF-based radar and communications have been no longer considered as two completely separate systems. The initial research efforts towards the joint design of radar and communications focus on how to superimpose communication information onto the common radar waveforms \cite{roberton2003integrated, saddik2007ultra, jamil2008integrated}. As communication technology develops, modern wireless communication systems including the fourth-generation (4G) and 5G apply the orthogonal frequency division multiplexing (OFDM) as an efficient modulation format. OFDM has been applied in radar due to its potential for high range resolution and low probability of intercept characteristics \cite{tigrek2012ofdm,hu2014radar}. More recently, the increasing popularity of future smart devices and the emergence of new business opportunities and application scenarios have led to a massive proliferation of telecommunication traffic. The bandwidth shortage experienced by wireless communications has motivated the use of the mmWave band with rich spectrum resources. To overcome serious issues such as high path loss and reduced scattering in mmWave spectrum, the massive multiple-input multiple-output (mMIMO) is regarded as one of the key solutions for 5G-and-beyond networks \cite{marzetta2010noncooperative} thanks to its high beamforming gain. Introducing MIMO techniques into radar, MIMO radar \cite{bliss2003multiple} has become a hot research issue around the world. As the hardware architectures, signal processing techniques, and channel characteristics of communication and radar systems are getting closer to each other, these two functionalities deeply interwove and evolved in the same direction over the past decades. The convergence between them has been a research issue from decades ago to the present. The relevant studies have been investigated for decades under different names, such as Radar-Communications (Rad-Com) \cite{commu1}, Joint Radar and Communications (JRC) \cite{liu2020joint}, Dual-functional Radar and Communications (DFRC) \cite{liu2020joint}, and Joint Communications and Radar Sensing (JCRS) \cite{zhang2020perceptive} (hereby referred to as RF-ISAC in this paper).
	
	\textbf{As the key candidate technology of 6G, RF-ISAC still faces certain limitations.} RF-ISAC limits the sensing functionality to radar sensing, resulting in the integration gain it achieved being only manifested at the hardware level. This limitation will hinder its application prospects in the emerging applications in the B5G/6G era where massive multi-modal environment information can be utilized for facilitating communication and sensing functionality. The reasons lie in the following two aspects. Firstly, as the frequency bands used by future communication systems increase, the relationships between the RF environment and the geometry of physical environment become more closely related. Unfortunately, RF-ISAC cannot characterize such relationships comprehensively and granularly through the parameters it measures. Secondly, most existing works on RF-ISAC are based on static environments, while mobility that comes along with the emerging application scenarios is a critical feature that cannot be ignored. Mobility can bring a significant performance impact on sensing and communication functionalities. For instance, the fast-varying characteristics of channels caused by dynamic environments significantly increase the communication overhead required by standard \cite{chengbookVCN} while RF-ISAC technology is unable to provide significant improvement in this regard. To summarize, due to the aforementioned limitations, RF-ISAC will inevitably face certain performance bottlenecks when encountering the emerging application scenarios and more stringent quality of service (QoS) requirements therein, e.g., higher throughput, lower latency, and smaller jitter, which will be detailed in Section \ref{ii1}.
	
	\textbf{From RF-ISAC to multi-modal sensing-communication integration.} In fact, there are various types of sensors that have completely different working principles and can collect sensory data with different properties, including but not limited to the radar, light detection and ranging (LiDAR), red-green-blue-depth (RGB-D) cameras, and the global positioning system (GPS), which are collectively referred to as \textit{multi-modal sensing}. Radar is only one of the sensors that operates in the RF frequency band. Harnessing a great potential for enhanced environmental awareness, multi-modal sensing holds many promises for developments and performance assurance in communication and sensing systems in the B5G/6G era. Non-RF sensors can obtain much more diverse communications-related features while RF-based radar sensing can solely achieve target detection and parameter measurement. Furthermore, multi-modal sensory data from multiple agents can jointly constitute a reinforced sensing system with the assistance of communication networks. Benefiting from these advantages, unlike RF-ISAC which restricts the ``integration" to the unified hardware or waveform, multi-modal sensing-communication integration aims to achieve mutual functional assistance or enhancement at various potential technical aspects of communication and sensing systems. For instance, the inclusion of multi-modal sensing can accelerate the beam selection process, such as LiDAR-aided beam prediction \cite{klautau2019lidar}, camera-GPS-aided beam prediction \cite{alkhateebDRONE}, and camera-aided beam prediction \cite{alkhateebDRONE2,alrabeiah2020millimeter}, where conventional methods mandate an exhaustive search across all candidate beam pairs. To summarize, the extensively studied RF-ISAC should be expanded to include non-RF sensors and evolve into multi-modal sensing-communication integration. In fact, in recent years, relevant research has already been conducted focusing on various topics but lacks a unified framework as well as a systematic review. In this survey, we define such a framework as \textit{\textbf{Synesthesia of Machines (SoM)}} inspired by human synesthesia for the first time. SoM explicitly gives the goals of intelligent multi-modal sensing-communication integration. Based on this, it differentiates and summarizes various approaches of achieving such goals by defining three distinct operational modes. Synthesizes these features, SoM can be regarded as the unified framework of this field. SoM is oriented to generic scenarios, whereas this survey focuses on research status and potential applications in dynamic scenarios due to their growing popularity in B5G/6G networks and numerous challenges to be addressed  (which will be discussed in detail in Section \ref{ii1}). Within the SoM framework, we conduct a comprehensive survey on recent advances in intelligent multi-modal sensing-communication integration in this paper.
	
\subsection{Related Work}
	
	\begin{table*}[!htp]
		\renewcommand\arraystretch{1.5} 
		\centering	
	
			\caption{Summary of Existing Surveys Related to SoM}
			\label{tab:comparedwithothersurvey}
			\resizebox{\textwidth}{!}{
				\begin{tabular}{c|c|c|c|c}
					\toprule[0.35mm]
					\makecell[c]{\textbf{Existing works}}	 &
					\textbf{Survey topic(s)}	 & \textbf{\makecell[c]{Relevant sections\\ in this survey}} & \textbf{Technology scope} &	\textbf{\makecell[c]{Reviewed multi-modal \\ sensing-related works}}\\
					\midrule[0.15mm]
					\cite{feng2020JRC,cui2021integrating,mishra2019toward,zhang2021overview,wei2023}	& \makecell[c]{The current research progress on the\\ ISAC signal design and signal processing.} & \ref{waveform} & RF-ISAC & None \\ 
					\midrule[0.15mm]		
					\cite{liu2022survey}	& \makecell[c]{The current research progress on the\\ fundamental limits of ISAC} &  \ref{waveform} & RF-ISAC & None \\ 
					\midrule[0.15mm]			
					\cite{liu2022integrated}	& \makecell[c]{The background, key applications, and \\ state-of-the-art approaches of ISAC  with\\ the focus on RF-based radar sensing} &  \ref{waveform}, \ref{som beamforming}& RF-ISAC & $<5$\\ 
					\midrule[0.15mm]			
					\cite{Chengisac}	& \makecell[c]{The ISAC possibilities and potential \\specifically for transportation applications \\ with discussions on multi-modal sensing} &  \ref{waveform}, \ref{som beamforming}, \ref{SoMConcert} & \makecell[c]{Multi-modal sensing\\-communication integration} & $<10$ \\ 
					\midrule[0.15mm]		
					\cite{CXISAC}	& \makecell[c]{Current status and development  trend \\ of ISAC in VCNs, where the multi-modal \\ sensing-related works are considered} &  \ref{smallscale_r}, \ref{part2}, \ref{SoMConcert} & \makecell[c]{Multi-modal sensing\\-communication integration} & $<20$ \\ 
					\midrule[0.15mm]		
					\textbf{This survey}	& \makecell[c]{Summarizing the paradigm, technical \\ details,  design principles, and research\\ progress of intelligent multi-modal \\ sensing-communication integration \\ for the first time} & $-$ & \makecell[c]{Intelligent multi-modal \\sensing-communication\\ integration} & \makecell[c]{$53$ (Research on \\specific technologies) \\ $+$ $11$ (Datasets)} \\ 
					\bottomrule[0.35mm]
				\end{tabular}	
			}

	\end{table*}
	
     Driven by the recent advances in ISAC-related topics, some research efforts have been made to review related works. Table~\ref{tab:comparedwithothersurvey} lists the relevant surveys that may align with our work. It can be concluded that \textbf{none of the existing surveys consider the growing significance of multi-modal sensing and focus their review on multi-modal sensing-communication integration.} Surveys \cite{liu2020joint,zheng2019radar,ma2020joint,feng2020JRC} introduced a state-of-the-art review of JRC technique according to four categories, namely, coexistence, cooperation, co-design, and collaboration. To describe the coupling of radar sensing and communications, Cui \textit{et~al.}~\cite{cui2021integrating} proposed a novel concept of signaling layer, which is the convergence result for communications and radar sensing. For different dual-function waveform design methods, surveys \cite{mishra2019toward} and \cite{zhang2021overview} reviewed their systems models and typical signal processing techniques. As a step further, Wei \textit{et al.} \cite{wei2023} thoroughly reviewed relevant techniques of ISAC signals from the perspective of its applications in 5G-A and 6G mobile communication systems. Liu \textit{et al.} \cite{liu2022survey} provided a systematic overview on the fundamental limits of ISAC system, aiming to guide researchers to better design the ISAC systems. A recent survey in \cite{liu2022integrated} gave a comprehensive review on the background, application scenarios (including dynamic and static ones), signal processing, and information theoretical limits of ISAC. Moreover, it sheds light on the deeper integration between sensing and communications, i.e., communication-assisted sensing and sensing-assisted communications. Although \cite{liu2022integrated} has provided a clear bird's-eye view of the ISAC technique, it still falls short in the summary of the existing works and challenges of ISAC in dynamic scenarios. Moreover, \cite{liu2022integrated} does not consider the potential and methodology of the application of non-RF sensors. The above surveys comprehensively review the waveform design and signal processing technology in ISAC systems as well as the fundamental limits, while \cite{zhang2021enabling} focuses on the enhancement of the sensing functionality. Based on JRC technology that integrates radio sensing into mobile communication networks, \cite{zhang2021enabling} gave a comprehensive review of the perceptive mobile network. Still, the perception functionality is limited to radar sensing while it is unclear how sensing enhances the communication system design.  
	
	Our previous work \cite{CXISAC} provided a review on the current status and development trend of ISAC in VCN scenarios, where non-RF sensors are considered and relevant works are reviewed.  On the basis of \cite{CXISAC}, our previous work \cite{Chengisac} first generalized the concept of synesthesia to ISAC in VCNs. However, the specific definitions and operating principles of synesthesia in the ISAC system are not presented. Although \cite{Chengisac} supplements the analysis of the challenges faced by joint design of sensing and communications in dynamic scenarios lacking in \cite{CXISAC}, it lacks a comprehensive review of current research progress regarding multi-modal sensing-communication integration.

\textbf{Contributions of this survey:} We point out that existing ISAC/Rad-Com/JRC/DFRC/JCAS-related surveys, such as \cite{zheng2019radar,zhang2020perceptive, liu2020joint, ma2020joint, mishra2019toward, feng2020JRC,zhang2021overview, cui2021integrating,liu2022survey, zhang2021enabling, liu2022integrated, wei2023}, equate sensing functionality with RF-based radar sensing, so the integrated design of sensing and communications they focused on is naturally limited to the RF-based equipment and the literature reviews mostly center on dual-function physical layer design and relevant techniques. Although surveys \cite{CXISAC, Chengisac} consider the growing significance of multi-modal sensing, they still fall short in offering technical details. Furthermore, although existing surveys \cite{liu2022integrated, CXISAC, Chengisac} have discussed dynamic application scenarios for illustrating the potential of ISAC for future wireless networks, relevant works tailored for such scenarios have not been systematically reviewed and a unified framework has not been summarized. In view of the prior works, there is still a lack of holistic discussion and systematic cognition on multi-modal sensing-communication integration in the open literature. These limitations motivate us to \textbf{conduct a comprehensive review on the recent advancements made}, and most importantly, to \textbf{establish a unified framework regarding multi-modal sensing-communication integration and detail its design paradigm}. To the best of our knowledge, this is the first survey centering on intelligent multi-modal sensing-communication integration that covers \textbf{motivations, mechanism, enabling technologies, current research status, and challenges}.

\vspace{1em}
\subsection{Contributions and Organization of This Paper}	
The contributions of this paper are summarized below.
\begin{itemize}
	\item
	We propose the concept of SoM by generalizing human synesthesia to machine senses, which is the first attempt of establishing a systematic framework for the intelligent multi-modal sensing-communication integration. Evolved from the extensively studied RF-ISAC, SoM emphasizes the role of multi-modal sensing, expands to more diverse objectives, considers new design methods, and facilitates more challenging scenarios.
\end{itemize}
\begin{itemize}
	\item
	Within the SoM framework, we summarize the paradigm in which multi-modal sensing and communications mutually enhance each other for the first time by proposing three operational modes of SoM: SoM-evoke, SoM-enhance, and SoM-concert. The above efforts offer the most comprehensive and systematic approach and the clearest cognition to intelligent multi-modal sensing-communication integration.
\end{itemize}

\begin{itemize}
	\item
	
	We provide a detailed and comprehensive discussion around SoM. We review the prerequisite of SoM research, i.e., mixed multi-modal (MMM) dataset, which serves as a ``dataset guide'' for researchers and is not covered in existing surveys. We overview the current research status and challenges of SoM-enhance-based and SoM-concert-based applications in communication and sensing systems, which have been achieved by the research community but have not yet been systematically summarized. This survey also covers the review of extensively studied RF-based radar sensing-communication integration, which is a special case of SoM. 
\end{itemize}

\begin{itemize}
	\item
	We present simulation results around SoM to illustrate its beneficial effects and shed light on future research, including the exploration of mapping relationships between multi-modal sensing and communications, dual-function waveform design, and SoM-enhance-based predictive beamforming design.  We also present detailed observations and insights centering on different aspects of SoM research. 
	
\end{itemize}

\begin{itemize}
	\item
	Within the SoM framework, we lay out a list of open issues and research directions from two perspectives, aiming to comprehensively inspire SoM research. These include: i) further solidifying its research foundation, which includes the construction of MMM datasets and exploration of mapping relationships; ii) exploring its future applications in communication and sensing systems, i.e., the future applications of the three operational modes.
\end{itemize}

\begin{figure*}[!t]
			\centering
			\includegraphics[width=1\linewidth]{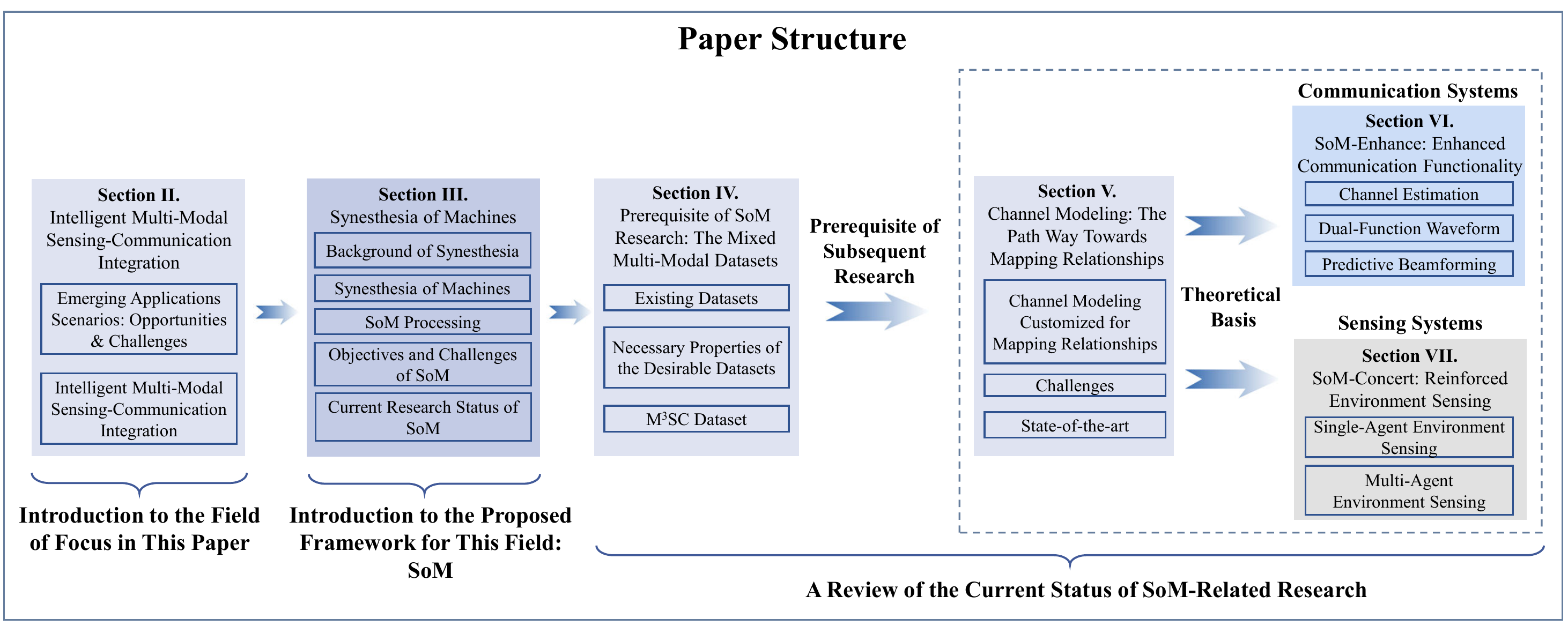}
			\caption{The structure of this paper.}
			\label{structure}	
	\end{figure*}

	The overall organization of this paper is illustrated in Fig.~\ref{structure}. Section \ref{details} gives the motivation and potential of multi-modal sensing-communication integration in detail. Section \ref{Intro of SoM} introduces SoM in detail, then gives its objectives, challenges, and current research state. Section \ref{Dataset} overviews the prerequisite of SoM research, i.e., mixed multi-modal (MMM) dataset. Section \ref{part1} gives the importance and challenges of the exploration on the mapping relationships between multi-modal sensing and communications, which can be supported by the channel modeling customized for SoM. Section \ref{part2} overviews the application of SoM-enhance in transceiver design and presents simulation results. Section \ref{SoMConcert} overviews the related works of SoM-concert for reinforced environment sensing in single-agent and multi-agent scenarios. Section \ref{F} gives the open issues and future research directions within the context of SoM. Finally, Section \ref{C} concludes this paper.
	
	\section{Intelligent Multi-Modal Sensing-Communication Integration: A Paradigm Shift in ISAC}
	\label{details}
	The extensively studied RF-based radar sensing-communication integration is calling for a paradigm shift confronted with the mobility feature of emerging application scenarios in the B5G/6G era and the abundant environment information (including multi-modal sensing and communication information) therein. In this section, we first discuss the great challenges that emerging dynamic application scenarios bring to communication and sensing systems, and analyze the potential of sensing-communication integration to address the challenges. Then, we analyze the necessity and superiority of incorporating multi-modal sensing in detail.
	
	\subsection{Emerging Application Scenarios: Opportunities \& Challenges}
	\label{ii1}
	6G communications will have numerous applications in large-scale machine-type communication scenarios with high mobility. Ultra-fast, low-latency, and high-reliability communication and sensing functionalities are required to support such scenarios \cite{mahmood2020white}. We provide several examples below:
	
	\textbf{Intelligent mobile manned platform}: Self-driving cars will not only free people from the burden of driving, but also optimize the passengers' experience through various high-quality onboard services \cite{chengbook}. In addition to the intelligent vehicle, more types of agents will join the intelligent mobile manned platform, providing people with more convenient transportation services. For example, unmanned aerial vehicles (UAV) will be widely used in industrial, agricultural, and military scenarios thanks to their outstanding agility. In order to bring about the space-air-ground multi-layer transportation services, 6G needs to meet the high communication and sensing requirements of diversified intelligent mobile manned platforms.
	
	\textbf{Mobile unmanned applications}: In the future mobile unmanned applications, such as intelligent logistics and intelligent factories, large-scale machine-type communications will be the mainstream of communications. In the process of manufacturing and product delivery, the actions of machines can simplify or even change the communication system design, thereby cutting down the signaling overhead. Therefore, accurate localization and sensing are the key technologies to assist the communications between machines and fulfill motion capture. The communication functionality for intelligent logistics and intelligent factories is expected to meet stringent QoS requirements.
	
	\textbf{Space-Air-Ground-Sea Integrated Network (SAGSIN)}: SAGSIN will provide ubiquitous, intelligent, and collaborative information assurance for various network applications, and support the access of space-based, air-based, sea-based, and ground-based terminals with speeds up to $1000$ km/h. 6G networks supporting the three-dimensional (3D) full coverage of space-air-ground-sea need to increase the breadth and depth of network coverage, so as to achieve global pervasive communications.

	\textbf{Smart cities}: Smart cities have emerged as a new paradigm, enabling the efficient and dynamic utilization of resources through ubiquitous devices to offer diverse services to citizens \cite{smartcity1}. Smart cities are anticipated to provide safer, more effective, and more intelligent services, ultimately enhancing the level of intelligence in smart cities. Such visions require the deployment of high-performance communication and sensing functionalities, massive computing resources, and big data analytics within the smart city network to provide users with real-time and high-quality experiences.

	\textbf{Industry 5.0}: The forthcoming Industry 5.0 strengthens the importance of adopting a ``human (user)-centric" approach in developing future technologies, services, and applications \cite{industry1}. The core objective is to meet the stringent QoS requirements, privacy, and security requirements of users, which raises higher demands for advanced communication and automation technologies, such as AI and machine-to-machine communications \cite{industry2}.

	\textbf{What are the challenges that come along with these emerging applications?} While the aforementioned applications are expected to change the production and lifestyles of human beings profoundly, they also pose great challenges to the communication and sensing system design. From the perspective of communication systems, the simultaneous high mobility of the receiver, transmitter, and scatterers will result in rapid changes in the wireless electromagnetic environment, causing significant frequency selectivity and time selectivity in wireless channels. Consequently, the channel model and channel impulse response (CIR) information need to be updated frequently due to the fast-varying channels, and the scheduling strategy of multiple communication resources (e.g., power, bandwidth, and beamwidth) needs to be adjusted in real time. Conventional methods commonly adopted in low-speed or static scenarios will fail to meet those of dynamic scenarios due to their high communication overhead and latency. From the perspective of sensing systems, the sensory data collected by the sensors that move with dynamic agents will be distorted. For example, the varying round-trip distance of the laser emitted by the LiDAR causes a distortion in the measured distance.  Moreover, the limited data acquisition frequency and sensing range of the sensors on a single dynamic agent hinder its ability to collect real-time data during high-speed movement and result in a short effective time of sensory data. As dynamic scenarios undergo accelerated changes, a larger sensing range is required to ensure intelligent agents detect objects in advance, thereby allowing them to have sufficient time to make informed behavior decisions.

	In summary, as 6G takes ISAC (RF-ISAC defined in this paper) as the potential key technology, its technical evolution towards better performance for emerging application scenarios with high mobility holds significant research value and has already led to widespread attention. In what follows, we analyze the potential of multi-modal sensing-communication integration in detail.
	
	\subsection{Intelligent Multi-Modal Sensing-Communication Integration}
	\label{ii2}
	\textcolor{black}{
		The aforementioned challenges in the emerging application scenarios will undoubtedly bring about a breakthrough in new technologies. Research on potential enabling technologies is extremely urgent to support high-quality communication and sensing applications in the dynamic environment. However, confronted with numerous new challenges, the existing RF-ISAC, the  potential key technology in 6G, has not yet been capable of fulfilling the above vision. RF-based radar sensing and communications both rely on electromagnetic waves to detect the target or transmit data. The integration gain is only realized at the hardware level and is limited to motion parameters measurement, while the performance enhancement of sensing and communication systems typically needs the global and fine-grained characterization for rapid changes in the environment. Therefore, RF-ISAC is significantly insufficient for addressing the aforementioned challenges. Nowadays, RF-ISAC is calling for a paradigm shift, where multi-modal sensing plays a crucial role.}
	
	\subsubsection{The Potential of Multi-Modal Sensing-Communication Integration}
	The introduction of multi-modal sensing enables the integration system to open its ``eyes" to see the physical environment through multiple domains, such as the electromagnetic wave propagation domain, visible or infrared light domain. \textcolor{black}{ Recently, such a trans-domain integration system has been proven to be reasonable and superior by several works, such as link selection \cite{fan2022radar}, beam alignment \cite{ali2020passive,xu2023computer, alouini21CVbeam}, and resource allocation \cite{xu2023multi}}, where multi-modal sensing shows the potential of extracting RF environment-related features in abstract formats to facilitate various applications in the communication system. 
	
	With the aid of multi-modal sensing-communication integration, we envision that the B5G/6G networks will have the native sensing ability \cite{zhang2020perceptive,Rahman2020frame} and achieve the integration and symbiosis of sensing and communication functionalities. The above vision is expected to offer great opportunities to tackle the challenges that sensing and communication systems face in dynamic scenarios mentioned in Section \ref{ii1}. From the perspective of sensing-assisted communications, the widely deployed sensors in the B5G/6G networks can reflect the parameters of the physical environment required by the communication system design, such as the velocity and position of the scatterers. By extracting the specific environmental features needed for the communication system design from the sensory data, the high overhead and latency caused by traditional methods in addressing serious problems like double-selectivity are expected to be significantly reduced. The rapidly developing wireless communication network will also bring a qualitative leap in the performance, robustness, and redundancy of sensing systems in dynamic scenarios. Relying on high-quality communication links, dynamic agents can share the sensory data to build a broader range of sensing networks and realize beyond-vision-range sensing. Secondly, the vision of constructing the next-generation wireless networks with native sensing ability is inseparable from intelligent multi-modal sensing-communication integration. The B5G/6G network will constantly sense and depict the physical environment through electromagnetic waves used for communications as well as the widely deployed multi-modal sensing-communication integrated base stations (BS) and terminals. It is only possible to model the physical environment based on the B5G/6G network when the terminals utilize a variety of multi-modal sensors and communication equipment operating in different domains. In conclusion, incorporating multi-modal sensing is of great significance for augmenting the sensing capability of wireless networks to the physical environment by using sensors operating in different domains and different frequency bands.

	\subsubsection{RF-ISAC Versus Intelligent Multi-Modal Sensing-Communication Integration}
	We illustrate the limitation of RF-ISAC in dynamic scenarios and the benefit of multi-modal sensing through a case study. Consider an ISAC-enabled dynamic VCN scenario where a road side unit (RSU) is serving a passing vehicle. The RSU needs to accurately estimate the channel state information (CSI) between itself and the passing vehicle, which serves as the foundation for the subsequent communication system design. This process needs to be executed frequently due to the fast-varying channel characteristics in VCN \cite{myCOMST}. Conventional channel estimation methods inevitably lead to excessive signaling overhead and latency when applied to such dynamic scenarios \cite{LS,mmse}. To address that concern, the sensing functionality is expected to provide additional information about the vehicle and the surrounding environment, thereby assisting the RSU in obtaining more accurate CSI with less signaling overhead. In the current RF-ISAC system, RSU can measure the vehicle's velocity and distance through the dual-function waveform, which is helpful for resisting dual-selectivity characteristics in dynamic scenarios. However, CSI is a joint result of various critical elements in the environment, such as the position and velocity of scatterers, receivers, and transmitters. Merely obtaining the measurements of the vehicle's motion parameters through the current ISAC system is insufficient to provide the necessary environmental information for accurately recovering CSI, especially in dynamic scenarios where the measurements generally contain large errors. Multi-modal sensing can bring an opportunity in this context. RGB cameras can provide two-dimensional (2D) visual features which can potentially reflect the potential scatterers in the surrounding environment. LiDAR can measure distances and create a detailed 3D map of the surrounding environment using laser beams, which naturally supplements the 3D structural information and helps RSU reconstruct the surrounding environment. Radar can provide measurements of targets that are directly related to channel characteristics. By appropriately processing and extracting features from multi-modal sensory data, multi-modal sensing can overcome the limitations of current RF-ISAC technology and bring significant performance gains to communication functionality in dynamic scenarios. In the subsequent sections, we make an initial attempt to establish a unified framework for the intelligent multi-modal sensing-communication integration inspired by human synesthesia, i.e., SoM, and conduct a comprehensive discussion around it.

	\section{Synesthesia of Machines}
	\label{Intro of SoM}
\begin{figure*}[!t]
		\centering
		\includegraphics[width=0.95\textwidth]{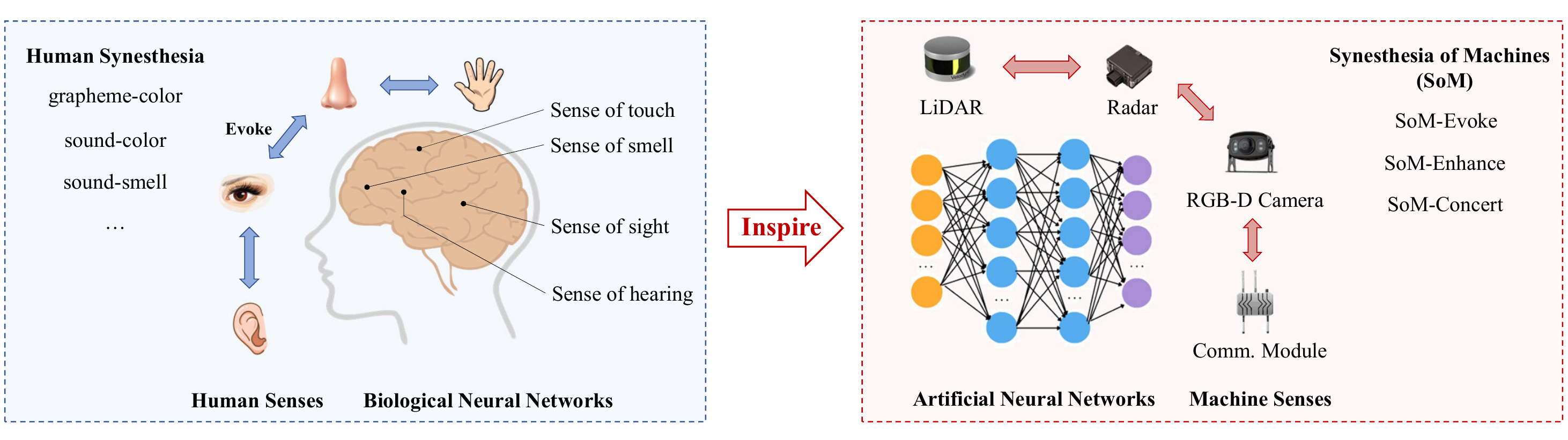}
		\caption{Illustration of human synesthesia and SoM.}
		\label{SoH}
	\end{figure*}

	{\color{black} The potential of multi-modal sensing in communication and sensing systems has already received widespread attention. Multi-modal sensing-communication integration has been studied for many years despite the lack of systematic cognition and a unified framework in the research community. In the field of communication system design, intrinsically, the channel is determined by the wireless propagation environment, which can be jointly captured and reconstructed by multi-modal sensing with different functions. Actually, such visions have already been proven feasible through extensive research. Specifically, some studies have already investigated the use of multi-modal sensing to aid in channel-related downstream tasks, e.g., resource allocation \cite{xu2023multi,fan2022radar}, channel estimation \cite{sensingaidedCE2022,sensingaidedCE2023}, channel prediction \cite{gao2021prediciton}, channel covariance prediction \cite{xu2021deep,alkhateebCCP}, beam selection \cite{feifeibeam,alkhateebLiDARBeam,chowdhuryV2I,gonzalez2016radar}, handover decision \cite{charan2021vision}, and codebook design \cite{chen2022computer}.
		In the field of sensing system design, by utilizing multi-modal sensors on multiple agents, a more comprehensive collection of environmental features can be achieved, thereby enhancing the sensing accuracy and expanding the sensing range. Related works have emerged in various scenarios and tasks, including the bird's-eye view object detection \cite{early1wirges2018object, early5}, 3D object detection \cite{early4, early6}, object tracking \cite{early2steyer2019grid}, and road segmentation \cite{latenew2}. 
		In fact, the aforementioned works all fall within the scope of intelligent multi-modal sensing-communication integration research. However, they are currently fragmented, and the field lacks a cohesive framework to date. 
		
		In this section, we present the unified framework for intelligent multi-modal sensing-communication integration, i.e., SoM, and provide a detailed introduction. Afterward, within the framework of SoM, we will review the existing relevant research in the subsequent sections.
		
		\subsection{Background of Synesthesia}}
	Synesthesia is an involuntary human neuropsychological trait in which the stimulation of one sense organ will automatically evoke another sense organ, as depicted in Fig.~\ref{SoH}. Human synesthesia can occur in various forms, for example, grapheme-color, sound-color, flavor-temperature, flavor-sound, sound-smell, and time units-colors. Among them, grapheme-color synesthesia is the most-studied one. In grapheme-color synesthesia, when an individual sees certain numbers or letters, he experiences corresponding colors. Recent studies have shown that human synesthesia may result from hyperconnected neurons in the human brain. Some researchers \cite{mattingley2001unconscious,beeli2005coloured,frangeul2016cross} proposed that if the neurons responsible for transmitting one sensory signal are activated by a real stimulus, they may then trigger spontaneous cross-activation of neurons responsible for transmitting another sensory signal. That is, the brain areas responsible for processing different types of senses have similar structures and are interconnected. 
	\subsection{Synesthesia of Machines}
	\subsubsection{Machine Senses}
	Recent advances in wireless communications have greatly propelled the development of the Internet of Things (IoTs) with ubiquitous sensing capabilities. Enabled by the great potential of 6G, future 6G-IoT applications \cite{niyatoIoT} have received much attention from both academia and industry, such as vehicular IoT and autonomous driving, UAVs, satellite IoT, and industrial IoT.  Various devices therein continuously sense the surrounding environment and accomplish various tasks. Just like human senses the surrounding environment through a variety of different organs, various sensors and communication devices equipped on these IoT devices can also collect massive and diverse environment information. Naturally, we analogously refer to them as ``\textit{machine senses}", including but not limited to LiDAR, radar, and RGB-D camera, as illustrated in Fig.~\ref{SoH}. Specifically, radar detects the surrounding environment and collects the clusters through electromagnetic waves, which contain the target's information (e.g., position, radar cross section, and Doppler information). LiDAR transmits laser pulses to the surrounding environment and then obtains dense point clouds, where each point contains the 3D coordinate, reflection intensity, and echo times. The communication module modulates data to the electromagnetic wave and transmits it to the target users through antennas. During this process, the communication module constantly detects and estimates the electromagnetic propagation environment. The wireless channel data and radar echoes in the wireless electromagnetic wave propagation domain, images and LiDAR point clouds in the visible or infrared light spectra domain all capture different properties of the physical environment from different domains. 
	
	\subsubsection{Definition}
\begin{figure}[!t]
	\centering
	\includegraphics[width=1\linewidth]{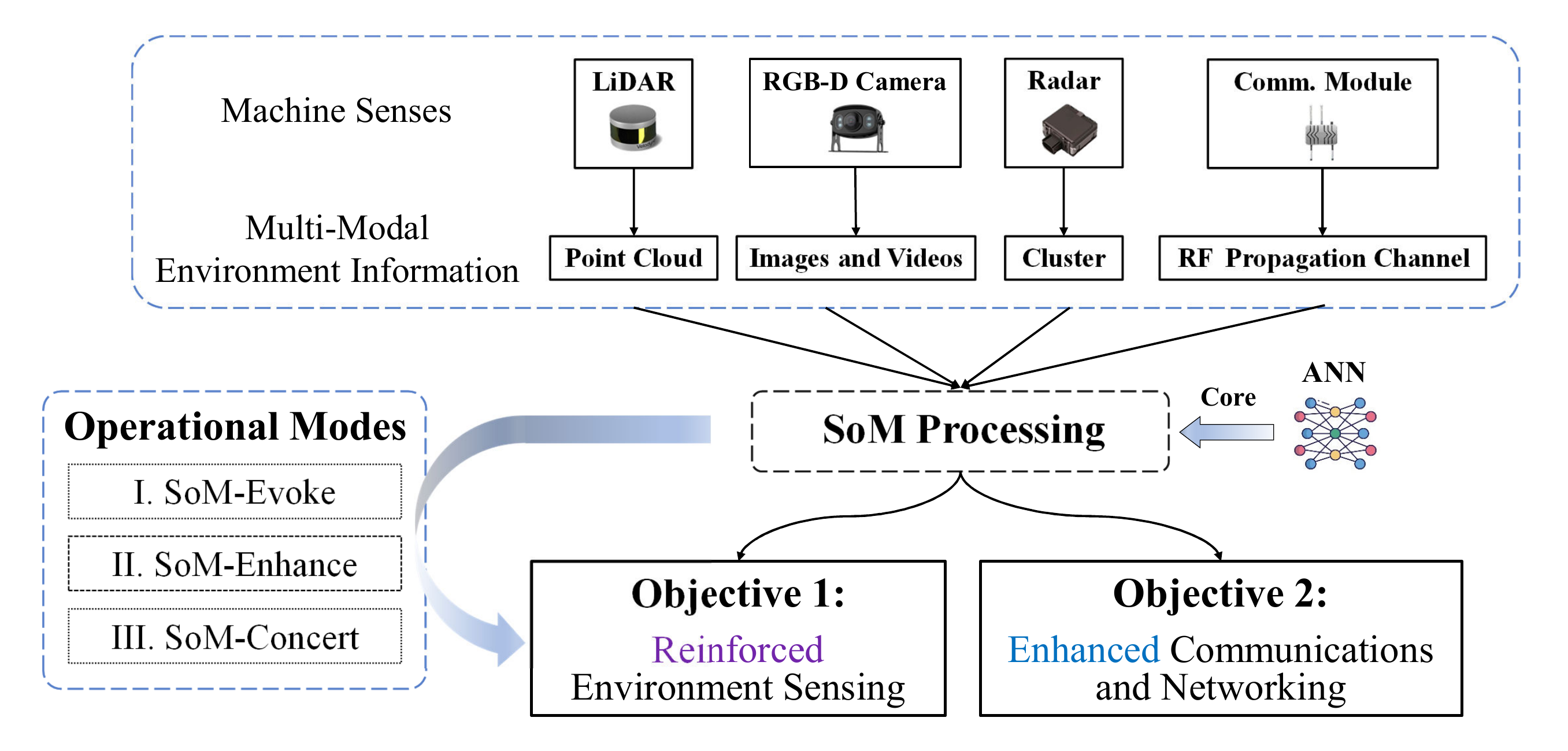}
	\caption{Illustration of SoM defined in this paper.}
	\label{gsom}
\end{figure}

Due to the similarity between the human senses and machine senses, it is possible to generalize the concept of human synesthesia to the field of machines, as illustrated in Fig.~\ref{SoH}. We first define \textit{SoM} as the synesthesia processing of various environment information collected by machine senses, which can lead to reinforced environmental sensing, as well as enhanced communications and networking functionality, as schematically illustrated in Fig.~\ref{gsom}. 

\subsubsection{Operational Modes}
SoM can operate in three distinct modes: SoM-evoke, SoM-enhance, and SoM-concert. Note that, different from human synesthesia which often involves no more than two senses, SoM can potentially involve multiple machine senses.

\begin{figure*}[!t]
	\centering
	\includegraphics[width=1\textwidth]{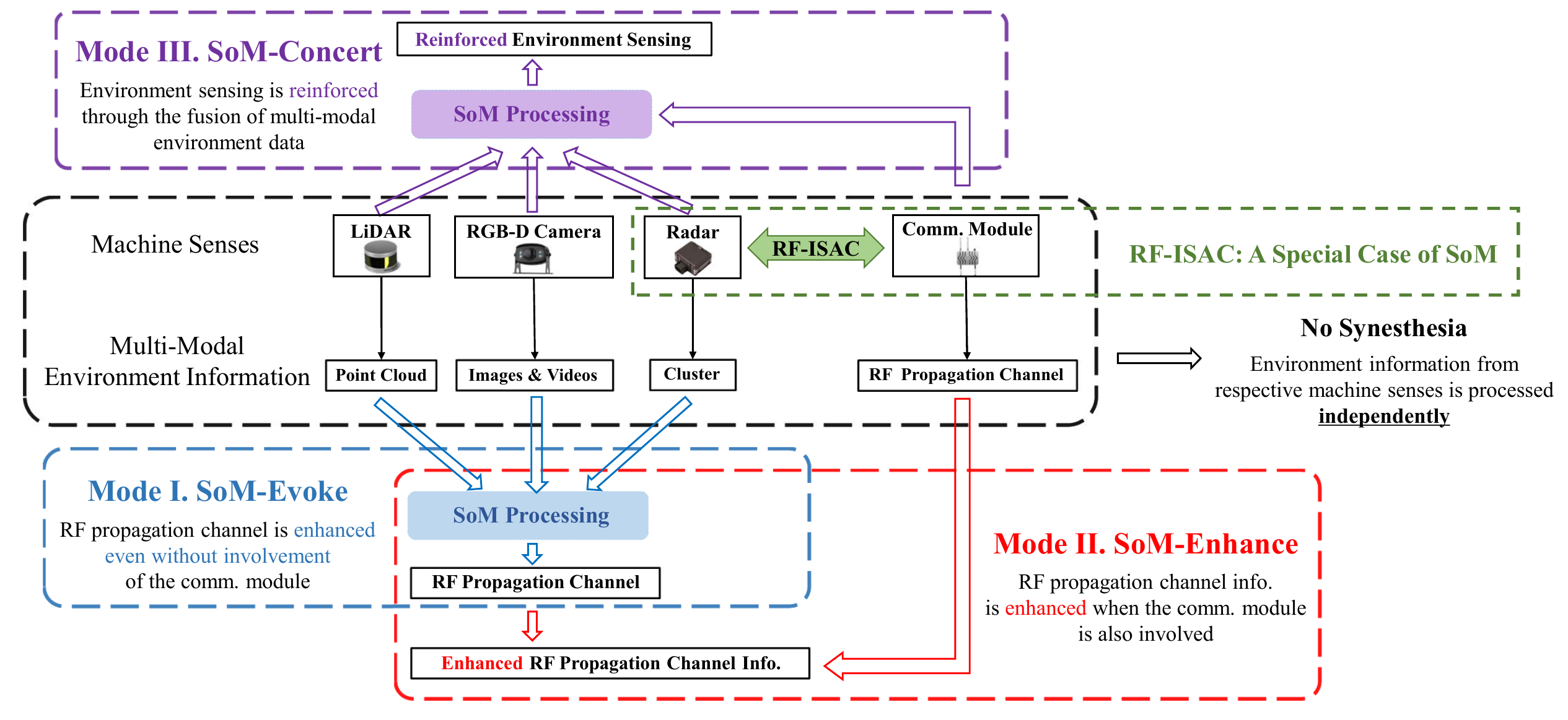}
	\caption{Illustration of no synesthesia case, RF-ISAC, and three distinct operational modes of SoM defined in this paper.}
	\label{som}
\end{figure*}

Specifically, if certain environment information can be enhanced even when the corresponding machine sense is inactive by jointly processing the environment information obtained from one or more other sensors, then  \textit{\textbf{SoM-evoke}} is taking place, as shown in Fig.~\ref{som}. We argue that SoM-evoke for communication functionality enhancement is of greater value in practical applications. That is, the performance of communication system is directly enhanced through SoM processing of various environment information when the communication module is not working. For example, CSI and channel statistics can be directly obtained based on multi-modal sensory data, based on which the channel quality can be determined and the link state can be predicted. Then, the link with an excellent state is selected for transmission, thereby improving the throughput of wireless network.

In practice, the communication module in Fig.~\ref{som} can operate in the regular mode to gather channel information.  Under this circumstance, the SoM processing of various environment information can bring significant enhancement to the communication functionality while markedly reducing the communication module's overhead. We refer to such a process as \textbf{\textit{SoM-enhance}}, as schematically illustrated in Fig.~\ref{som}. For example, the precoding design and mmWave beam alignment of the communication systems can be optimized by jointly processing the 3D coordinates of the target provided by the LiDAR and the velocity information of the target obtained from the radar.

Furthermore, the fusion of various environment information can boost the specific sensing tasks, thus reinforcing the overall environment sensing capability through SoM processing, despite that the source environment information per se is not optimized. In that case, \textbf{\textit{SoM-concert}} is taking place, as illustrated in Fig.~\ref{som}. For example, the dense semantic information in RGB images and the precise structural information in LiDAR point clouds can jointly improve the performance of object detection and classification, thereby enhancing the reliability of environment sensing.

It is noteworthy that the extensively studied RF-ISAC is a special case of SoM. RF-ISAC integrates radar sensing and communication functionalities through the unified waveform and hardware, in an attempt to improve resource utilization and reduce hardware size, as illustrated in Fig.~\ref{som}. However, RF-ISAC realizes the integration of merely two machine senses in RF format with no guarantee of performance gain for them, especially in dynamic scenarios. 

\subsection{SoM Processing}
	SoM processing of multi-modal environment information is the core of realizing reinforced environmental sensing, as well as enhanced communications and networking functionality, as illustrated in Fig.~\ref{som}. However, traditional model-based methods, such as signal processing algorithms in communication systems, face challenges in achieving such objectives. Traditional model-based methods have limited capability to incorporate other information (such as visual information in RGB images) and effectively fuse them. They also typically make assumptions and simplifications to simplify computations or obtain analytical solutions, which contradicts the complex and nonlinear nature of intelligent integration between multi-modal sensing and communications. In contrast, artificial neural network (ANN) methods can effectively address these issues.  Serving as one of the main tools used in machine learning, ANNs are brain-inspired systems which mimic the way that biological neurons transmit signal to one another. They are excellent tools for finding features that are far too complicated and numerous for conventional methods to model. Furthermore, multi-modal environment information collected through various machine senses exhibits inherent differences in data structures, semantic information, and operating frequency bands. For instance, RGB images consist of the intensity values of the three primary colors, LiDAR provides the coordinates and intensity of the reflection points, and communication devices obtain signal propagation characteristics of RF environments. Given such natural challenges, the core task of SoM processing is to learn complex feature representations from such diverse multi-modal data and to model the extremely complex and nonlinear relationships among them, thereby achieving mutual assistance and enhancement among them. Confronted with such a vision, ANN naturally becomes a necessary and fundamental tool. Therefore, SoM processing can be considered to be carried out in an intelligent way as ANN serves as the fundamental tool. However, it is noteworthy that the so-termed intelligent way does not refer to solely relying on large amounts of training data to solve problems, treating ANNs as black boxes, and selecting proper networks based on practical experiences, i.e., data-driven approaches. In contrast, we argue that the intrinsic requirements for utilizing ANNs in SoM processing are reasonably designing network architectures and setting network parameters based on theoretical research of communication and sensing systems and making the designed networks interpretable for specific applications, that is, model-enhanced data-driven approaches.

	It is also worth noting that while ANNs are utilized as a fundamental tool in SoM, other machine learning approaches need to be incorporated to address specific challenges that ANNs may encounter in certain cases. Deep reinforcement learning can be used to address the issue of requiring extensive labeled datasets in ANNs.  To improve the generalization of ANN models and lower the requirement for large amounts of task-specific training data, meta-learning can be considered as an efficient tool for its ability of ``learning to learn". It aims to extract general information about the learning process on individual tasks in order to improve adaptability to novel tasks. Federated learning can be used to deal with the data privacy issue in ANN models by training them using distributed data across multiple devices, without centralizing the data in a single location. Currently, researchers have applied the above approaches to SoM-related research, such as fast beamforming in mMIMO \cite{jsac_DRL_mimo} and beam selection \cite{gu2023meta,chowdhuryFLASH,chowdhurymultimodal}.
	
	Under the guidance of the above principles, SoM is expected to fulfill the intelligent multi-modal sensing-communication integration in a trustworthy and interpretable manner with ANNs serving as the fundamental tool.	
	\subsection{Objectives and Challenges of SoM}
	One may think that the study of SoM has no basis as the underlying mechanism of its counterpart, human synesthesia, remains unknown. However, we argue that the study of SoM is feasible due to the following reasons. Firstly, one of the difficulties that human synesthesia research faces is that the information collected by various human senses is almost impossible to express, understand, and process explicitly. However, the operating principles of various machine senses are known, and the corresponding sensory data obtained is much easier to understand and process. The availability and interpretability of multi-modal environment data collected by machine senses make it possible to study various research topics of SoM. Secondly, human synesthesia depends on subjective descriptions and cannot be objectively presented in the form of data or observable phenomena. {\color{black}However, the concept of SoM is derived from a large number of scientific observations and research}, providing relevant research with an objective basis and authentication methods. Thirdly, human synesthesia is closely related to the human brain and may be caused by hyperconnected neurons. The extremely complex biological structure of the human brain makes it challenging to understand its operating principles. However, the architecture and mechanism of the ANNs that SoM relies on have been extensively studied and have a sufficient theoretical basis. In summary, different from its human counterpart, SoM can be studied based on the availability of interpretable data, implementation tools, and a large number of existing scientific observations. All these render SoM a feasible and promising candidate technology in the B5G/6G era.
	
	Given the above premises, SoM is expected to attain the following objectives, which we believe hold significant application values to B5G/6G networks. Firstly, enhance the RF propagation channel information through SoM processing of multi-modal sensing information without the communication module's involvement. For example, SoM-evoke has the potential of directly obtaining large-scale fading characteristics of channels through multi-modal sensing. On this basis, the received signal power at different positions within the coverage range of the cell is estimated to organize the distribution of macro/micro sites to reduce energy consumption and improve communication capacity \cite{ruan2019evoke}. Secondly, when the communication module operates with limited functionality, SoM aims to bring significant enhancements to it by inferring useful electromagnetic environment characteristics from multi-modal sensory data, without consuming additional communication resources (e.g., spectrum, power, and time). For example, SoM has the potential of reducing the high communication overhead and latency caused by conventional channel estimation methods in mmWave communication systems by directly obtaining the receiver's position and motion state information from multi-modal sensory data \cite{xu2021deep}. Thirdly, the reinforced and expanded environment sensing capability achieved by integrating the multi-modal environment information collected by various machine senses happens to be the goal of SoM as well. For example, by utilizing reliable communication links, multi-modal sensory data from multiple vehicles can be shared and fused to achieve beyond-view-range sensing \cite{early3zheng2022multi, early7}, which has the potential to augment vehicles' sensing capability.
	
	However, the realization of the above objectives faces unprecedented challenges. Multi-modal environment data collected by various machine senses has distinct data structures and semantic information. Due to the huge gap among the various machine senses in terms of functionality, operation principles, and frequency bands, the SoM processing has to reveal the mapping relationship among the various machine senses to ensure that the subsequent relevant research has a theoretical basis. For example, radar and communication modules rely on electromagnetic waves, but RGB-D cameras and LiDAR rely on visible or infrared light. The multi-modal environment data collected from distinct domains has to be processed and transformed into a decipherable format for specific purposes based on the mapping relationship. RF-based sensors that operate at different bands or bandwidths also sense different environmental features. When the radar operates at $77$ GHz and the communication system operates at $28$ GHz, the SoM processing between radar and communications relies on the mapping relationship between the electromagnetic environments at different frequency bands. The basis of revealing the above mapping relationships is the aligned multi-modal sensory data and wireless channel data. However, such datasets are rarely researched and constructed, posing a huge challenge for SoM research. In Section \ref{Dataset}, we will survey existing datasets for sensing and communication research, and analyze their limitations. Moreover, we will introduce a simulation dataset containing multi-modal sensory data and wireless channel data. Afterward, we will elaborate on how such a dataset can support SoM research.
	
    \subsection{Current Research Status of SoM}
\begin{figure}[!t]
		\centering
		\includegraphics[width=1\linewidth]{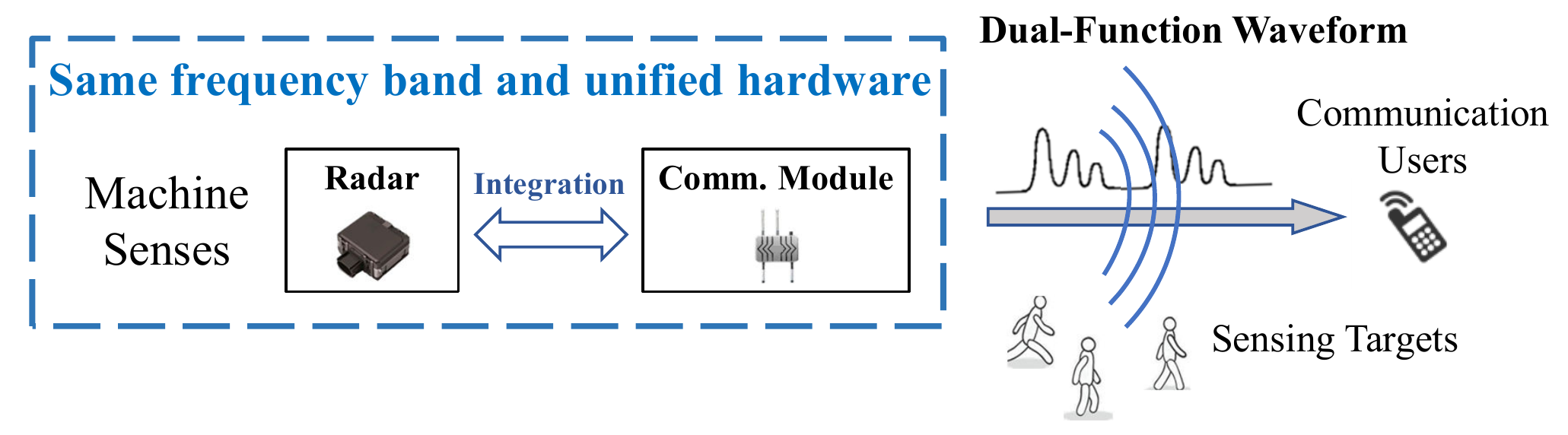}
		\caption{Illustration of RF-ISAC, which is a special case of SoM.}
		\label{som4}
	\end{figure}

	Currently, the research on SoM is mostly limited to RF-ISAC. As discussed in Section \ref{Background}, RF-ISAC has been studied extensively due to the similar operating frequency band, hardware architecture, and signal processing techniques of radar sensing and communications. {\color{black}The research on RF-ISAC focuses on dual-function waveform designs \cite{kumari2017ieee, grossi2017opportunistic,liu2017multiobjective}, signal processing \cite{li2017joint,zheng2017adaptive}, and fundamental limits \cite{nartasilpa2018communications}.} As discussed above, RF-ISAC can be regarded as a special case of SoM, as illustrated in Fig.~\ref{som4}. The reasons are as follows: i) RF-ISAC aims at using a unified waveform to simultaneously realize two functionalities. That is, only two machine senses (radar and communications) are considered, ii) the radar and communication functionalities operate at the same frequency band, which only achieves a limited degree of integration gains. Despite the potential advantages like higher resource utilization, less power consumption, and smaller hardware size, RF-ISAC still has limitations. Firstly, the SoM processing of RF-ISAC is marginal. In fact, the dual-function waveform is mostly realized only by modulating the communication data on  radar waveform \cite{saddik2007ultra,hass2016dual,ma2021frac} or detecting the target through the characteristics of communication waveform (such as OFDM) \cite{commu1}. Secondly, there is no guarantee of the respective performance enhancement of radar and communication functionalities. Under certain circumstances, the dual-function waveform may even cause performance degradation. In summary, radar sensing and communications are only integrated at the hardware level and their respective functionalities are not necessarily enhanced.
 
	As of today, the research on SoM-evoke is still an open field since it requires an in-depth understanding of the complex mapping relationship between multi-modal sensing and communications, whereas this still lacks research. Only a few studies have made a preliminary attempt at SoM-enhance, targeting relatively limited research topics (e.g., beam prediction, handover, and blockage prediction). The research on SoM-enhance-based improvement that multi-modal environment information brings to the communication module is still not systematic and thorough, the study of the whole area is still at an early stage. The fundamental reason for the infancy of SoM-evoke and SoM-enhance is the lack of datasets with aligned multi-modal sensory data and wireless channel data. Without such datasets, it is hardly possible to systematically study the mapping relationships between multi-modal sensing and communications. Research on SoM-concert has already received widespread attention, specifically in the area of multi-modal sensor fusion and multi-vehicle cooperation, as SoM-concert currently concentrates on fusing multi-modal sensory data with communications serving as a media of information transmission and is not an in-depth integration of sensing and communications. It has been proven that vehicles equipped with multi-modal sensing and communication devices can improve their sensing capabilities, including range, accuracy, and reliability, through SoM processing of the collected multi-modal environment data.
	
	\vspace{1em}

	\subsection{Key Takeaways}
	
	In this section, we introduce the unified framework that we establish for the intelligent multi-modal sensing-communication integration, SoM, and detail its design paradigm. The major features of SoM and key takeaways learned are summarized as follows:
	\begin{itemize}
		\item
		SoM fully considers the role of \textbf{multi-modal sensing} except for RF-based sensing and expands its research objectives to improve the performance of communication and sensing systems from much more aspects compared to the existing RF-ISAC techniques.
	\end{itemize}
	
	\begin{itemize}
		\item
		SoM processing of multi-modal environment data is the core of realizing SoM's goals and the specific implementation methods are classified into three operational modes, i.e., \textbf{SoM-evoke}, \textbf{SoM-enhance}, and \textbf{SoM-concert}.
	\end{itemize}
	
	\begin{itemize}
		\item
		Different from RF-ISAC which mostly relies more on traditional signal processing techniques and mathematical optimization methods, SoM attaches more importance to \textbf{ANNs} to unveil the complex mapping relationships between multi-modal sensing and communications, and achieve multi-layer and multi-objective performance enhancement in the model-enhanced data-driven manner.
	\end{itemize}
	
	\begin{itemize}
		\item
		SoM is not only oriented to \textbf{generic scenarios}, but also particularly suitable for tackling challenges arising from \textbf{more challenging dynamic scenarios}.
	\end{itemize}

	 In summary, the goal of extensively studied RF-ISAC is merely to unify radar sensing and communication functionalities, and to pursue tradeoffs between them. SoM aims to optimize the overall communication and sensing systems by enhancing the performances of various technical components therein. Unlike RF-ISAC, the integration between multi-modal sensing and communications can be conducted in more diverse ways and offers performance gains for diverse system applications. In this regard, SoM establishes a systematic framework of intelligent multi-modal sensing-communication integration, rather than being a new technology without any research foundation. In the following sections, we will provide a detailed review and discussion of the enabling technologies, current research status, challenges, and future directions of SoM.

	\section{Prerequisite of SoM Research: The Mixed Multi-Modal (MMM) Datasets}
	\label{Dataset}	
	
	A high-quality dataset is the key to the successful design and application of many systems. Take the communication system design as an example. The design of communication module is closely related to the channel characteristics, which are analyzed from a large amount of wireless channel data. Likewise, the design principle of the sensing system is to accurately perceive the surrounding environment by utilizing data from multiple sensors, which is also the basis of performance measurement. The vision of SoM is to realize reinforced environmental information as well as enhanced communications and networking functionality, with the SoM processing of various environment information serving as the core. Consequently, multi-modal environment information is a prerequisite of SoM research. Moreover, as noted in Section \ref{Intro of SoM}, ANN is at the core of SoM to deal with multi-modal sensory data, unveil the mapping relationships, and subsequently realize the intelligent multi-modal sensing-communication integration in a trustworthy and interpretable manner. A high-quality dataset is the lifeblood of the ANN's learning capacity and plays a pivotal role in ensuring the accuracy of its outputs. Consequently, SoM research is inseparable from high-quality training data. {\color{black}However, real-world datasets cannot meet the requirement of flexibly customizing desired scenarios attributed to the labor and cost concerns.} {\color{black}Different from real-world datasets, simulation datasets achieve a decent trade-off between complexity and quality by advanced simulation platforms, which have been validated for accuracy through measurement campaigns \cite{sss-h1,sss-h2,sss-h3}. Therefore, they have received extensive attention and possess the ability to support SoM research.} In this section, we develop discussions around the datasets for sensing and communication research, as well as the desirable datasets for SoM research. Afterward, we introduce a simulation MMM dataset constructed for SoM research.
	
	\subsection{Existing Datasets for Sensing and Communication Research}
	\label{class}
\begin{figure*}[!t]
			\centering
			\includegraphics[width=1\linewidth]{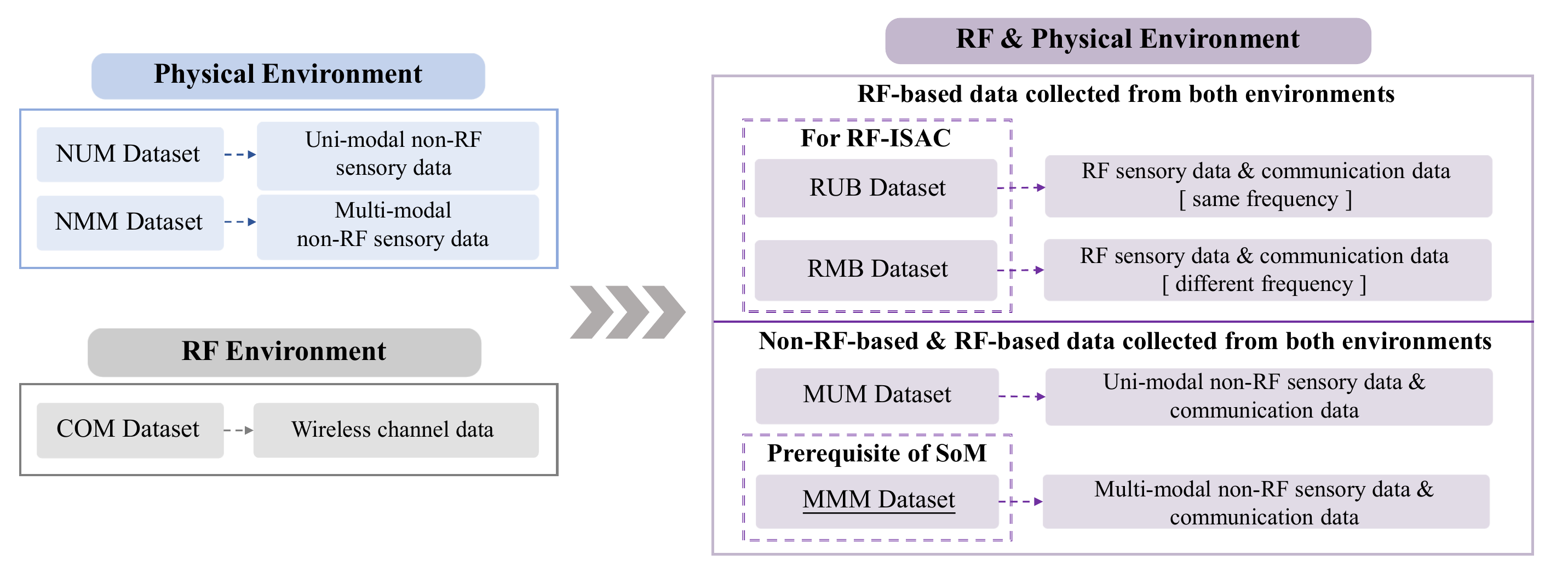}
			\caption{Classification of existing datasets for sensing and communication research.}
		\label{diagramofdataset}
    \end{figure*}

	In this subsection, we classify the datasets for sensing and communication research based on the type of data contained in the datasets, \textcolor{black}{as shown in Fig.~\ref{diagramofdataset}.}
	
	 \subsubsection{Dataset Classification} 
  \label{classification}
       \
       \newline
	\noindent	\underline{\textbf{Communications Only (COM)}:}	Dataset that only contains wireless channel information in the electromagnetic wave propagation environment, such as CIR, direction of departure (DoD), and direction of arrival (DoA). 
	
	\noindent	\underline{\textbf{Non-RF Uni-Modal (NUM)}:}	Dataset that only contains uni-modal non-RF sensory data collected from the physical environment.
	
	\noindent	\underline{\textbf{Non-RF Multi-Modal (NMM)}:} Dataset that only contains multi-modal non-RF sensory data collected from the physical environment.
	
	\noindent	\underline{\textbf{RF-only Uni-Band (RUB)}:} Dataset that contains the wireless channel data as well as RF sensory data collected from electromagnetic wave propagation environment and physical environment, where the communication system and the radar sensing system operate at the same frequency band.
	
	\noindent	\underline{\textbf{RF-only Multi-Band (RMB)}:} Dataset that contains the wireless channel data as well as RF sensory data collected from electromagnetic wave propagation environment and physical environment, where the communication system and the radar sensing system operate at different frequency bands.
	
	\noindent	\underline{\textbf{Mixed Uni-Modal (MUM)}:} Dataset that contains the wireless channel data as well as uni-modal sensory data collected from electromagnetic wave propagation environment and physical environment.
	
	\noindent	\underline{\textbf{Mixed Multi-Modal (MMM)}:} Dataset that contains the wireless channel data as well as multi-modal sensory data collected from electromagnetic wave propagation environment and physical environment.

 \subsubsection{Existing Datasets}
	\label{existing data}
	
	In order to enable the readers to have a grasp of the current status of datasets for sensing and communication research, we give the following examples that cover the seven types of datasets defined in Section \ref{classification}.
	
	\noindent	\underline{\textbf{COM Datasets}:}
	\begin{itemize}
		\item DeepMIMO \cite{alkhateeb2019deepmimo}: The DeepMIMO dataset aims at promoting the application of machine learning (ML) in mmWave/mMIMO research. The DeepMIMO dataset is constructed based on accurate ray-tracing data and is parameterized. Therefore, the users have the ability to modify the channel parameters to suit their needs in a flexible manner.
		
		\item 5GMdata \cite{klautau20185g}: The 5GMdata dataset presents a framework that combines a vehicle traffic simulator and a ray-tracing simulator to construct scenarios where transceivers and objects are moving.
	\end{itemize}
	
	\noindent	\underline{\textbf{NUM Datasets}:}
	\begin{itemize}	
		\item  LiDARsim \cite{manivasagam2020lidarsim}: The LiDARsim dataset is able to produce realistic simulations of LiDAR point clouds with the aid of real-world data and a deep neural network (DNN). It is also enhanced by the intensity simulation and conditional generative modeling for different weather conditions.
	\end{itemize}
	
	\noindent	\underline{\textbf{NMM Datasets}:}
	\begin{itemize}		
		\item  V2X-Sim \cite{li2022v2x}: V2X-Sim is a simulated multi-agent sensing dataset where the RSU and multiple vehicles are equipped with multi-modal sensors. It also provides diverse ground truths for multi-agent and multi-modal sensing tasks in the field of autonomous driving. Furthermore, it builds an open-source testbed for evaluating collaborative sensing algorithms. 
		
		\item  OPV2V \cite{xu2022opv2v}: The OPV2V dataset is a large-scale simulated benchmark dataset for distinct vehicle-to-vehicle (V2V) sensing algorithms where the connected automated vehicles are equipped with cameras, LiDAR, and GPS/inertial measurement unit (IMU) sensors.
		
		\item  SHIFT \cite{sun2022shift}: The SHIFT dataset is a multi-task synthetic dataset developed for autonomous driving. It utilizes realistic discrete and continuous shifts in terms of several driving safety-related parameters to simulate the sensory data under different times of day, the strength of cloud, rain, and fog.
		
	\end{itemize}
	
	\noindent	\underline{\textbf{RUB Datasets}:}
	\begin{itemize}	
		\item WALDO \cite{waldo}: The WALDO dataset is constructed for indoor human localization by actual measurement. In WALDO, IEEE 802.11ay PHY software and the multi-static radar are used to obtain channel data and radar data. 
	\end{itemize}

	\noindent	\underline{\textbf{RMB Datasets}:}
	\begin{itemize}
		\item Radar-COM \cite{ali2020passive}: Ali \textit{et al.} constructed a simulation dataset in \cite{ali2020passive} to verify the proposed radar-assisted mmWave beamforming scheme, where the frequency modulated continuous wave (FMCW) radar and communication systems operate at different frequency bands.
	\end{itemize}
	
	\noindent	\underline{\textbf{MUM Datasets}:}
	\begin{itemize}	
		\item  LiDAR-COM \cite{klautau2019lidar}: Klautau \textit{et al.} \cite{klautau2019lidar} constructed a dataset that contains LiDAR data and ray-tracing simulations, where the vehicles are equipped with LiDAR equipment and communication receiver.
	\end{itemize}	
	\noindent	\underline{\textbf{MMM Datasets}:}
	\begin{itemize}	
		\item Vision-Wireless (ViWi) \cite{alrabeiah2020viwi}: The ViWi datasets provide co-existing visual data and parameterized wireless channel data. The sensors are only mounted on the RSU in ViWi datasets. Moreover, they do not contain wireless channel data between vehicles, which makes them unable to support research on V2V communications.
		{\color{black}
			\item e-FLASH \cite{gu2022multimodality}: The e-FLASH dataset is a large multi-modal dataset consisting of camera image, GPS, and LiDAR data supported with RF ground truth, which is developed for facilitating research on mmWave band communication. It contains seven unique real-world line-of-sight (LoS) and NLoS vehicular network scenarios.}	
		\item DeepSense 6G \cite{DeepSense}: The DeepSense 6G dataset is a large-scale real-world multi-modal dataset with communication and sensory data. The DeepSense 6G dataset plans to contain more than 30 scenarios at different times of the day. However, constructing datasets through actual measurements brings high costs, and consumes a lot of labor and material resources when increasing the complexity of scenarios.
	\end{itemize}	
	
	For clarity, Table \ref{tab4} summarizes the basic information of several existing datasets for sensing and communication research. Note that our work listed in Table \ref{tab4} will be introduced in Section \ref{ammm}.\footnote{The proposed M$^3$SC datasets can be accessed and applied for through: http://pcni.pku.edu.cn/cn/dataset\_1.html. The specific data types provided are available in Table \ref{tab4}.} Considering the advantage of simulation datasets, it can be observed from Table \ref{tab4} that many simulation datasets have been constructed. For example, the Radar-COM  dataset in \cite{ali2020passive} and the DeepMIMO dataset in \cite{alkhateeb2019deepmimo}  utilize the ray-tracing-based Wireless InSite$^\circledR$ \cite{WI}, the ViWi dataset in \cite{alrabeiah2020viwi} utilizes the popular game engine Blender$^\mathrm{TM}$, and the OPV2V dataset in \cite{xu2022opv2v} utilizes the precise CARLA simulator \cite{CARLA}. Note that the aforementioned datasets are of high reliability and accuracy due to the utilization of precise simulation platforms, whose reliability and accuracy have been validated by available measurement campaigns in  \cite{sss-h1,sss-h2,sss-h3}.
	
	
	\begin{table*}[!t]
		\footnotesize
		\renewcommand\arraystretch{1.5} 
		\caption{Comparison of Existing Datasets for Sensing and Communication Research}
		\centering
		\label{tab4}
		\resizebox{\textwidth}{!}{
			\begin{tabular}{c|c|c|c|c|c|c|c|c|c|c|c}
				\toprule[0.35mm]
				\multirow{2}{*}{\textbf{Datasets}} &    \multicolumn{5}{c|}{\textbf{Sensory Data}} & \multirow{2}{*}{ \textbf{\makecell[c]{\textbf{Wireless}\\\textbf{data}}}}  & \multirow{2}{*}{ \textbf{\makecell[c]{\textbf{Weather}\\\textbf{condition}}}}  & \multirow{2}{*}{ \textbf{\makecell[c]{\textbf{Time}\\\textbf{of day}}}} & \multirow{2}{*}{\textbf{Type}} & \multirow{2}{*}{\textbf{\makecell[c]{\textbf{Data}\\\textbf{sources}}}}  & \multirow{2}{*}{\textbf{Scalability}} \\ 
				\cmidrule[0.15mm]{2-6}
				&    \makecell[c]{RGB\\image} & LiDAR &  \makecell[c]{Depth\\map} & IMU & Radar & & & & & &\\
				\midrule[0.15mm]   
				\makecell[c]{DeepMIMO \cite{alkhateeb2019deepmimo}} &$\times$	&$\times$	&$\times$	&$\times$	&$\times$ &  \checkmark & $\times$ & $\times$ & COM & Simulation & High	\\ 
				\midrule[0.15mm] 
				\makecell[c]{5GMdata  \cite{klautau20185g}} &$\times$	&$\times$	&$\times$	&$\times$	&$\times$ &  \checkmark & $\times$ & $\times$ & COM & Simulation & High	\\ 
				\midrule[0.15mm]
				\makecell[c]{LiDARsim  \cite{manivasagam2020lidarsim}} & $\times$	& \checkmark	&  $\times$ &  $\times$	& $\times$ &  $\times$ & \checkmark & $\times$ & NUM & Simulation & High	\\ 
				\midrule[0.15mm]
				\makecell[c]{V2X-Sim  \cite{li2022v2x}} & \checkmark & \checkmark & \checkmark & \checkmark & $\times$  &   $\times$ & $\times$ & $\times$ & NMM & Simulation & High \\
				\midrule[0.15mm]
				\makecell[c]{OPV2V \cite{xu2022opv2v}} &\checkmark	& \checkmark	& \checkmark & \checkmark	& $\times$ &  $\times$ & $\times$ & $\times$ & NMM & Simulation & High	\\ 
				\midrule[0.15mm]
				\makecell[c]{SHIFT  \cite{sun2022shift}} &\checkmark	& \checkmark	& \checkmark & \checkmark	& $\times$ & $\times$  & \checkmark & \checkmark  & NMM & Simulation & High	\\ 
				\midrule[0.15mm]
				\makecell[c]{WALDO  \cite{waldo}} & $\times$	&$\times$	&$\times$	&$\times$	& \checkmark &  \checkmark & $\times$ & $\times$ & RUB & Real World & Low	\\ 
				\midrule[0.15mm]
				\makecell[c]{Radar-COM  \cite{ali2020passive}} &$\times$	& $\times$	& $\times$& $\times$	& \checkmark & \checkmark & $\times$  & $\times$ & RMB & Simulation & High	\\ 
				\midrule[0.15mm]
				\makecell[c]{LiDAR-COM \cite{klautau2019lidar}} &$\times$	& \checkmark	& $\times$ & $\times$	& $\times$	 &  \checkmark & $\times$	& $\times$ & MUM &Simulation	& High\\ 
				\midrule[0.15mm]
				\makecell[c]{ViWi \cite{alrabeiah2020viwi}} &\checkmark	& \checkmark	& \checkmark & $\times$	& Limited &  \checkmark & $\times$ & $\times$ & MMM 	& Simulation  & High 	\\ 
				\midrule[0.15mm]
				{\color{black}\makecell[c]{e-FLASH \cite{gu2022multimodality}}} & {\color{black}\checkmark}	& {\color{black}\checkmark}	& {\color{black}$\times$} & {\color{black}$\times$}	& {\color{black}$\times$} &  {\color{black}\checkmark} & {\color{black}\checkmark} & {\color{black}$\times$} &  {\color{black}MMM} 	& {\color{black}Real World } & {\color{black}Low} 	\\ 
				\midrule[0.15mm]
				\makecell[c]{DeepSense 6G  \cite{DeepSense}} &\checkmark	& \checkmark  & $\times$ & $\times$	& \checkmark	 & \checkmark & \checkmark & \checkmark & MMM 	& Real World  & Low	\\
				\midrule[0.15mm]
				\textbf{Our work: M$^3$SC \cite{M3C} } &\checkmark	&\checkmark	&\checkmark	 & $\times$	& \textcolor{black}{\checkmark} & \checkmark & \checkmark	& \checkmark & MMM  &Simulation & High \\ 
				\bottomrule[0.35mm]
				
		\end{tabular}}
	\end{table*}

	\subsection{Necessary Properties of the Desirable MMM Datasets}
	\label{necessary}
	According to the definition and objectives of SoM, we point out that the desirable MMM datasets for SoM research should possess the following properties:
	
	\begin{itemize}
		\item \textbf{Aligned multi-modal sensory data and wireless channel data.} SoM processing of various environment information is the core of realizing the mutual promotion between multi-modal sensing and communication functionalities. Therefore, only datasets that contain aligned multi-modal sensory data and wireless channel data can meet the basic demand of SoM research. 
		
	\end{itemize}
	
	\begin{itemize}
		
		\item \textbf{Multi-modal environment data collected from different agents.} Multi-modal environment data collected from different agents ensures that the scope of SoM research in a certain scenario is comprehensive enough. Moreover, such data represents the characteristics of a certain scenario more comprehensively, which improves the performance and scalability of SoM-based applications. Additionally, it provides flexibility for SoM research towards multiple targets.
		
	\end{itemize}
	
	\begin{itemize}
		\item \textbf{High-fidelity scenarios close to reality.} Scenarios containing details close to reality can provide more realistic simulation data, such as lighting, weather conditions, object material, and model complexity. Such high-quality and realistic data is conducive to revealing the mapping relationship between multi-modal sensing and communications as accurately as possible, thereby making SoM research more practical.
		
	\end{itemize}
	
	{\color{black} To construct desirable MMM datasets, there are some available simulation platforms, such as Wireless InSite$^\circledR$ \cite{WI}, CARLA \cite{CARLA}, AirSim \cite{AirSim}, WaveFarer$^\circledR$ \cite{WaveFarer}, and Simulation of Urban MObility (SUMO) \cite{SUMO}. Specifically, Wireless InSite$^\circledR$ utilizes ray-tracing technology based on geometrical optics and the uniform theory of diffraction to obtain accurate channel information. CARLA  covers various high-fidelity scenarios, such as urban cities, highways, and rural areas, and accurately models the mobility of vehicles. In AirSim, the simulation scenarios are constructed by utilizing advanced 3D modeling software, which can accurately render intricate structures. WaveFarer$^\circledR$ is an advanced mmWave radar simulation software, which can efficiently and accurately analyze radar interactions with structures, targets, and other features. SUMO can generate realistic and accurate traffic scenarios by simulating various aspects of urban mobility, including vehicles, pedestrians, public transport, and traffic signals.
    However, despite the availability of these simulation platform options, there is still a large gap between the existing datasets and the desirable MMM datasets. On one hand, the existing MMM simulation dataset has relatively rough scenario rendering and lacks weather/time simulation. On the other hand, although a few real-world datasets can meet the demands of desirable MMM datasets, they consume a lot of labor and material resources, and cannot flexibly and timely construct the target scenarios.} In summary, MMM simulation datasets that support flexibly building any target scenario with superior rendering effects are in urgent need.

	\subsection{M$^3$SC: A Generic Dataset for Intelligent Multi-Modal Sensing-Communication Integration}
	\label{ammm}
	
	As discussed in Section \ref{necessary}, the existing simulation datasets cannot meet the requirements for SoM research. Aiming at filling this gap, we develop a simulation dataset named M$^3$SC \cite{M3C} in a dynamic VCN crossroad scenario. The M$^3$SC dataset contains aligned multi-modal sensory data and wireless channel data, and considers the mMIMO and mmWave communications. {\color{black}To construct the M$^3$SC dataset, we utilize efficient and high-fidelity simulation platforms, i.e., Wireless InSite$^\circledR$, AirSim, and WaveFarer$^\circledR$. To be specific, Wireless InSite$^\circledR$ exploits the ray-tracing technology to generate accurate channel information. AirSim simulates multi-modal sensory data based on simulation scenarios exported from advanced 3D modeling software. WaveFarer$^\circledR$ can accurately model radar interactions with structures, targets, and other features. In fact, there are several other alternative simulation platforms available, such as CloudRT \cite{CloudRT}, CST Studio Suite \cite{CST}, and CARLA \cite{CARLA}. However, these simulation platforms have their own limitations for constructing the M$^3$SC dataset. CloudRT, which can collect wireless channel information, relies on a cloud-based GPU-accelerated simulation platform, thus requiring more computing resources. CST Studio Suite, capable of collecting RF sensory information, is notably complex in terms of modeling the geometry environment. As a result, it requires more simulation time to generate large-scale models in the M$^3$SC dataset. CARLA, which can collect non-RF sensory information, has limited available road conditions and longer rendering times compared to AirSim.} In this subsection, we present the construction principles of the M$^3$SC dataset with the help of Wireless InSite$^\circledR$, WaveFarer$^\circledR$, and AirSim.
	
\begin{figure*}[!t]
			\centering
			\includegraphics[width=0.92\linewidth]{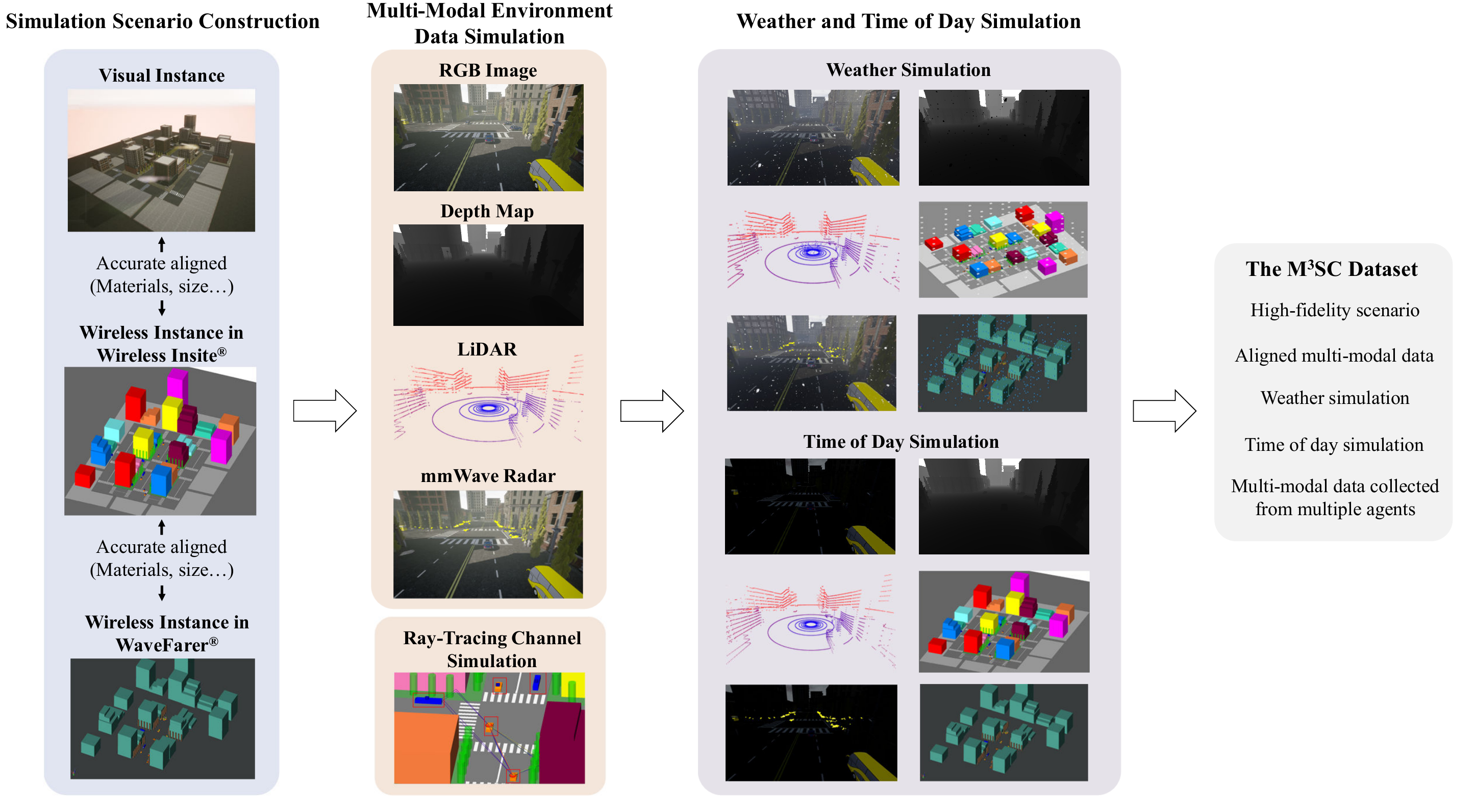}
			\caption{An overview of the proposed M$^3$SC dataset construction workflow, which shows the generation and processing methods of the aligned multi-modal sensory data and wireless channel data at a certain snapshot. In the dynamic VCN crossroad case, the visual instance has very fine details. For the convenience of ray-tracing simulation, simplified buildings, trees, RSUs, and other infrastructures are used in the wireless instances.}
		\label{platform}
	\end{figure*}

	In terms of visual data, we utilize an urban traffic environment produced by PurePolygons$^\text{TM}$, namely Modular Building Set, as the visual instance for constructing the M$^3$SC dataset. As depicted in Fig.~\ref{platform}, the environment contains a wide variety of buildings, roads, and infrastructures, with a resolution of 2048 $\times$ 2048. We simulate multiple vehicular movement trajectories and different velocities of dynamic vehicles to mimic a VCN crossroad scenario. Building upon this environment, we then use the AirSim \cite{AirSim}, a plug-in that is built on the 3D game engine Unreal Engine$^\circledR$, to add vehicles and RSUs into the environment and obtain multi-modal sensory data, including RGB images, depth maps, and LiDAR point clouds. The parameters of these sensors can be flexibly adjusted, such as the field of view (FoV), the resolution, the range of LiDAR, and the number of LiDAR channels.
	
	{\color{black}In terms of mmWave radar data, we utilize WaveFarer$^\circledR$ simulation platform \cite{WaveFarer}. The 3D model  aligned with AirSim, such as  buildings, trees, pedestrians, and vehicles, is exploited. In the VCN crossroad scenario, ten mmWave radars are introduced and each radar is equipped with eight Tx antennas and four Rx antennas.  The radar uses the linear FMCW with $77-81$ GHz frequency range and the azimuth information can be collected  via the digital beamforming.  Furthermore, the maximum detection range of radar devices is $74.9$ m and the range resolution is $0.1499$ m. The Doppler velocity range is $\pm47.42$ m/s and the velocity resolution is $0.939$ m/s. According to the requirement of different investigation tasks,  the aforementioned parameters can be adjusted and user-defined methods can be leveraged to properly process mmWave radar data to obtain environmental features.}
	
	In terms of wireless data, the M$^3$SC dataset contains wireless channel data of 1500 snapshots, covering V2V and vehicle-to-infrastructure (V2I) communication links. In this dataset, the wireless instance of the environment is simplified from the visual instance by omitting the fine visual details that do not affect the wireless channel data. The wireless instance is imported into the ray-tracing software Wireless InSite$^\circledR$ by Remcom \cite{WI} to generate wireless channel data, as shown in Fig.~\ref{platform}.  The vehicles are equipped with MIMO arrays that contain $32$ antennas and the RSUs are equipped with MIMO arrays that contain $128$ antennas, which all use half-wave dipoles operating at a frequency of $28$ GHz. The system bandwidth is $2$ GHz.
	
	In terms of the weather simulation and time of day simulation, AirSim supports adjusting the amount of snow, rain, and fog in the scenario. It also supports manually setting a certain time of day and collecting sensory data at that time. {\color{black}To ensure the accurate matching of visual instance in the AirSim and wireless instance in the Wireless InSite$^\circledR$ and WaveFarer$^\circledR$, we manually place snowflakes or water drops in the wireless instance.} Moreover, the time settings of AirSim are carefully tuned to optimize the rendering effects of the sensory data, such as sky brightness and reflected light intensity. 
	
	{\color{black}Furthermore, it is essential to validate the accuracy of the developed simulation dataset. For the developed M$^3$SC simulation dataset, the validation process and the corresponding challenge are given below. Firstly, to construct the M$^3$SC simulation dataset, we utilize efficient and high-fidelity software, i.e., AirSim, WaveFarer$^\circledR$, and Wireless InSite$^\circledR$, whose efficiency and accuracy are validated by the measurement campaign. Secondly, based on the efficient and accurate software, we appropriately configure the simulation parameters based on the realistic propagation environment and parameters used in commercial products. Thirdly, an experimental hardware platform needs to be constructed to validate the  simulation dataset. However, since it is time-consuming and challenging to construct the measurement platform, e.g., joint calibration and synchronization of measurement devices, we are currently working on developing an experimental hardware platform for the validation of the M$^3$SC simulation dataset. }
	
\begin{figure*}[!t]
		\centering
		\subfloat[snapshot 1: sunny weather (vehicle side)]{\includegraphics[width=1.5in]{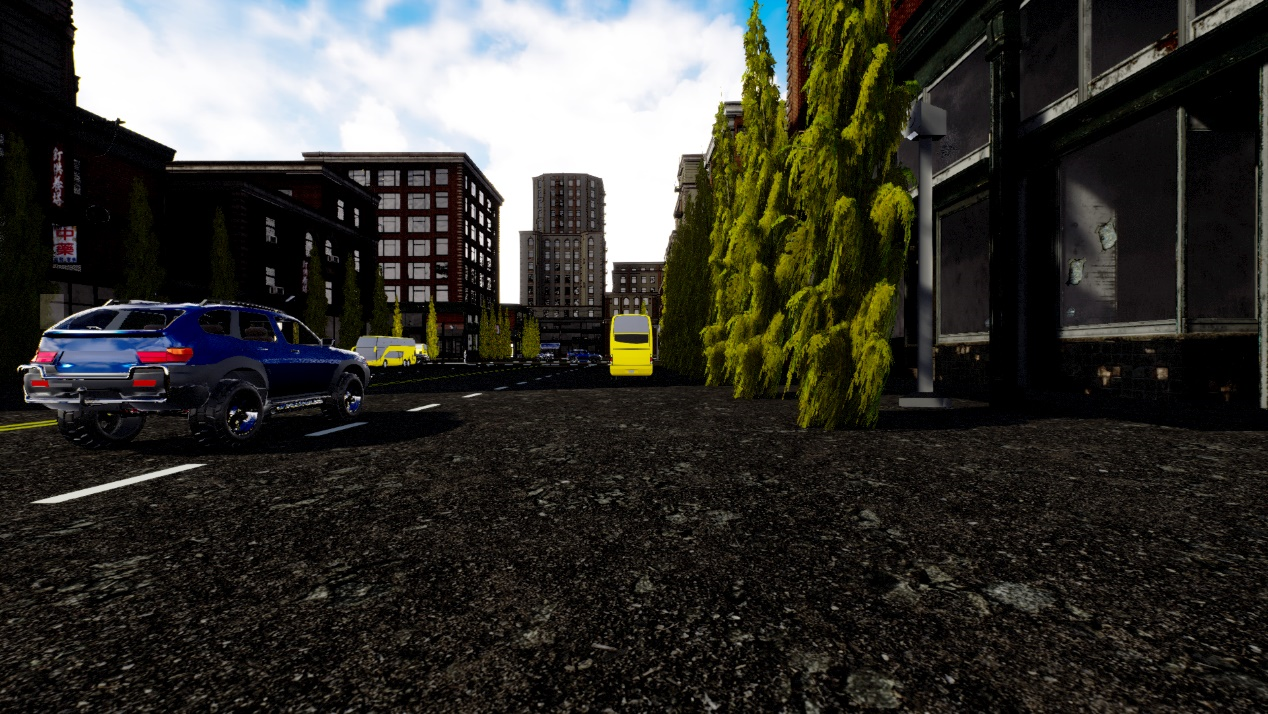}%
			\label{v1}}
		\hfil
		\subfloat[snapshot 1: snowy weather (vehicle side)]{\includegraphics[width=1.5in]{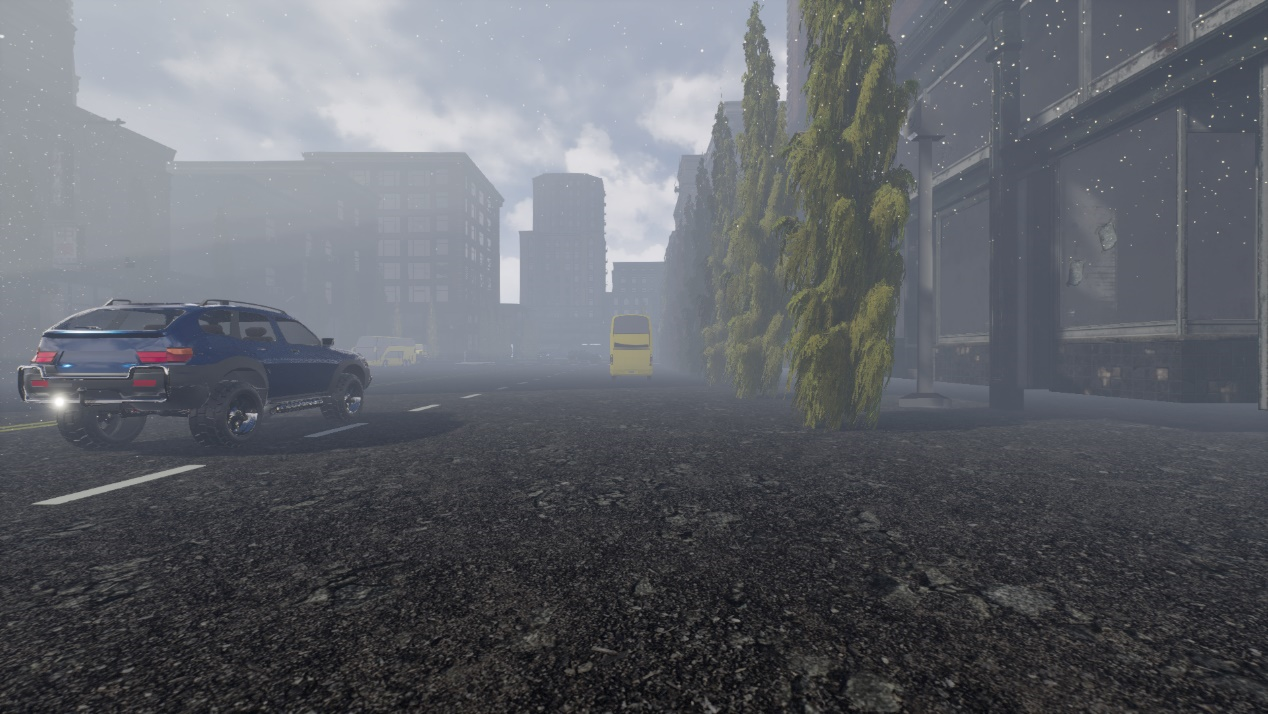}%
			\label{s1}}
		\hfil
		\subfloat[snapshot 1: sunny weather (RSU side)]{\includegraphics[width=1.5in]{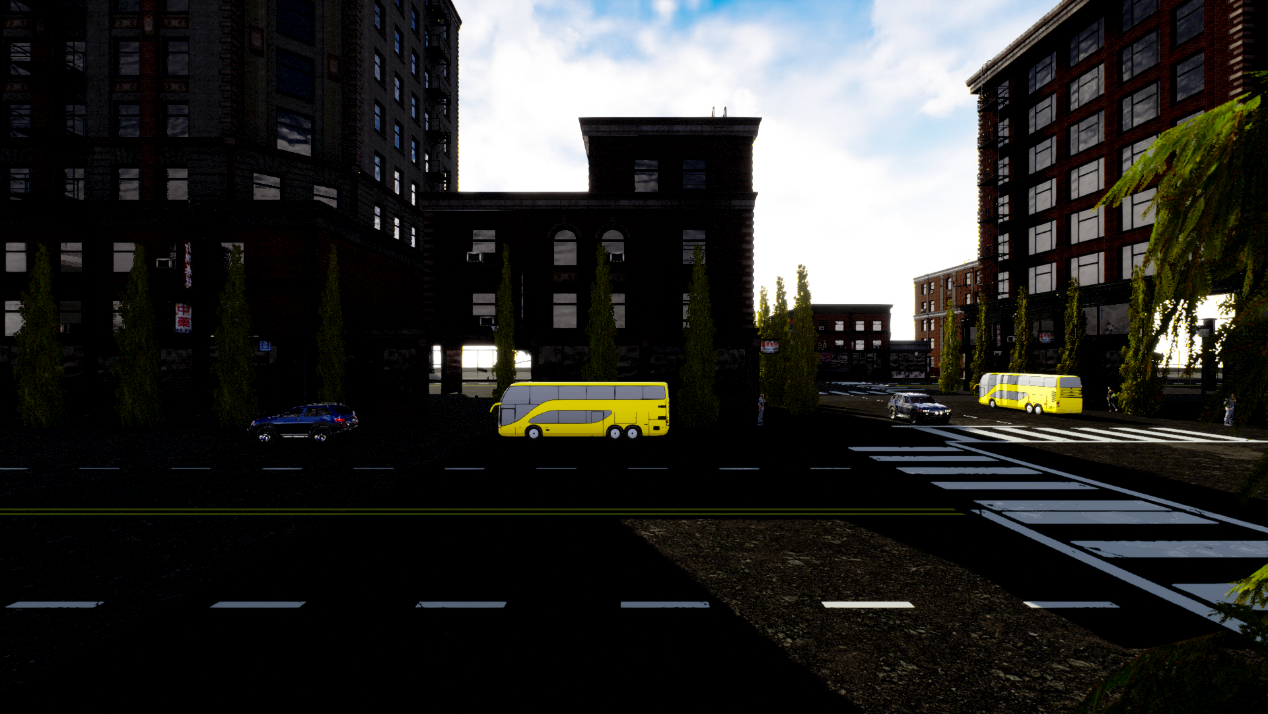}%
			\label{r1}}
		\hfil
		\subfloat[snapshot 1: night of the day (RSU side) ]{\includegraphics[width=1.5in]{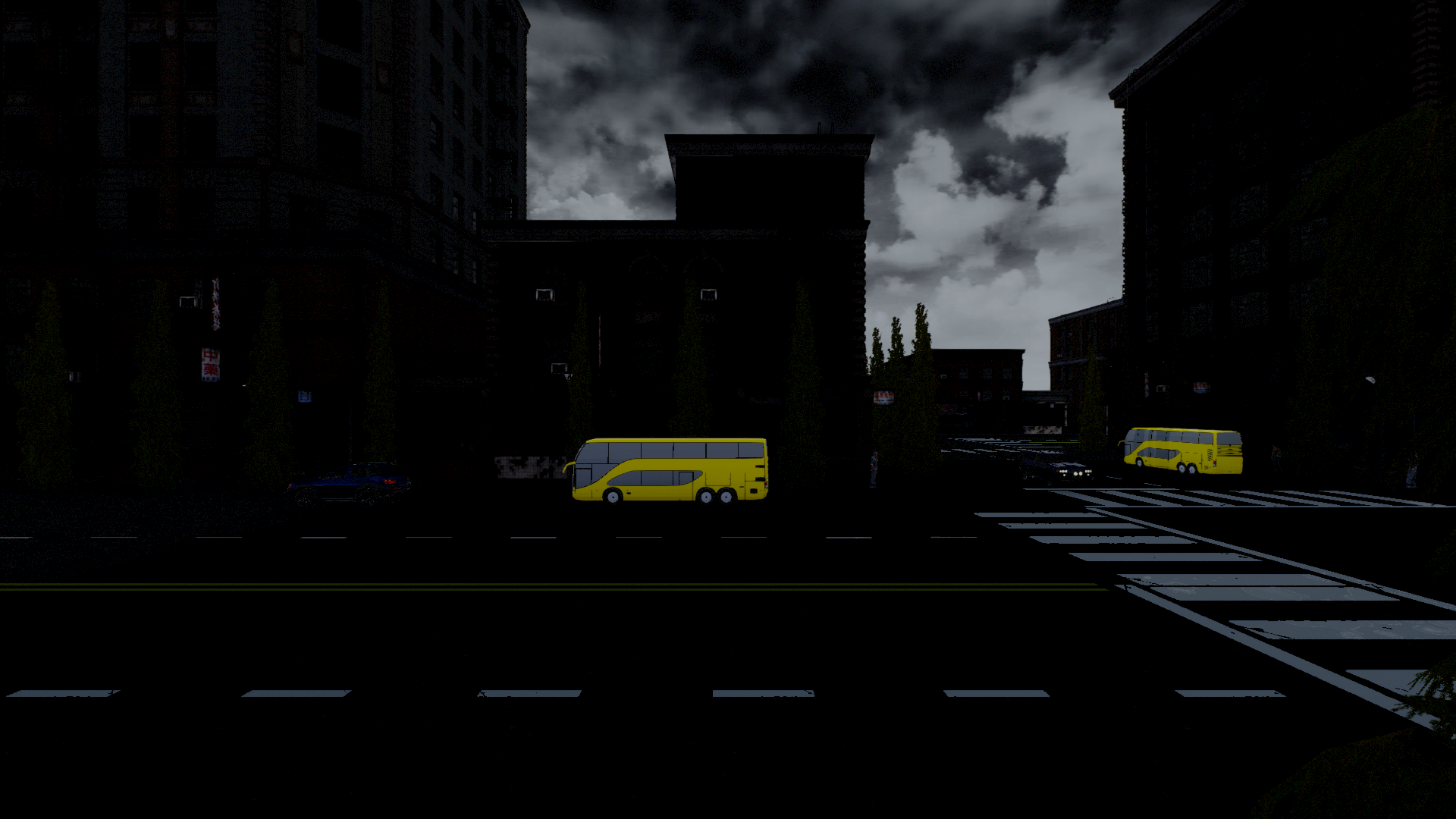}%
			\label{n1}}
		\hfil
		\subfloat[snapshot 601: sunny weather (vehicle side)]{\includegraphics[width=1.5in]{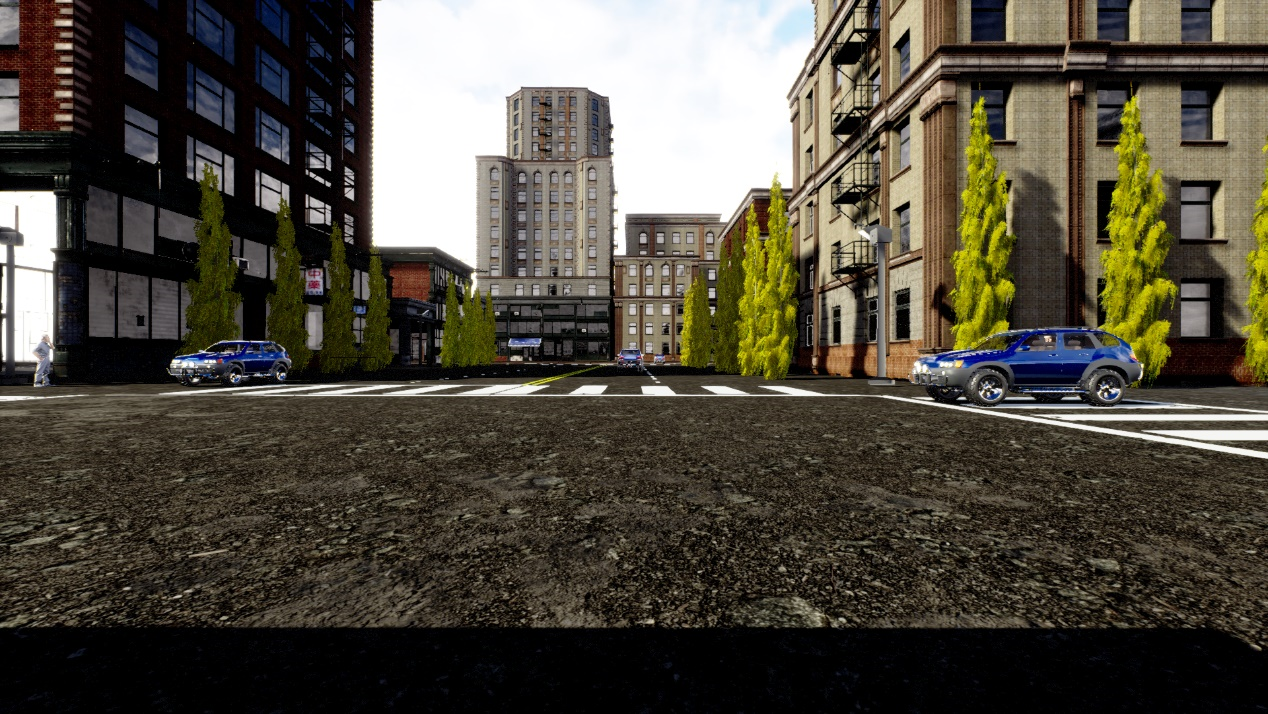}%
			\label{v601}}
		\hfil
		\subfloat[snapshot 601: snowy weather (vehicle side) ]{\includegraphics[width=1.5in]{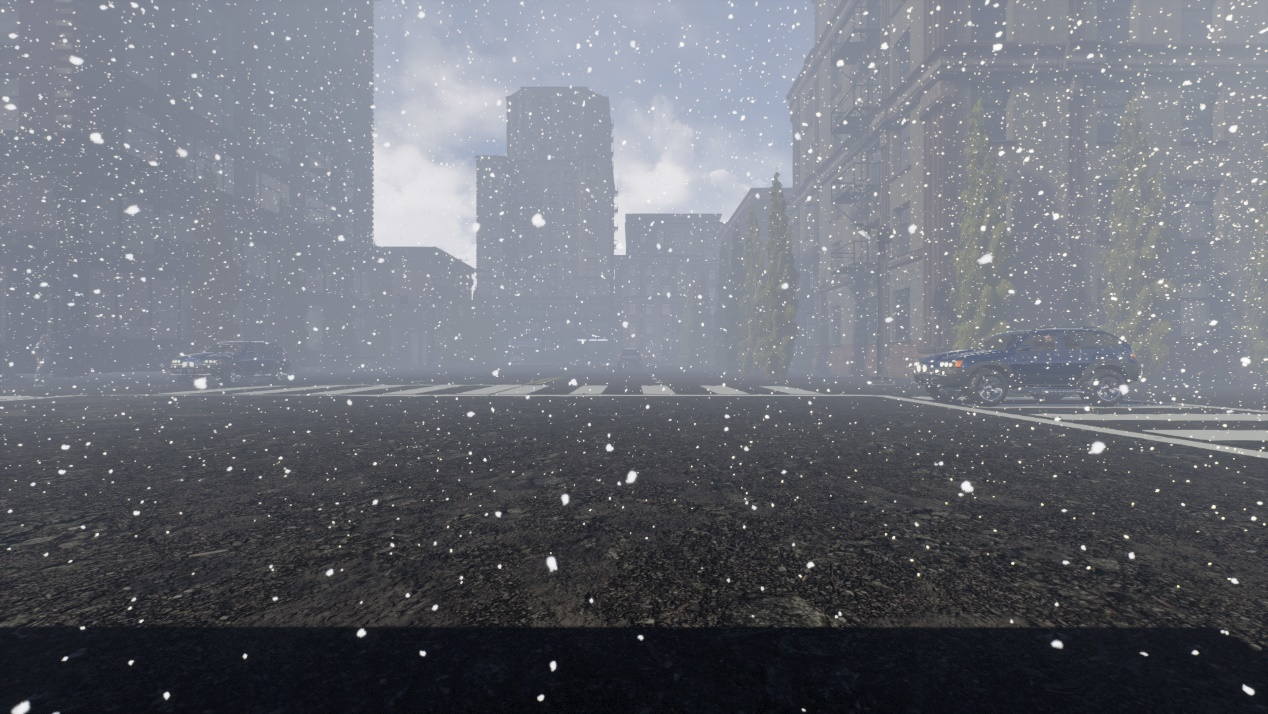}%
			\label{s601}}
		\hfil
		\subfloat[snapshot 601: sunny weather (RSU side)]{\includegraphics[width=1.5in]{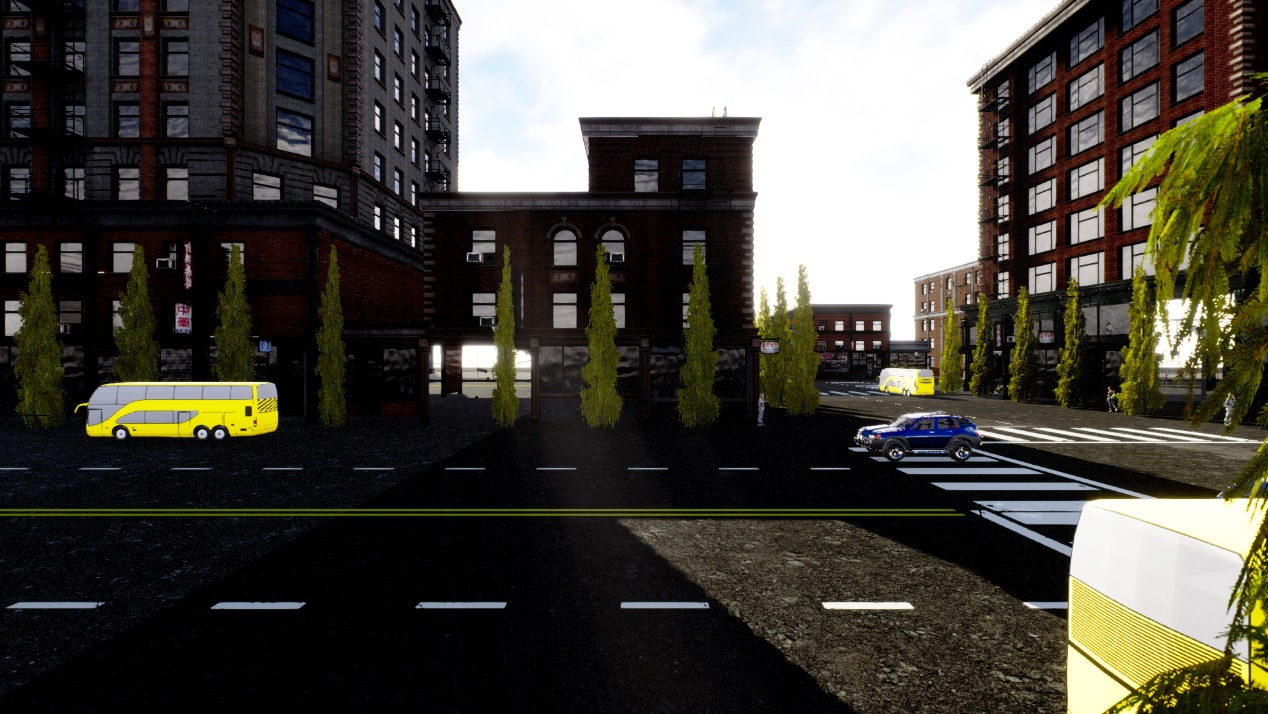}%
			\label{r601}}
		\hfil
		\subfloat[snapshot 601: night of the day (RSU side)]{\includegraphics[width=1.5in]{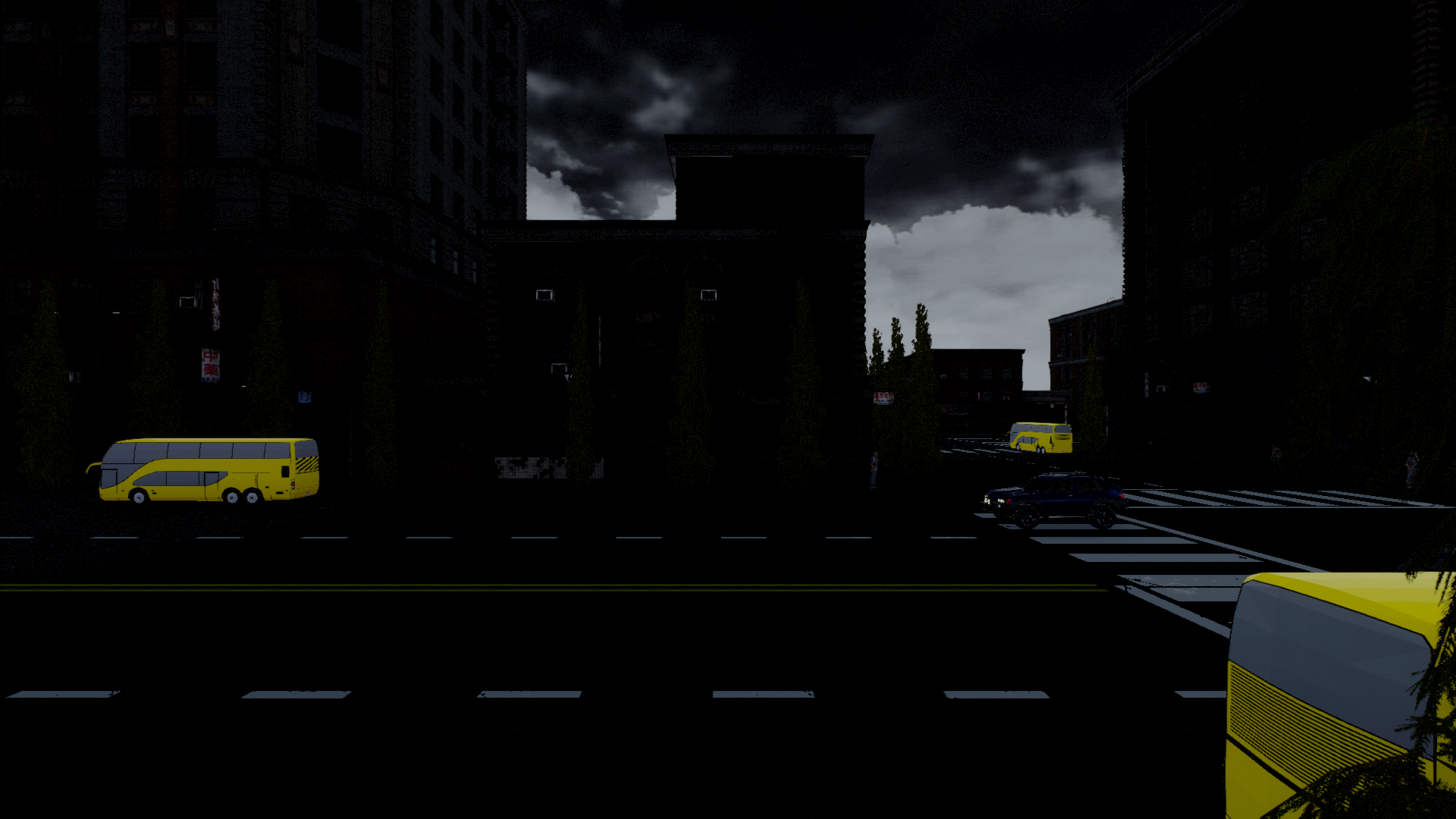}%
			\label{n601}}
		\caption{The RGB images taken by the vehicle and RSU at {\color{black}snapshots} 1 and 601. Due to the page limit, only RGB images taken in sunny weather, snowy weather, and night are shown in this survey.}
		\label{example}
	\end{figure*}

	To the best of our knowledge, this is the first MMM simulation dataset that supports flexible customization of scenarios and contains the aligned multi-modal environment data under different weather conditions and times with adjustable degrees. To illustrate, Fig.~\subref*{v1}-Fig. \subref*{s1} and Fig.~\subref*{v601}-Fig.~\subref*{s601} show the RGB images taken by a vehicle in sunny and snowy weather conditions at snapshot 1 and snapshot 601, respectively. Fig.~\subref*{r1}-Fig. \subref*{n1} and Fig.~\subref*{r601}-Fig.~\subref*{n601} show the RGB images taken by an RSU at the day and at night at snapshot 1 and snapshot 601, respectively.
	
	{\color{black}\subsection{Key Takeaways}
		In this section, we discuss the current research status of datasets for communication and sensing system research. Furthermore, we discuss the necessary properties of the MMM datasets and briefly introduce our work, i.e., the M$^3$SC dataset. Key takeaways learned are as follows.
		
		\begin{itemize}
			\item Datasets that contain aligned multi-modal sensory data and wireless channel data are scarce. Due to the limitation of cost, labor, and material resources, the real-world MMM datasets cannot meet the requirements of SoM research flexibly and timely. Consequently, a simulation dataset construction platform that can support flexible customization of specific research scenarios and adjustment of equipment parameters is of great importance. 
			
			\item We point out that available MMM dataset is one of the fundamental conditions for exploring the mapping relationships between multi-modal sensing and communications. For example, we can utilize the developed object detection ANNs to detect the objects based on sensory data. We can also utilize the clustering algorithm to obtain the scattering clusters with the help of wireless channel data. By doing so, we can explore the complex mapping relationship between objects in the physical environment and scattering clusters in the electromagnetic environment with the assistance of suitable ANNs. In addition to supporting the exploration of mapping relationships, MMM dataset also plays an indispensable role in supporting the subsequent research on various SoM-based applications in communication and sensing systems. Therefore, to promote further research on SoM, the construction of MMM datasets urgently calls for more research efforts.
			
		\end{itemize}
	}
	
	\section{Channel Modeling: The Pathway Towards Mapping Relationships Between Multi-Modal Sensing and Communications}
	\label{part1}
	The research on SoM requires an in-depth understanding of complex mapping relationships between multi-modal sensing and communications. Such complex mapping relationships can be explored by channel modeling customized for SoM, i.e., the characterization of large-scale fading and small-scale fading. In this section, we first discuss the complex mapping relationships between multi-modal sensing and communications, then analyze the challenges that the exploration of such mapping relationships via channel modeling faces, and finally overview the state-of-the-art channel modeling for SoM.

	\subsection{Channel Modeling Customized for Mapping Relationships}
	The natural differences between multi-modal sensing and communications in  terms of original functionalities, operation frequency bands, data structures, and semantic information bring the inherent requirements of SoM research, i.e., the exploration of complex mapping relationships between these two functionalities. It is not so challenging to explore the mapping relationships among multi-modal sensory data as the way in which multi-modal sensors obtains the physical environment feature is similar. Consequently, the mapping relationships among multi-modal sensory data are often reflected in coordinate transformation and semantic information matching. However, due to the significant differences in the acquisition frequency band and data representation of multi-modal sensory data and RF wireless channel information, it is extremely challenging to explore the mapping relationships between them. Fortunately, channel modeling is an effective solution to realistically characterize the RF wireless propagation environment and support the understanding of the RF wireless propagation effect. Therefore, channel modeling can be utilized as the pathway to explore the complex mapping relationships between multi-modal sensing and wireless communications, which is the foundation of SoM research.
	
	Channel modeling includes the characterization of large-scale fading characteristics and small-scale fading characteristics. Channel large-scale fading characteristics represent signal variation over long distance between transceivers in the environment, including path loss and shadowing. Multi-modal sensors capture the physical environment information between transceivers. As a result, utilizing channel modeling as a bridge, the complex mapping relationship between large-scale fading characteristics and the physical environmental information collected by multi-modal sensors can be explored. Furthermore, the accurate characterization of channel small-scale fading characteristics, i.e., multipath effect, requires a proper modeling of scattering clusters in the propagation environment \cite{TITS_my,fsfef,TWCMY}. Note that the scattering clusters are certain objects in the physical environment, such as vehicles, buildings, and trees \cite{sensor-4,Rappaport1-4,newpaper,huangj1}. For the mapping relationship between channel small-scale fading and multi-modal sensing, the premise of accurately exploring the mapping relationship is to realistically model scattering clusters. The reason is that the positions and trajectories of objects in the physical environment can be obtained with the help of multi-modal sensory data. Since scattering clusters in the electromagnetic environment have practical physical meaning, i.e., they correspond to specific objects in the physical environment, the evolution of scattering clusters needs to be modeled in a smooth and realistic way. For example, the smooth and continuous movement of mobile vehicles in the physical environment can be obtained by multi-modal sensory data. In the electromagnetic environment, only by modeling the smooth evolution of dynamic scattering clusters and mapping them to the mobile vehicles in the physical environment can we explore the mapping relationship between mobile vehicles and dynamic scattering clusters. It is noteworthy that the modeling of smooth evolution of scattering clusters can be achieved by capturing channel consistency. With the employment of mmWave-Terahertz (THz) and mMIMO technologies, the necessity of capturing channel consistency is further underlined \cite{myCOMST}. 
	
	In summary, proper channel modeling methods customized for SoM, which are capable of exploring the mapping relationships between multi-modal sensing and large/small-scale fading, need an in-depth investigation.
	
	\subsection{Challenges}
	As the foundation of SoM research, the mapping relationships are of paramount importance. However, the exploration of such relationships via channel modeling customized for SoM faces huge challenges. The challenges are given as follows.
	\begin{itemize}
		\item  \textbf{Significant acquisition frequency difference between non-RF sensory data and RF wireless channel data:} The propagation effects under non-RF and RF are significantly different. Specifically, compared to RF propagation, the divergence of non-RF propagation is smaller and the directionality is stronger. Also, unlike RF propagation, there is generally no diffraction in non-RF propagation due to the small wavelength. In addition, attributed to the great difference between RF and non-RF bands, their reflection and scattering mechanisms are significantly different. As a consequence, the correlation between non-RF sensory data collected in physical environment and RF wireless channel data obtained in electromagnetic environment is significantly low. {\color{black}The frequency band also has a significant impact on RF channel attenuation. This poses a huge challenge for exploring the mapping relationship between non-RF sensory data and RF wireless channel data under different frequency bands. Regarding the large-scale fading of the channel, the path loss in the high frequency band is significantly greater than that in the low frequency band. As for the small-scale fading of the channel, in the mmWave frequency band with an ultra-large bandwidth, the uncorrelated scattering assumption that is valid in the sub-6 GHz frequency band no longer holds, resulting in channel frequency non-stationarity.}
		
		\item  \textbf{Significant data representation difference between non-RF sensory data and channel large-scale fading data:} 
		Multi-modal sensory data, such as RGB images, depth maps, mmWave radar point clouds, and LiDAR point clouds, has a 2D or 3D structure with discrete values.  However, channel large-scale fading, such as path loss and shadowing, has a one-dimensional structure with continuous values.
		
		\item \textbf{Complex mapping between objects and clusters:}  The RGB-D camera can obtain the position of objects, i.e., potential   clusters. Radar and LiDAR can obtain the corresponding movement or velocity information. As a result, those potential   clusters can be reliably identified, enabling real-time tracking of the transmission environment. The detailed relationship between any pair of the  cluster and antenna can be precisely calculated. The complicated geometry information of each  cluster can also be acquired in real time. In this case, it is necessary to characterize the smooth evolution of  clusters along the space/array and time axes, and supply the spatially consistent and smoothly temporal-evolving CIR for SoM  \cite{Rappaport1-4,3GPP}. However, the capturing of smooth  cluster evolution is of huge challenge owing to the complex appearance and disappearance of scatterers \cite{myCOMST}. Therefore, to explore the mapping relationship between multi-modal sensing and small-scale fading, it is essential and challenging to imitate the smooth  space-time cluster evolution, which can be achieved by capturing space-time channel consistency.  
	\end{itemize}

	\subsection{State-of-the-Art}
	\subsubsection{Mapping Relationship Between Multi-Modal Sensing and Large-Scale Fading}
\begin{figure}[!t]
		\centering
		\subfloat[]{\includegraphics[height=2cm,width=3cm]{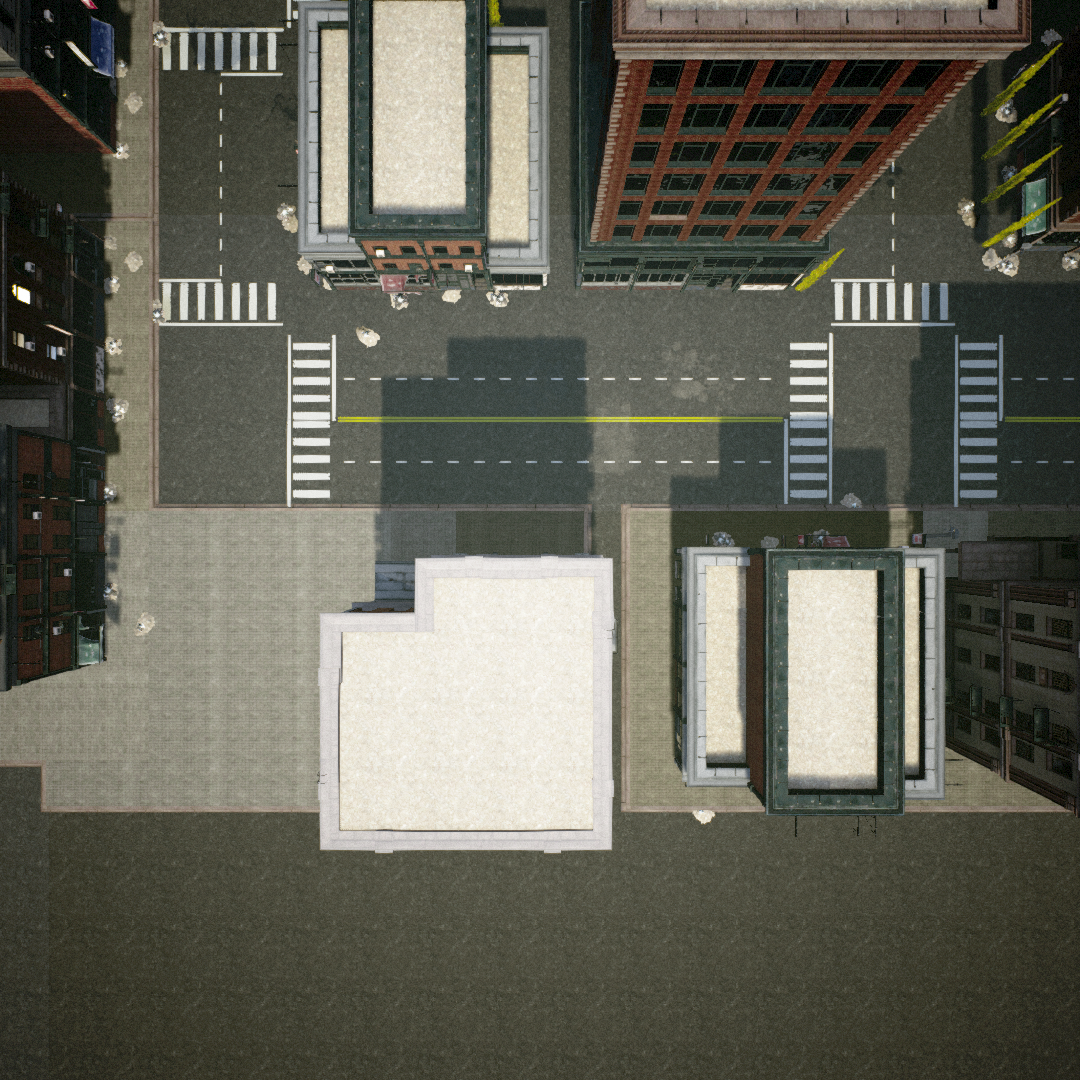}%
			\label{rgbpl}}
		\hfil
		\subfloat[]{\includegraphics[height=2cm,width=3cm]{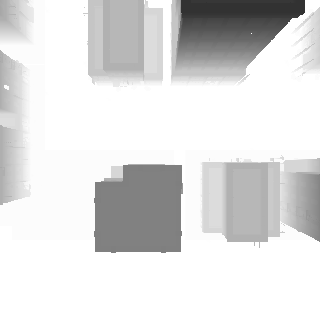}%
			\label{dppl}}
		\hfil
		\subfloat[]{\includegraphics[height=3.3cm,width=6.8cm]{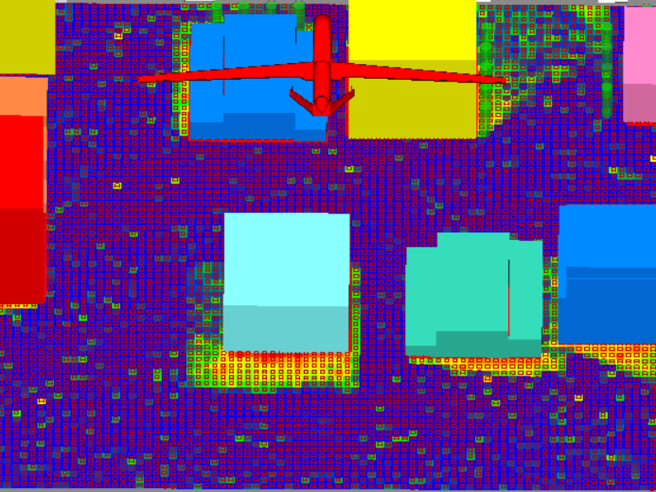}%
			\label{hpl}}
		\caption{Multi-modal sensory data and communication data collected in AirSim and Wireless InSite$^\circledR$: (a) RGB image collected by AirSim; (b) depth map collected by AirSim; (c) heat map of path loss collected by Wireless InSite$^\circledR$.}
		\label{pl1}
	\end{figure}

\begin{figure}[!t]
		\centering			
		\includegraphics[width=0.45\textwidth]{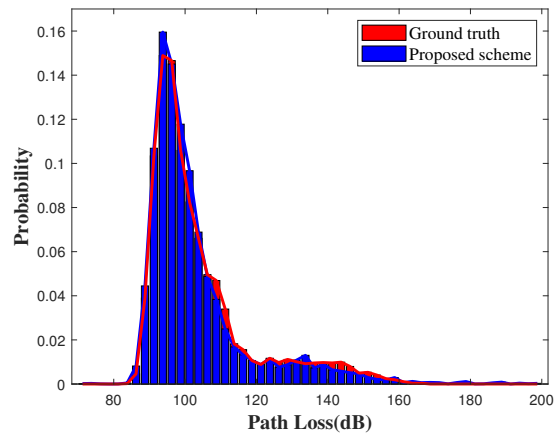}
		\caption{RGB-D-based path loss histogram distribution prediction results.}
		\label{pl2}
	\end{figure}

	Research on the mapping relationship between channel large-scale fading and multi-modal sensing is still an open field due to the lack of MMM datasets and consideration of multi-modal sensing. To provide a reference point for research in this field, we carry out preliminary research on the exploration of the mapping relationship between path loss and multi-modal sensing. Specifically, we propose an RGB-D-based path loss prediction scheme in the UAV scenario. Firstly, the developed scheme utilizes the YOLO object detection network \cite{yolo}, a popular object detection model known for its speed and accuracy, to detect buildings in the RGB images taken by the UAV. The detection results are matched with the aligned depth map to obtain the density, height, and relative position of buildings. Then, a fully connected neural network is designed to predict the path loss distribution. For the dataset construction, we use Wireless InSite$^\circledR$ to collect the ray-tracing-based path loss data.  Since ray-tracing-based technology generates channel parameters and channel data based on the geometrical optics and uniform theory of diffraction, the ray-tracing-based data closely resembles the real-world transmission propagation \cite{sss-h1,sss-h2,M3C}. Specifically, the accuracy of ray-tracing-based data is validated in \cite{sss-h1,sss-h2} by demonstrating a close agreement between channel properties obtained from measurements and those derived from Wireless InSite$^\circledR$, including path loss, angle of arrival, and root-mean-square delay spread. Furthermore, we use AirSim to collect the aligned RGB images and depth maps based on the M$^3$SC dataset, as shown in Fig.~\ref{pl1}. Fig.~\subref*{rgbpl} and \subref*{dppl} present the RGB image and depth image collected by AirSim, respectively. Fig.~\subref*{hpl} presents the heat map of path loss collected by Wireless InSite$^\circledR$ at the same snapshot. Fig.~\ref{pl2} shows the prediction result of the path loss distribution through the proposed scheme as well as the ground truth. It can be observed from Fig.~\ref{pl2} that the path loss distribution is accurately predicted based on RGB images and depth maps, demonstrating the practicability of the exploration on the complex mapping relationship between multi-modal sensing and channel large-scale fading. Considering the high accuracy of ray-tracing-based data collected by Wireless InSite$^\circledR$, we regard it as the ground truth and calculate the normalized mean square error of the predicted results to be $5.32\times10^{-4}$. Based on the developed RGB-D-based path loss prediction scheme, it has been demonstrated that there exist mapping relationships between physical space and electromagnetic space, which can be effectively explored using ANNs. In practical applications, the real-time predicted path loss result of a specific region can provide guidance for UAV system design and development. For instance, in terms of UAV networking, real-time path loss prediction in UAV-to-ground channels contributes to reducing communication energy loss and enhancing the overall energy efficiency of UAV networks.
	
	\subsubsection{Mapping Relationship Between Multi-Modal Sensing and Small-Scale Fading}
	\label{smallscale_r}
\begin{figure}[!t]
		\centering	
		\includegraphics[width=0.48\textwidth]{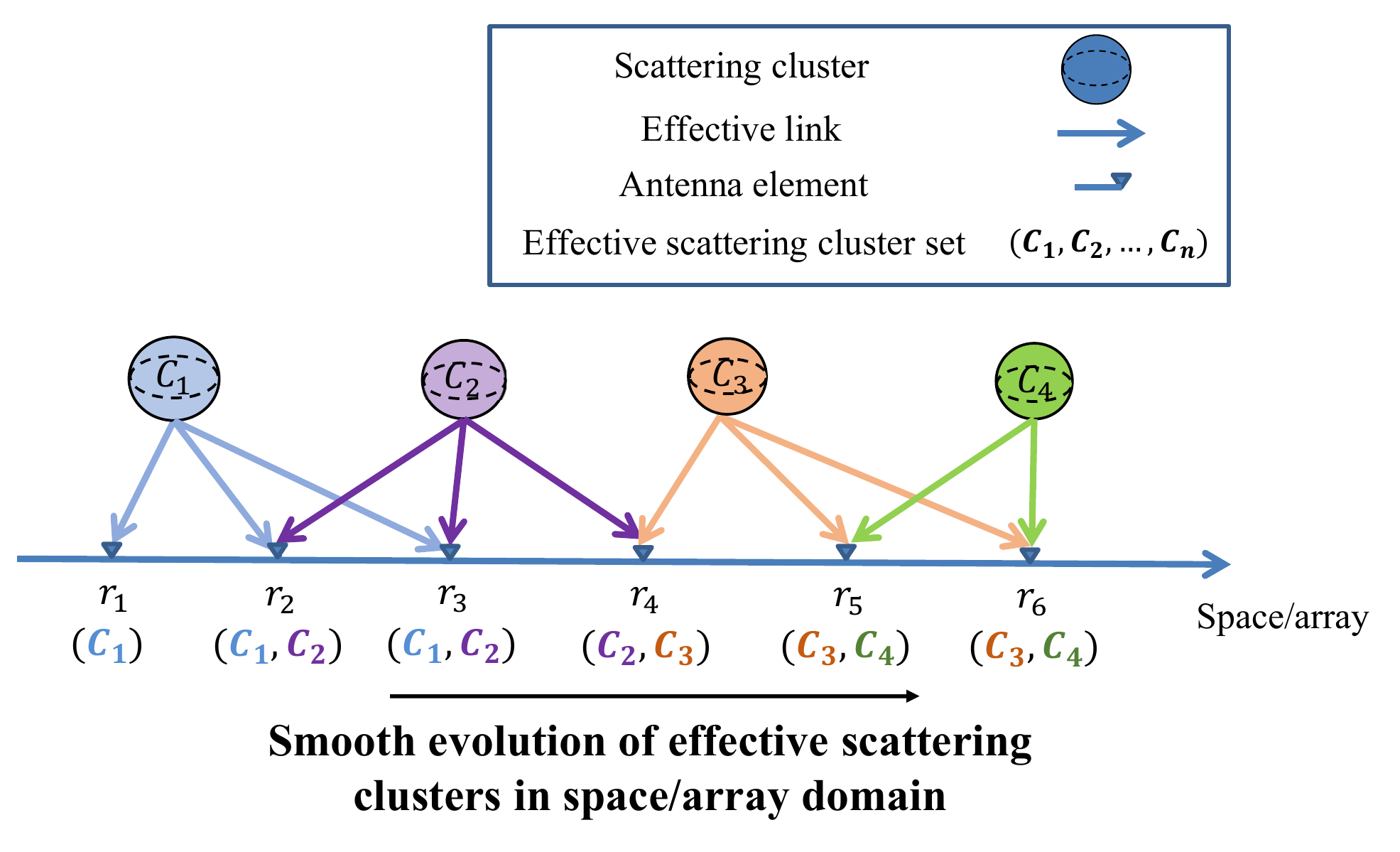}
		\caption{Smooth evolution of effective scattering clusters along the antenna array axis in channels with space consistency.}
		\label{space_consistency}
	\end{figure}

\begin{figure}[!t]
		\centering
		\includegraphics[width=0.5\textwidth]{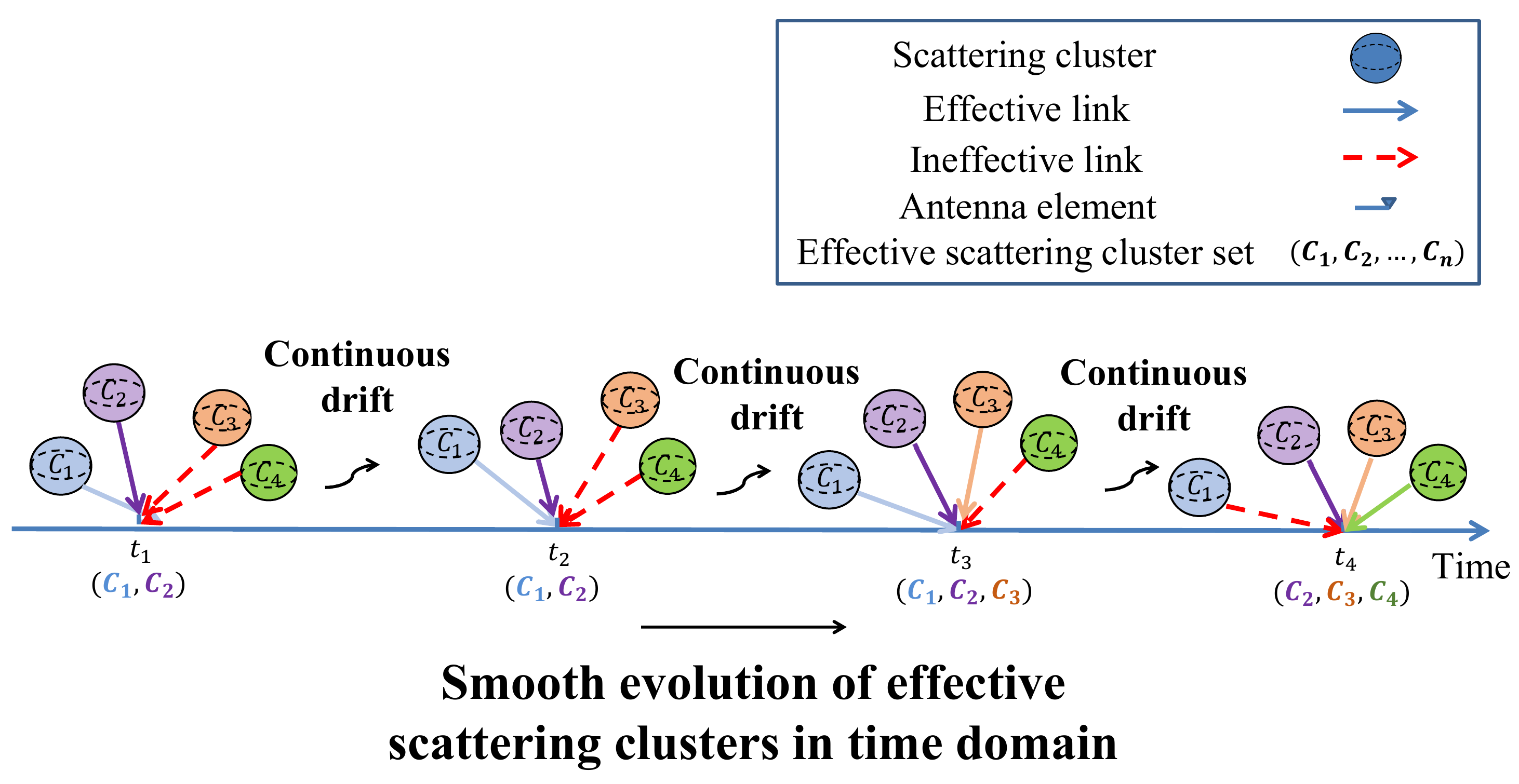}
		\caption{Smooth evolution of effective scattering clusters along the time axis in channels with time consistency.}
		\label{time_consistency}
	\end{figure}

	To support the exploration of mapping relationships between channel small-scale fading and multi-modal sensing, the state-of-the-art channel modeling for SoM focuses on capturing channel consistency. Channel consistency can be regarded as a typical channel physical feature, which inherently exists in each channel.  Channel consistency in the space/array domain or time domain, i.e., channel space consistency or channel time consistency, indicates that the channel varies smoothly and consistently as either the space/array or the time evolves \cite{myCOMST}. The mechanism underlying channel consistency is discussed as follows. In the channel with space consistency, two adjacent antennas closely spaced share many common effective scattering clusters in transmission \cite{Gao1,Gaonew}. In such a condition, the evolution of effective scattering clusters along the antenna array in the channel with space consistency is consistent and smooth, not random, as shown in Fig.~\ref{space_consistency}. Two adjacent antennas possess similar or even the same sets of effective scattering clusters in the propagation environment. For the channel time consistency, channels at two adjacent moments experience similar transmission scenarios, and they possess many common effective scattering clusters \cite{liuy-4}.  Fig.~\ref{time_consistency} indicates that, as effective scattering clusters continuously drift, their evolution over time in the channel with time consistency is also consistent and smooth, not random, resulting in similar or even the same sets of effective scattering clusters. {\color{black}As effective scattering clusters and transceivers, e.g., vehicles in V2V communications,  continuously drift in the transmission scenario, channels evolve smoothly and consistently over time. Furthermore, different vehicular movement trajectories, e.g., quarter turn, U-turn, and curve driving, have a significant impact on channel statistical properties \cite{xiong,myTVT}. Specifically, more complex vehicular movement trajectories result in lower temporal correlation of vehicular channels, necessitating the capture of channel time consistency.} 
	
	Currently, researchers have proposed some preliminary  channel models for SoM that can capture channel consistency. For clarity, the typical methods of mimicking SoM channel consistency are summarized in Table~\ref{tab:comparisonr1}. Based on the modeling approach, current channel models can be classified into geometry-based deterministic models (GBDMs) \cite{5393069,dewd}, non-geometry stochastic models (NGSMs) \cite{ZTECom,NGSMDavid,7473100}, and geometry-based stochastic models (GBSMs) \cite{8108211,e3e3,xilaren,myTVT}. Since the location of the scattering clusters in the transmission scenario can be identified in site-specific deterministic models, channel consistency can be mimicked in them \cite{5GCM}. The authors in  \cite{8647188,XUESONG1} developed two GBDMs at the mmWave band. These two GBDMs characterize the smooth and continuous drift of the scattering cluster and update the channel parameter consistently along the time axis. Consequently, by smoothly updating the scattering propagation environment over time, channel time consistency can be naturally imitated in the deterministic channel models for SoM \cite{8647188,XUESONG1}. This method is of high accuracy by considering the underlying propagation environment. However, its modeling complexity is high based on the deterministic channel modeling approach. Note that, to the best of our knowledge, there is currently no work dedicated to capturing channel consistency based on the NGSM modeling approach for channels for SoM \cite{myCOMST}.

	\begin{table*}[!htp]
		\renewcommand\arraystretch{1.6} 
		\centering
		\caption{Typical related works of capturing channel consistency of channels for SoM.}
		\label{tab:comparisonr1}
		\resizebox{\textwidth}{!}{
			\begin{tabular}{c|c|c|c|c|c|c|c}
				\toprule[0.35mm]
				\makecell[c]{\textbf{Ref.}}	 &
				\makecell[c]{\textbf{Modeling  }\\\textbf{approach}}	 & 	\makecell[c]{\textbf{Capturing method }} &	\makecell[c]{\textbf{Accuracy}}& 	\makecell[c]{\textbf{Complexity}}&
				\makecell[c]{\textbf{Scalability}}&	\makecell[c]{\textbf{Time  }\\\textbf{consistency}}&	\makecell[c]{\textbf{Space  }\\\textbf{consistency}}\\
				\midrule[0.15mm]
				\cite{8647188,XUESONG1}	& GBDM &\makecell[c]{Smoothly updating\\ propagation environment} & High & High & Low & \checkmark& $\times$\\ 
				\midrule[0.15mm]
				\makecell[c]{\cite{3GPP,5GCM,METIS,IMT}}	& GBSM &\makecell[c]{Generating temporally consistent\\ channel parameters} & Low & Low & High & \checkmark& $\times$\\ 
				\midrule[0.15mm]			\cite{COST2100f}	& GBSM &\makecell[c]{VR-based method} & High & High & Low & \checkmark& $\times$\\ 
				\midrule[0.15mm]			\cite{Gao1,Gaonew}	& GBSM &\makecell[c]{VR-based method} & High & High & Low & \checkmark& \checkmark\\ 
				\midrule[0.15mm]			\cite{newpaper}	& GBSM &\makecell[c]{Combination of VR-based method \& \\generating temporally consistent\\ channel parameters } & High & Medium & Medium & \checkmark& \checkmark\\ 
				\midrule[0.15mm]			\cite{xietongUGV}	& GBSM &\makecell[c]{ ML-based method } & High & Medium & High & \checkmark& \checkmark\\ 
				\bottomrule[0.35mm]
		\end{tabular}	}
	\end{table*}
	
	Because of the decent trade-off between modeling accuracy and computational complexity, the GBSM modeling approach has been extensively leveraged for current channel modeling and is also employed in the standardized channel modeling, e.g., Third Generation Partnership Project (3GPP) \cite{3GPP}, Mobile and wireless communications Enablers for the Twenty-twenty Information Society (METIS) \cite{METIS}, IMT-2020 channel model \cite{IMT}, and European COoperation in the field of Scientific and Technical research (COST) 2100 \cite{COST2100f}. In \cite{3GPP}, 3GPP Release 14 proposed a general channel model 3GPP TR38.901 via GBSM modeling. The proposed 3GPP TR38.901 channel model in \cite{3GPP} has been standardized and can support operating frequency from $0.5$~GHz to $100$~GHz and the configuration of large-scale antenna arrays. To mimic time consistency in channels for SoM, the 3GPP TR38.901 channel model in \cite{3GPP} generated temporally consistent channel parameters with the help of the 2D random process. The generated temporally consistent channel parameters are parameter-specific-correlation-distance-based, which can imitate the smooth evolution of channels when updating channels, thereby capturing the time consistency of channels for SoM. Similarly, the standardized channel models, including METIS channel model \cite{METIS} and IMT-2020 channel model \cite{IMT}, also leveraged the method of generating temporally consistent channel parameters to enable time consistency of channels for SoM. Different from 3GPP TR38.901 channel model \cite{3GPP},  METIS channel model \cite{METIS}, and IMT-2020 channel model \cite{IMT}, the standardized channel model, named as 5G Channel Model in \cite{5GCM},  proposed a new method to characterize the time consistency of channels for SoM. Specifically, the 5G Channel Model defined some grids and further generated four complex normal distributed independent and identically distributed random variables on four vertexes of each grid. Also, 5G Channel Model \cite{5GCM} generated temporally consistent random variables and determined temporally consistent channel parameters in channel realization, thus mimicking time consistency of channels for SoM. However, the aforementioned standardized channel models in \cite{3GPP,METIS,IMT,5GCM} ignored the capturing of space consistency of channels for SoM. In summary, generating temporally consistent channel parameters can mimic channel consistency in a low-complexity manner and remain applicable in various scenarios via adjusting key parameters. Nonetheless, this method is of low accuracy for neglecting the underlying geometry of the transmission environment.
	
	In addition to the generation of temporally consistent channel parameters, channel consistency can also be characterized by capturing the smooth and consistent evolution of effective scattering clusters and underlying geometry transmission environment. The standardized COST 2100 channel model in \cite{COST2100f} proposed a visibility region (VR) method and assigned each scattering cluster to a VR. Once transceivers are in a VR, the scattering cluster assigned to the VR is an effective scattering cluster, which contributes to CIR matrix. As transceivers and dynamic scattering clusters smoothly drift, transceivers enter and leave VR assigned to the corresponding scattering cluster, leading to the smooth and consistent evolution of effective/ineffective scattering clusters over time. Consequently, COST 2100 channel model \cite{COST2100f} can mimic the time consistency of channels for SoM. However, space consistency of channels for SoM was ignored. To further capture space consistency therein, the authors in \cite{Gao1,Gaonew}  extended the VR method in \cite{COST2100f} to the space/array domain and developed a 2D GBSM for the mMIMO channel. The authors in \cite{Gao1,Gaonew} assigned each scattering cluster to two VRs, one at the mobile station, named MS-VR, and the other at the BS, named BS-MR. If an antenna is in the VR of a scattering cluster, the scattering cluster is observable to the antenna and will be identified as effective. As adjacent antenna elements are closely spaced, their VRs are also closely spaced. In such a case, adjacent antenna elements have similar effective scattering clusters, therefore mimicking the space consistency of channels for SoM in the GBSMs \cite{Gao1,Gaonew}. In summary, since the smooth and consistent variation of the propagation environment is adequately characterized from the perspective of scattering clusters, the VR-based method is of high accuracy. Nonetheless, the complexity in real-time and realistic tracking of continuous drift of scattering clusters is prohibitive via the VR-based method. 
	
	Aiming at having an excellent compromise between complexity and accuracy, the authors in  \cite{newpaper}  proposed a mMIMO mmWave GBSM. The proposed GBSM \cite{newpaper} developed a new method, which combined the VR-based method and the method of generating temporally consistent channel parameters. To capture the space and time consistency of channels for SoM, the method in \cite{newpaper} constructed space-VR and time-VR with each cluster and each antenna element as the center, respectively. Additionally, with the aid of the uniform continuity theorem and bounded variation function, a visible factor was introduced to enable the soft cluster power handover. Consequently, the cluster power smoothly and consistently varied as the scattering cluster evolved on the time and space/array axes. This further enabled the capturing of time-space consistency of channels for SoM.
	
	Currently, ML has received extensive attention and is the core of AI and data science \cite{science}. Owing to its outstanding efficiency, our previous work in \cite{xietongUGV} leveraged the spectral clustering algorithm to capture the time-space consistency of channels for SoM and proposed a GBSM at the mmWave band for high-mobility scenarios. The developed method constructed a graph based on the 3D coordinates of scattering clusters and transceivers/terminals in the environment. Suppose the 3D coordinate of a scattering cluster is in the same clustering group as the transceiver/terminal. In this case, the scattering cluster is effective and contributes to the CIR of the sub-channel associated with the transceiver/terminal. Due to the continuous drift of the scattering cluster and transceiver/terminal, their 3D coordinates and the constructed graphs at two adjacent moments are similar. As a result, effective scattering cluster sets are similar at two adjacent moments, and time consistency of channels for SoM can be captured. Moreover, two close transceivers have similar graphs, and thus have similar clustering results and effective scattering cluster sets. In this case, the space consistency of channels for SoM can be mimicked. In terms of the ML-based method for capturing channel consistency, it offers a good trade-off between accuracy, scalability, and complexity. However, ML-based approaches heavily rely on parameter settings. Currently, MMM datasets, which serve as the source and core of ML-based methods, are scarce. As a result, the ML-based method for capturing channel consistency is still in its infancy. In the upcoming B5G/6G era, this aspect of research is expected to be greatly strengthened with {\color{black}more available MMM datasets}.
	
	{\color{black}
		\subsection{Key Takeaways}
		In this section, we delve into the challenges of exploring complex mapping relationships between multi-modal sensing and communications. We also overview the current research status of channel modeling customized for exploring such mapping relationships and present simulation results. Key takeaways learned are as follows.
		\begin{itemize}
			\item The exploration of relationships between multi-modal sensory data and wireless channel data faces huge challenges. First, the typical operation frequency difference between non-RF sensing and communication functionality results in different propagation effects. Second, there are significant differences in the representation between non-RF sensory data and wireless channel data. Third, the mapping between objects and clusters is complex and nonlinear. Summarizing and studying the aforementioned three challenges is of significant importance for properly designing the scheme for exploring mapping relationships, e.g., choosing a suitable tool and extracting similar features.  Specifically, due to the extremely complex nonlinearity of the relationship between multi-modal sensory data and wireless channel data, ANNs naturally become a suitable and efficient tool. Furthermore, since the representations of multi-modal sensory data and wireless channel data are significantly different, it is crucial to search for features that could establish relationships between them and model such relationships with the help of ANNs.
			
			\item For the mapping relationship between multi-modal sensing and channel large-scale fading characteristics, we conduct a preliminary work with the help of the YOLO object detection network and a customized fully connected neural network. Although the current path loss prediction performance is decent, more investigation is needed to further obtain a more refined path loss value by exploiting richer geometric topological information in the physical environment. Furthermore, since the mapping relationship between multi-modal sensing and channel small-scale fading characteristics is complex and nonlinear, we can explore the mapping between objects in the physical environment and clusters in the electromagnetic environment by ANNs as a preliminary attempt. Subsequently, based on the clusters obtained by the multi-modal sensory information, their smooth and consistent evolution can be characterized to capture channel consistency, thus supporting the subsequent SoM research.
			
			
		\end{itemize}
	}
	
	\section{SoM-Enhance: Enhanced Communication Functionality}
	\label{part2}	
	One of SoM's visions is to enhance the functionality of communications and networking. To be specific, SoM aims at inferring communication-related side information from multi-modal sensing to streamline the communication systems whilst reducing the overhead needed as compared with the conventional paradigm. As the wireless channels stem from electromagnetic radiation between the transceivers, a comprehensive knowledge of the electromagnetic environment is indispensable to the operation of communication systems. SoM-enhance-based channel estimation is envisioned to offer more concrete CSI information to guide transceiver design. Built upon the concrete CSI information, SoM-enhance can be further utilized to boost the communication system performance, where waveform design and beamforming design jointly constitute the key factors. In light of their unique importance, this section will delve into their current research state and implementing challenges.
	\subsection{SoM-Enhance-Based Channel Estimation in Dynamic Scenarios}
	\label{Estimation}
	Acquiring accurate CSI is critical to communication system design, including but not limited to precoding, demodulation, and equalization. Channel estimation has been a fundamental research topic for a few decades \cite{alkhateeb2014channel} and numerous methods have been proposed. 
	It is envisioned that MIMO-OFDM will still be the core technology for B5G/6G. Channel estimation for such systems faces unprecedented challenges \cite{cecomst}. Firstly, mmWave mMIMO systems generally adopt hybrid (analog/digital) precoding to cut down hardware cost and power consumption \cite{el2014spatially,gao2021model,gaoHcodebook}, thus the design flexibility is much weaker than digital precoding. Secondly, as the number of antennas increases, the number of pilot sequences needed for non-blind channel estimation also increases proportionally, leading to an excessive communication overhead.

	When it comes to dynamic scenarios, channel estimation for mmWave MIMO-OFDM communication systems becomes even more challenging. The fundamental cause lies in the strong dynamics of propagation environment \cite{ozdemir2007channel}. Channels in dynamic scenarios often exhibit more significant Doppler and delay spread than those in static scenarios \cite{gao20estimation}. The commonly adopted scheme based on parameter estimation and fixed channel model becomes no longer applicable to dynamic scenarios due to the changes in network topology or channel characteristics. In order to acquire accurate CSI without excessive signaling overhead, it is imperative to resort to comprehensive environmental information collected from multi-modal sensors. Multi-modal sensing is expected to provide side information matching the channel propagation characteristics through dedicated processing, thus greatly improving the channel estimation performance. In this subsection, we first delineate the challenges in SoM-enhance-based channel estimation and provide some insights into the connection between multi-modal sensing and mmWave channels, then overview the relevant channel estimation methods that may inspire the SoM-enhance-based ones.
	
	\subsubsection{Challenges}
	\label{estimation chan}
	Currently, the research efforts towards SoM-enhance-based channel estimation are well underway. The challenges lie in two aspects. Firstly, SoM-enhance-based channel estimation aims at mining different physical properties from multi-modal sensory data instead of purely relying on transceiver processing by convention. Unfortunately, the MMM dataset that contains aligned multi-modal sensory data and wireless channel data is largely lacking, hindering a deeper exploration of SoM-enhance-based channel estimation.  
	
	The second challenge lies in how to infer electromagnetic environment characteristics from multi-modal sensory data, which belongs to the research category of mapping relationships discussed in Section \ref{part1}. Only by taking the deep learning (DL) model as a black box and transforming the multi-modal sensory data into abstract features through ANNs on the basis of MMM datasets, the accuracy of channel estimation may improve under certain conditions. However, this is far from enough for ensuring scalability and adaptability. In order to develop a general and interpretable design paradigm and unleash the utmost potential of multi-modal sensing, the mapping relationships between multi-modal sensory data and the electromagnetic environment and propagation channels need to be found. Multi-modal sensory data in different scenarios, weather conditions, and times of the day needs to be analyzed simultaneously with aligned wireless channel data, such as DoD, DoA, and CSI. ANNs may benefit the exploration of such mapping relationships, which tend to be relatively abstract and may not have an analytic solution in some cases. For example, when multi-modal sensing is expected to be utilized for channel estimation of mmWave communication systems, the limited scattering property \cite{brady2013beamspace} that mmWave channels exhibit is a breakthrough to infer electromagnetic environment characteristics from multi-modal sensory data. The object position information in visual space may be closely related to the scattering experienced in the mmWave propagation. Unfortunately, research on such mapping relationships is still in its infancy and lacks a complete framework.

	\subsubsection{State-of-the-Art}
	
	As discussed in Section \ref{estimation chan}, mmWave channel model aligns better to the physical geometry due to its limited scattering property, which is potentially related to object information implied in multi-modal sensory data such as position and velocity. Therefore, the SoM-enhance-based channel estimation for mmWave channels is of great research potential and could be initially explored. {\color{black}}To provide some insights into such explorations, state-of-the-art ANN-based channel estimation methods that can potentially inspire the SoM-enhance-based channel estimation research are reviewed in this subsection despite the lack of multi-modal sensing.

	CSI can be regarded as a perceptible function of different key elements in the environment (such as the positions of scatterers, receivers, and transmitters), which are naturally related to environmental features characterized by multi-modal sensing. In light of this, researchers have begun to study how to estimate CSI with the aid of ANNs' ability to extract features and fit any non-linear functions, especially in imperfect environments. Although these ANN-based works do not take advantage of environment sensing, they can still demonstrate the feasibility of SoM-enhance-based channel estimation and inspire the research of SoM-enhance-based channel estimation paradigm. We first review the ANN-based channel estimation approaches for general channels to provide broader ideas and visions on the usage of ANNs in the channel estimation task. Ye \textit{et al.} \cite{ye2017power} explored the application of ANN to channel estimation in OFDM systems. The authors prove that the DNN-based method is superior to traditional least square and minimum mean square error methods when the number of pilots is limited through a 5-layer DNN. Although mMIMO and dynamic scenarios are not considered in this scheme, it successfully illustrates the feasibility of using ANNs to learn the complex characteristics of wireless channels. To address the complexity and cost concerns brought by hybrid processing architecture in mmWave mMIMO system, Dong \textit{et al.} \cite{dong2019deep} proposed a spatial-frequency-temporal convolutional neural network (CNN) model which incorporates different types of channel correlations and saves the spatial pilot overhead remarkably. These correlations can also be regarded as a bridge by which multi-modal sensing and wireless channels can be processed to align with each other and then boost the estimation accuracy. The above works \cite{ye2017power,dong2019deep} are designed for static scenarios and will experience performance loss in dynamic scenarios. To this end, Liao \textit{et al.} \cite{liao2019chanestnet} proposed a ChanEstNet which can better adapt to the fast-varying features of high-speed channels. Specifically, the ChanEstNet combines CNN and bidirectional long short-term memory (Bi-LSTM) network to learn the characteristics of fast time-varying and non-stationary channels from a large amount of training data. Similarly, Moon \textit{et al.} \cite{moon2020deep} proposed to use DNN to learn the mapping relationship between omni-received signals and CSI in VCN scenarios where the users only need to transmit only one uplink training sequence. Particularly, \cite{moon2020deep} also utilized the Bi-LSTM network to promote the prediction of the user's channel in real time after obtaining the estimated channel. In terms of channel prediction, multi-modal sensing can potentially provide more abundant environment features and offer a more solid basis for more accurate estimation results by SoM-enhance.
	
	Recently, ANN-based channel estimation methods tailored for beamspace mmWave channels have been proposed to further enhance the estimation accuracy. These methods are more in line with SoM-enhance-based designs since the beamspace channel aligns better with the geometry of environment. More importantly, we may gain some insights into how multi-modal environment information can be processed, matched with beamspace channel, and then boost the estimation accuracy. Ma \textit{et al.} \cite{ma2020sparse} proposed to use ANN to estimate the beamspace channel amplitude for mmWave mMIMO communication systems. Then, the least square algorithm is used for reconstructing the sparse channel estimation results into the original channel. Simple DNN models are adopted in \cite{ma2020sparse} to solve the beamspace channel estimation problem, which may bring performance bottlenecks in certain cases (e.g., scenes of low signal-to-noise ratio or insufficient training data). In light of this, \cite{wei2019ampbeamspace, liu2021beamspace, wei2021beamspace} adopted ANNs with more complicated architecture or exploited the prior information of the beamspace channel to achieve better estimation accuracy. To better exploit the sparsity structure of mmWave channels in the angular domain, Gao \textit{et al.} \cite{gao2022beamspace} proposed to segment the entire beamspace into many small regions with the aid of GPS information and designed a dedicated ANN for each region. Then, the channel estimation accuracy is enhanced by jointly optimizing the region-specific measurement matrix and channel estimator. The role of location information provided by GPS in \cite{gao2022beamspace} demonstrates that side information from multi-modal sensing in SoM can potentially bring benefits to channel estimation, for example, in terms of beamspace division. The above works rely on simulation datasets for network training since real channel data is not available before the scheme is deployed, which may lead to performance loss when applied in realistic scenarios. To address this issue, He \textit{et al.} \cite{he2023beamspace} proposed a model-driven unsupervised learning network which can be trained with limited measurements and applied in new environments. Moreover, the proposed network can also be trained with available real channel data to further improve the estimation performance. The aforementioned ANN-based works demonstrate the feasibility of utilizing ANNs to learn the RF signature of the environment and then recover the beamspace channel. However, only wireless channel data is used as the input data, which is not powerful enough to deal with complicated dynamics. {\color{black}Recently, some researchers have explored the feasibility of using the side information of communication environments provided by multi-modal sensory data to aid channel estimation. In terms of channel statistical characteristics estimation, Xu \textit{et al.} \cite{xu2021deep} designed a DL model that uses the multi-view environmental images collected by the user and user speed information to estimate channel covariance matrix in the absence of the user's location. In terms of real-time CSI estimation, Jiang \textit{et al.} \cite{sensingaidedCE2022} proposed to utilize radar to obtain the distance, velocity, and direction information of the mobile user and scatters, which is directly related to the delay, Doppler, and angle of departure/arrival information. By this means, the pilot signaling overhead required by conventional channel estimation methods for orthogonal time frequency space systems can be effectively reduced. 
		
	To date, extensive channel estimation schemes have been proposed for different communication system settings utilizing DL methods. Although they can achieve satisfactory performance in some cases, simply relying on the limited and coarse environment features obtained from a few pilot signals can lead to inevitable performance degradation when facing complex and dynamic RF environments. We point out that SoM-enhance-based methods, such as \cite{xu2021deep, sensingaidedCE2022}, can effectively improve this defect by injecting RF environment-related features obtained from off-the-shelf multi-modal sensing into channel estimation process. In summary, multi-modal sensing, as one of the key features of SoM, can bring more potential possibilities to break through the performance bottleneck encountered by existing methods in communication systems.}

	\subsection{Dual-Function Waveform Design in Dynamic Scenarios: A Special Case of SoM}
	\label{waveform}
	\subsubsection{Dual-Function Waveform Design}
	Dual-function waveform for sensing and communications is a special case of SoM at the physical layer. Dual-function waveform design focuses on achieving sensing and communication functionalities on a unified hardware platform and a shared spectrum band \cite{liu2020joint}. By allocating the overlapping spectrum resource for both communications and sensing, the signal layer of SoM will tightly couple communications and sensing and bring significant improvements to software flexibility, hardware resources, system size, weight, and power consumption \cite{zhang2021overview,Chengisac}. To this end, RF-ISAC inherently meets the objectives of SoM in the physical layer and enables simultaneous implementation of RF-based sensing, information transmission, and aggregation of environment information. Therefore, RF-ISAC can be considered as a special case of SoM in physical layer and dual-function waveform can serve as one of the sensing modalities.
	
	Existing dual-functional waveforms can in general be classified into three types: communication-centric \cite{commu1,commmu2,commu3,sensingplus1,sensingplus2,sensingplus3}, radar-centric \cite{sensing1,sensing2}, and jointly optimized \cite{joint1,joint2,joint3,joint4}. In communication-centric design, recent studies focus on enhancing sensing capability based on the current communication waveforms. Built upon OFDM, Braun \textit{et al.} \cite{sensingplus1} designed the spacing of subcarriers according to the maximum unambiguous distance and unambiguous speed, but the spacing available for tuning in OFDM is limited. Under a constant transmission power, the power allocation among subcarriers can be optimized for the specific requirement. Liu \textit{et al.} \cite{sensingplus2} optimized the power allocation of subcarriers to maximize the weighted sum of the channel capacity and mutual information of target sensing. Li \textit{et al.} \cite{sensingplus3} took the target detection probability and the communication capacity as indicators to guide power distribution in OFDM. However, the sensing capability is inherently limited due to the unregulated autocorrelation properties of the communication symbols. In radar-centric design, by using radar signals as carriers, communication symbols can be modulated in time, frequency, or other domains. In recent years, more attention has been paid to the opportunistic design based on index modulation, which naturally embeds digital information into radar waveform parameters, such as carrier frequency, time slot, and antenna assignment. For example, the MAJoRCom model proposed in \cite{sensing2} randomly changes the carrier frequency between different pulses, and randomly assigns these frequencies among its antenna units, introducing agility of both frequency and space. Radar-centric design, restricted by its pulse repetition rate, is difficult to be applied in scenarios having high-bandwidth, low-latency, and high-reliability requirements. The joint design works aim at balancing two wireless operations by shaping the radar beam pattern across all directions. Specifically, Liu \textit{et al.} \cite{joint1} achieved the target beam pattern by optimizing the precoding matrix, which satisfies the signal-to-noise ratio (SNR) constraint of each downlink communication receiver. By targeting communication performance as the optimization goal, Liu \textit{et al.} \cite{joint2} designed a waveform under a sensing constraint to minimize interference among multiple users.
	
	\subsubsection{Case Study: A Dual-Function Waveform Design in Dynamic Scenarios}
	
\begin{figure}[!t]
		\centering
		\includegraphics[width=\linewidth]{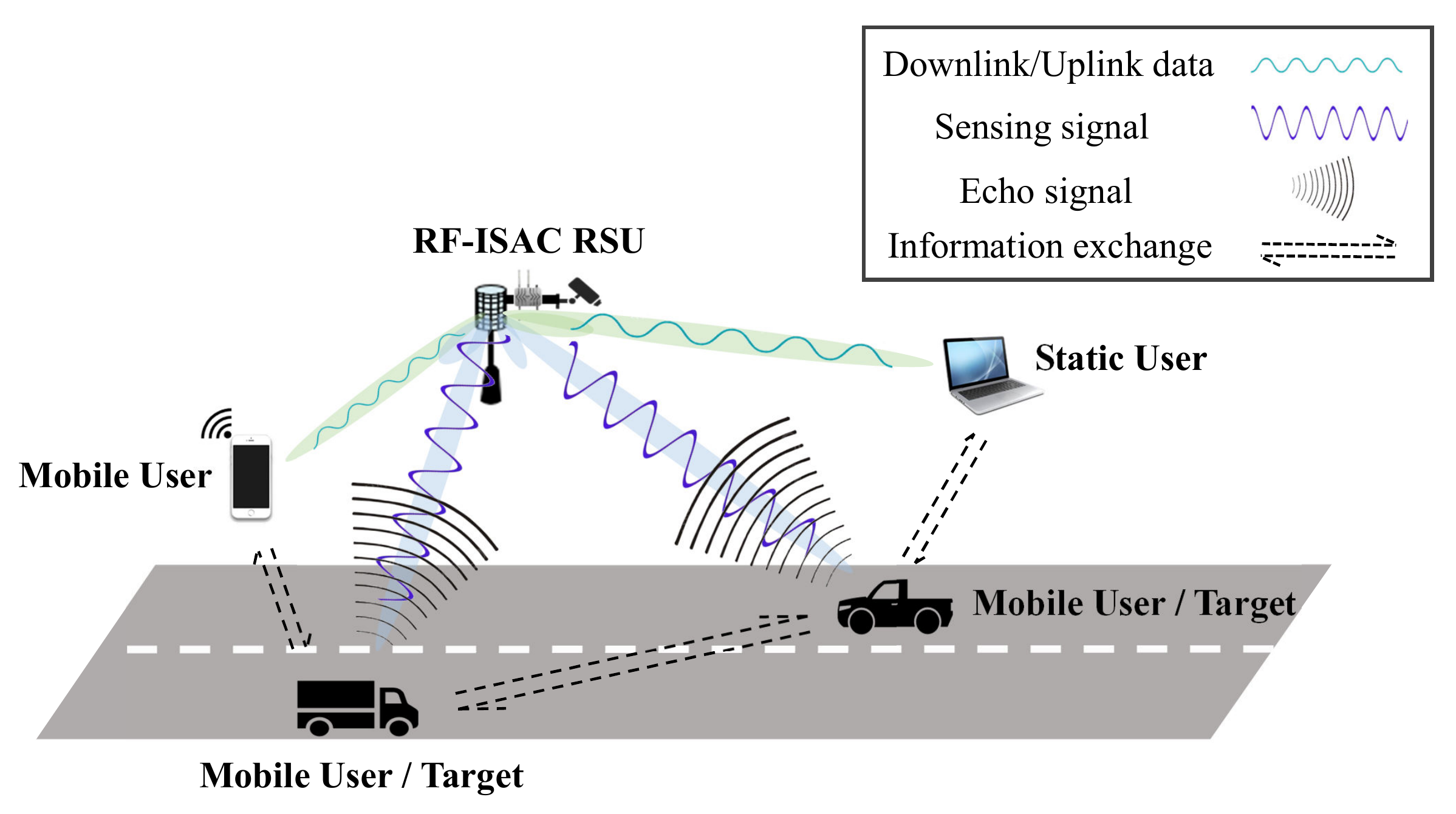}
		\caption{An example of dynamic VCN scenario enabled by RF-ISAC.}
		\label{System}
	\end{figure}

	There are still bottlenecks of dual-function waveform design in dynamic scenarios. When wireless communications move to mmWave band or even sub-Terahertz, the much weaker penetration ability along with the much higher attenuation at surfaces will render unreliable connection, such as an intermittent loss of LoS connection. Also, current RF-ISAC technologies have quite a weak adaptability in dynamic scenarios. As illustrated in Fig.~\ref{System}, the dual-function waveform is transmitted from the RF-ISAC RSU, received by the mobile user, and reflected from the user back to the RSU. RF-ISAC has to meet the demands of simultaneous efficient information transmission to the users and robust mobile target sensing in this dynamic system. However, in that case, significant Doppler shifts caused by the mobility and complicated movement of the targets put forward strict requirements for dual-function waveform design. Moreover, the complicated behavior of mobile users reduces the timeliness of parameter estimation. Several works focused on mmWave transceiver design in time-varying channels. Although the beamspace sparsity has been exploited for offsetting Doppler \cite{shijian1,shijian2,shijian3,shijian4}, the sensing functionality was not considered.
	
	We propose a novel superimposed waveform to fuse the benefits from both communications and sensing. Taking advantage of the high spectral efficiency and energy efficiency of index modulation \cite{im}, the proposed waveform overlays the index modulation-OFDM (IM-OFDM) with radar signal in the frequency domain, with the same power consumption as that of OFDM. Compared to communication-centric designs, the proposed waveform can boost sensing accuracy without sacrificing communication performance. With proper power allocation, the superimposed waveform can decently balance sensing and communication performance to accommodate different applications. 
	
	\begin{figure}[!t]
		{\color{black}
			\centering
			\includegraphics[width=\linewidth]{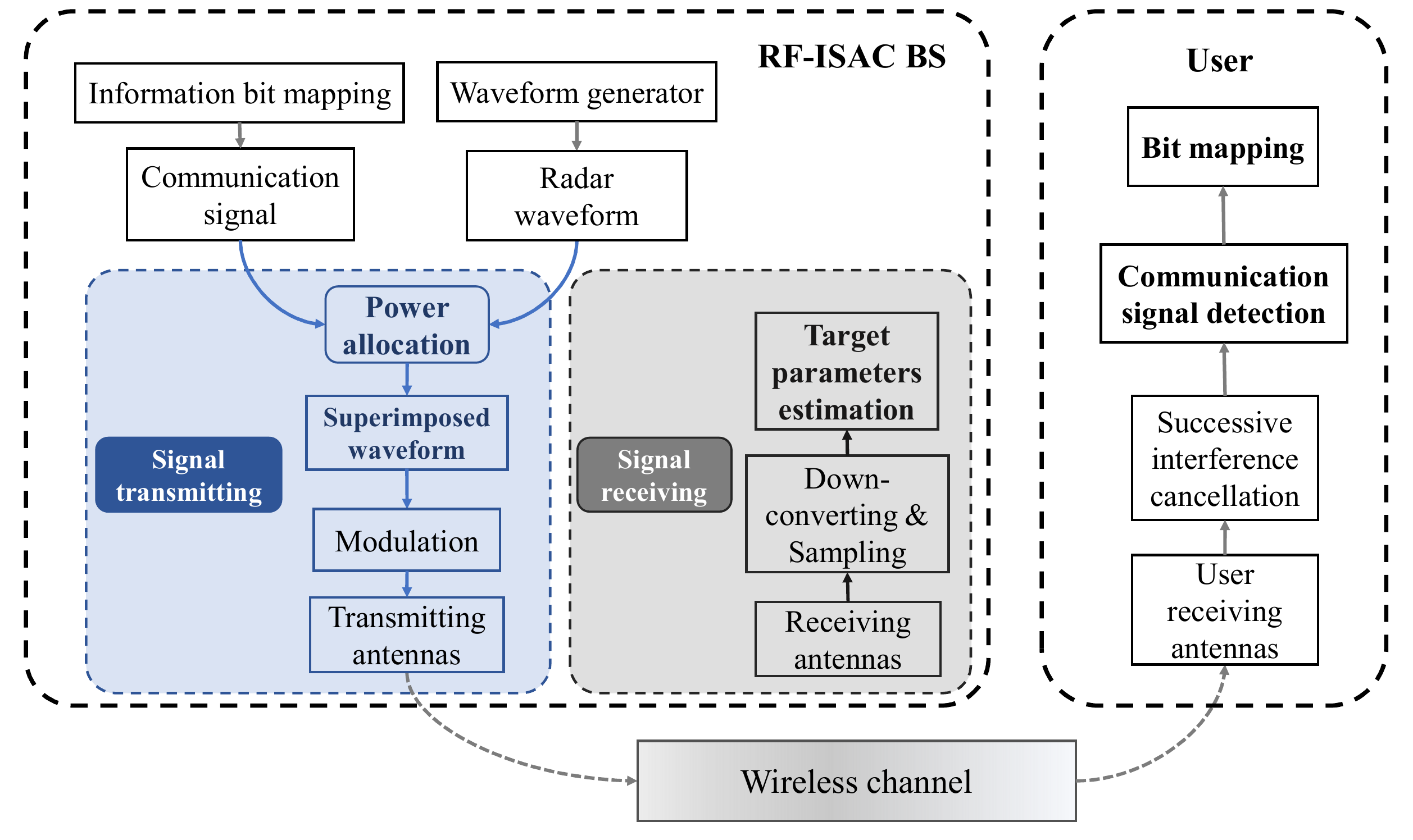}
			\caption{Signal processing framework of the proposed superimposed dual-function waveform.}
		\label{framework}
  }
	\end{figure}
	
\begin{figure}[!t]
		\centering
		\subfloat[]{\includegraphics[width=0.48\linewidth]{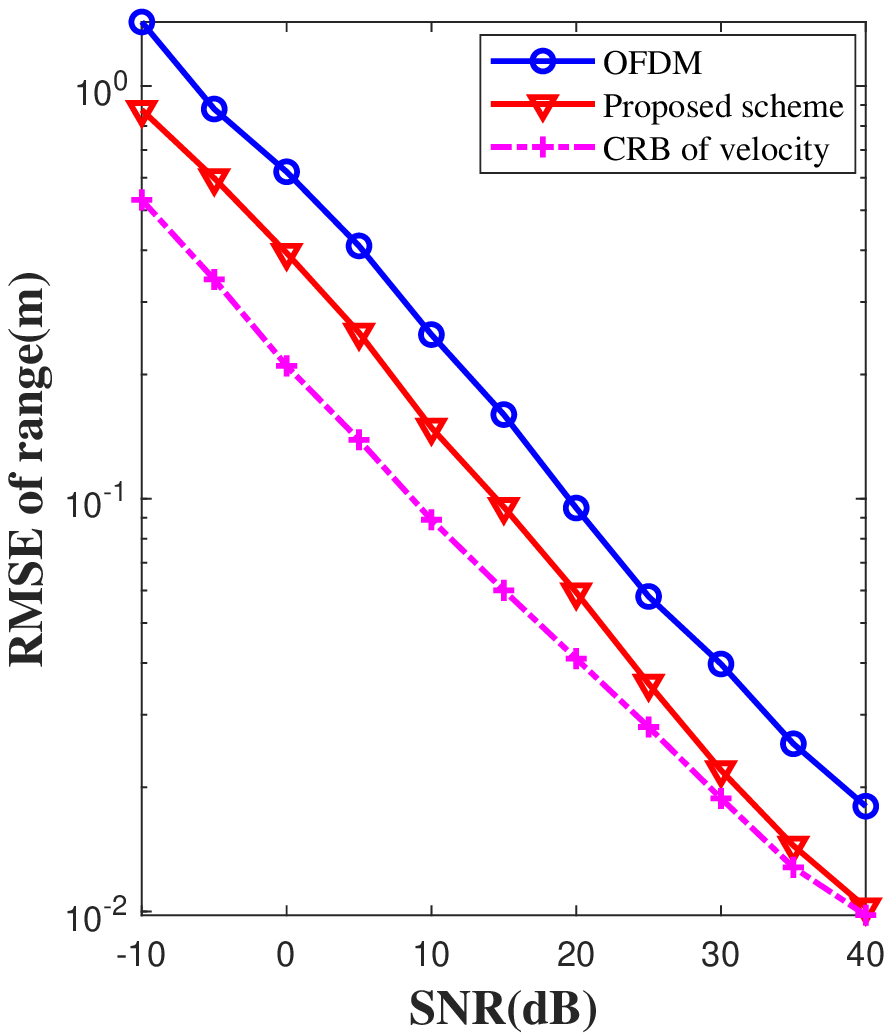}
			\label{simulation_waveform_1}}
		\hfil
		\subfloat[]{\includegraphics[width=0.48\linewidth]{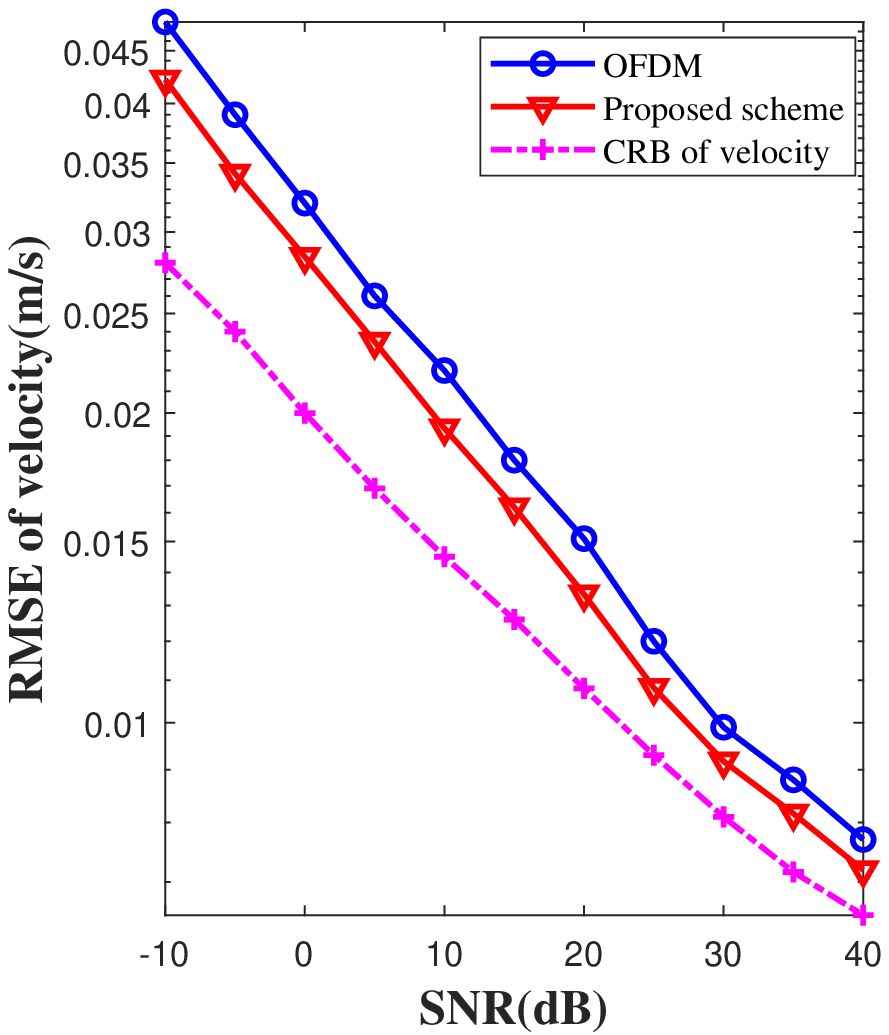}
			\label{simulation_waveform_2}}
		\caption{The range and velocity estimation results comparison of the proposed superimposed dual-function waveform and conventional OFDM: (a) range; (b) velocity. }
		\label{waveformsimu}
	\end{figure}

	The design framework of the superimposed waveform is shown in Fig.~\ref{framework}. In Fig.~\subref*{simulation_waveform_1}, we plot the variation of root mean square error (RMSE) of range estimation versus SNR with the communication-dedicated signal accounting for $40\%$ of the overall power. Under the same transmission power, the superimposed waveform gains up to $5$dB estimation gain over OFDM at high SNR, and the estimation performance is very close to Cram\'{e}r-Rao bound (CRB). In Fig.~\subref*{simulation_waveform_2}, the RMSE of velocity estimation is illustrated. At low SNR, the sensing accuracy with the proposed design can improve $15\%$. At high SNR, the RMSE has a much smaller gap to the CRB bound compared to OFDM.
	
	\subsubsection{Dual-function Waveform Design Considering Multi-Modal Sensing}
	
	When communication and sensing services co-exist in the same resource blocks, it is inevitable to make a compromise between both functionalities. Meanwhile, more and more data is generated by and transmitted across machines. As a result, a wise use of historical data will be beneficial to communications and sensing. In this regard, data can be viewed as a new source for signal design at the physical layer.
	
	\textcolor{black}{For waveform design in RF-ISAC, the use of data can overcome the model deficits such as the array calibration errors, the mutual coupling effects, the power amplifier nonlinearity, and the quantization noise, which facilitates the development of efficient algorithms for addressing sophisticated optimization problems.} Taking advantage of DL algorithms, the physical environment and wireless channel features can be extracted without introducing an additional in-band burden. Therefore, many designs are considering releasing the potential of data to achieve better communication-sensing trade-offs in the waveform design. In \cite{AE}, the authors presented a novel approach for data-driven ISAC using an auto-encoder structure including the proposal of the auto-encoder architecture, a novel ISAC loss function, and the training procedure. To further avoid the difficulty of data annotation in supervised learning, Liu \textit{et al.} \cite{learningbeam} proposed an unsupervised DL-based predictive beamforming scheme to craft well-suitable dual-functional waveforms to underpin mobile communications. 
	
	\textcolor{black}{SoM-enhance-based methods offer promising new opportunities by leveraging multi-modal sensing information such as vision and location inputs collected by various sensors. By incorporating these diverse inputs, the methods acquire a comprehensive set of features that effectively capture and represent the environment. Such a SoM-enhance-based method facilitates real-time monitoring of the environment's state and optimizes the communication-sensing performance in dual-function waveforms. }

	\subsection{SoM-Enhance-Based Beamforming in Dynamic Scenarios}
	\label{som beamforming}
	In the era of 5G, the mmWave band has received wide deployment thanks to its abundant frequency resources. Nevertheless, how to increase the received SNR is an outstanding issue. Through a large number of radiating elements, MIMO enables transmitters to steer the antenna in a certain direction and the beams become narrower. However, when the MIMO beamforming technique is applied in dynamic scenarios, the mobility management of narrow beams becomes a prominent issue. For instance, the rapid movement of UAVs leads to frequent changes in the relative positions between UAVs. Therefore, frequent beam alignment is required to ensure stable communications between UAVs \cite{zeng2019accessing,yang2019beam}. In VCN, the mmWave beam between the high-speed vehicle and the RSU also needs to be frequently configured to ensure the stability of the communication link \cite{kose2021beam}. This is crucial for VCN applications with stringent safety and reliability concerns.
	
\begin{figure}[!t]
		\centering	
		\includegraphics[width=1\linewidth]{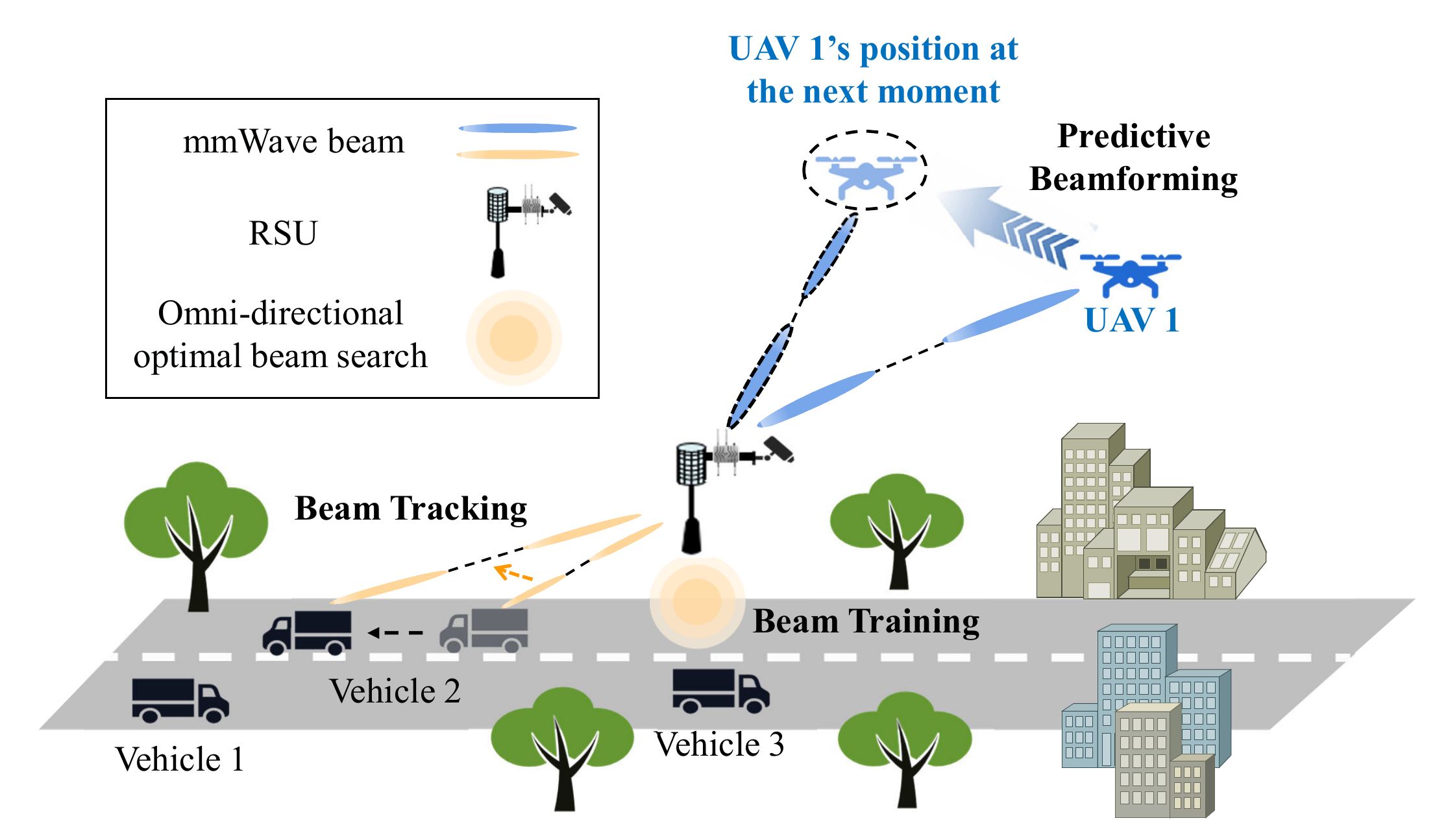}
		\caption{Illustration of beam training, beam tracking, and predictive beamforming in VCN and UAV networks.}
		\label{Beam tracking}
	\end{figure}

	To tackle the challenges in prompt beam management, many efficient beam alignment methods have been proposed in recent years. \textit{Beam training} \cite{wang2009beam,alkhateeb2014channel,tsang2011coding} is a conventional method to find the optimal mmWave beam pairs, which refers to the process of using an omni-directional transmit beam and receive beam to find the beam pairs with the largest received power. The omni-directional exhaustive search leads to large communication overhead and latency. To overcome this limit, the temporal correlation of beamforming angles is expected to reduce the range of exhaustive search since beamforming angles on two consecutive time slots change little, which is known as \textit{beam tracking}  \cite{zhang2019position,xu2021predictive}. However, beam training and beam tracking need to use a certain amount of bits on each data frame to sound channel changes. To further reduce communication overhead, researchers propose to directly perform the beamforming for the next time slot by predicting the position of the moving object in advance, which is known as \textit{predictive beamforming} \cite{liu2020radar,shaham2019fast}. Fig.~\ref{Beam tracking} shows the application of beam training, beam tracking, and predictive beamforming in dynamic scenarios. The goal of predictive beamforming is to directly obtain the target's future position in the absence of exhaustive beam pairing or frequent target detection. In view of this goal, applying SoM to predictive beamforming becomes a natural solution since there is a strong correlation between target's position and environmental features. Therefore, different from SoM-enhance-based channel estimation, it might not be necessary to mine the implied electromagnetic environment features from multi-modal sensing in SoM-enhance-based beamforming. Instead, SoM-enhance-based beamforming needs to extract the 2D position information as well as the motion state information of the target from multi-modal sensing. Since such features are relatively explicit compared with ones needed for channel estimation, challenges faced by SoM-enhance-based beamforming lie in different aspects. We argue that the challenges are reasonable and effective feature extraction methods, as well as the generalization of schemes in different scenarios, such as ad-hoc networks and V2V communications. Some recent works have explored sensing-assisted predictive beamforming but with only a single-type sensor coming into play. In this subsection, we start with a relevant literature review, followed by a case study on SoM-enhance-based beamforming in VCN.
	
	\subsubsection{State-of-the-Art}
	\label{review of beamforming}
	{\color{black}In recent years, extensive works have been carried out to fulfill the predictive beamforming in dynamic environments.} In this subsection, given abundant sensors and stronger dynamics in VCN and UAV networks, we focus on the SoM-enhance-based beam prediction design tailored for VCN and UAV scenarios as well as the SoM-related ones. 
	
	Extended Kalman filter (EKF) is often regarded as a useful tool to dynamically track the states of the moving objects  \cite{liu2020radar,shaham2019fast}. In \cite{liu2020radar,shaham2019fast}, EKF-based predictive beamforming schemes are proposed where DFRC signals are utilized as measurements. Note that angles are not regarded as state variables in \cite{shaham2019fast} since the corresponding calculation of the Jacobians in the EKF algorithm brings high complexity, which guarantees better tracking performance in complex scenarios. However, the sensing accuracy will be affected when the DFRC signals are applied in dynamic scenarios, thereby leading to errors in measurement values and reducing the accuracy of angle prediction. Although these works can roughly reach the goal of tracking state parameters of the passing vehicles, they can only handle naive state evolution models. The high nonlinearity of the vehicle's motion and the subjective behavior of drivers naturally lead to an extremely complicated evolution model that can hardly be described via mathematical models. In light of this, Liu \textit{et al.} \cite{liu2020tutorial} proposed a maximum likelihood-based scheme utilizing the vehicle's historical trajectory points and Xu \textit{et al.} \cite{xu2021predictive} proposed an EKF-based scheme where intelligent vehicles cooperate with the RSU and proactively provide abundant motion information about themselves. Through dedicated designs, \cite{liu2020tutorial,xu2021predictive} can be used to predict beam directions in arbitrary roadway scenarios. Nevertheless, periodic measurements are adopted in \cite{liu2020radar,shaham2019fast,liu2020tutorial,xu2021predictive}, which inevitably results in the accumulation of errors and brings excessive signaling overhead as well as latency. Therefore, new methods that do not depend on mathematical motion models and extra spectrum resources for measurements are worthy of research.
	
	The integration of ANNs and multi-modal environment data in SoM has the potential to fulfill the motion parameters prediction without the need for mathematical state evolution models. Moreover, wide-deployed multi-modal sensors can offer more comprehensive and diverse features while avoiding consuming extra spectrum resources. In VCN, various sensors mounted on the RSUs continuously sense the surrounding traffic environment and the multi-modal environment data captures the position, velocity, and angle information of passing vehicles. In this subsection, {\color{black}we review the recent advances of multi-modal or uni-modal sensing-based beam prediction works} to shed light on SoM-enhance's powerful ability to characterize environmental features. Mu \textit{et al.} \cite{mu2021integrated} proposed to use the fully-connected network to extract simple features from ISAC signals and predict the relative angle. However, the vehicles' motion feature implied in the signal echoes is rough and insufficient to guarantee accurate prediction, especially when a small antenna array is adopted. As the most widely deployed non-RF sensor, cameras can collect 2D visual information with rich environmental information embedded, which is of great help for target positioning. Benefiting from the rapid development of computer vision in recent years, \cite{alrabeiah2020millimeter,xu2023computer} proposed the vision-aided predictive beamforming design schemes based on existing mature DL algorithms or customized ones. In \cite{alrabeiah2020millimeter,xu2023computer}, the problem of optimal beam pairs prediction from the pre-defined codebook is converted to the problem of image classification. Specifically, \cite{alrabeiah2020millimeter} adopted a single-view camera equipped on RSU, which brings signaling overhead for vehicle identification and may violate the privacy of the customer. Although Charan \textit{et al.} \cite{charan2021vision} further enriched the input feature by combining images with beam index information to predict the incoming blockages, they still adopted the single-view camera on the RSU. In light of this, \cite{xu2023computer} proposed to use the multi-view cameras equipped on the intelligent vehicle as an alternative for beam alignment, and utilized 3D detection technique to obtain the 3D spatial distribution information of the vehicle's surrounding environment. LiDAR is one of the key sensors for autonomous vehicles and is capable of visualizing the 3D world by continuously emitting laser beams. With the wider application of LiDAR, its powerful ability to distinguish target objects inspires researchers to explore how it can be applied in predictive beamforming design \cite{klautau2019lidar,dias2019position,zheng2021deep}. Klautau \textit{et al.} \cite{klautau2019lidar} transformed the LiDAR raw data into a 3D histogram and designed a DL algorithm to extract simple features to complete the LoS decision and beam selection tasks. Instead of using LiDAR raw data as a 3D histogram \cite{klautau2019lidar}, Dias \textit{et al.} \cite{dias2019position} proposed to convert the histogram elements into binary values to indicate whether the histogram counter is larger than 0, thereby reducing the storage resource requirement. In order to further enhance the scalability of classification algorithms in large-scale antenna configuration, Zheng \textit{et al.} \cite{zheng2021deep} combined the latest DL design philosophy and the LiDAR data, and designed two DL-based beam pair selection schemes that are based on 3D convolutional neural networks and 2D convolutional neural networks. In addition to cameras and LiDAR, radar is also a widely-deployed sensor in autonomous driving. Radar has the ability to measure the range, velocity, and azimuth of multiple targets simultaneously. However, due to its poor atmospheric transmission characteristics, it is typically insufficient to track motion parameters in high precision. In \cite{demirhan2022radar}, FMCW radar is utilized to sense the environment. The raw radar data is first processed by the three different pre-processing approaches to obtain the range-angle maps, range-velocity maps, and radar cubes, which are regarded as the input of the designed ANN. {\color{black}Considering the dynamic deployment conditions of DL models, Gu \textit{et al.} \cite{gu2023meta} proposed to use meta-learning to address the generalization issues of the pre-trained beam selection model in non-line-of-sight (NLoS) scenarios. 
		
    The above works merely utilize one modality of sensory data to aid in beam selection, while the utilization of multiple sensor modalities enhances prediction performance by offering a more comprehensive representation of the environment. Salehi \textit{et al.} \cite{chowdhuryFLASH} proposed a multi-modal federated learning framework to enable vehicles to collaboratively participate in learning the shared prediction model while the raw multi-modal sensory data is not required to be uploaded to a central server. The utilization of multi-modal data makes the framework resilient to situations where certain data modalities are missing due to practical issues. Likewise, Salehi \textit{et al.} \cite{chowdhurymultimodal} proposed to leverage multi-modal sensory data collected from both vehicles and BS to aid in beam selection. To reduce the communication overhead caused by raw data transmission and the end-to-end latency, the authors adopted a distributed inference scheme to compress the raw data into high-level features at the vehicle. Yang \textit{et al.} \cite{feifeisemantic} also noticed the storage, computation, and privacy issues related to raw sensory data. In \cite{feifeisemantic}, Yang \textit{et al.} conducted an in-depth analysis of extracting environmental semantics from multi-modal sensing and proposed a framework for channel-oriented semantic communications. Based on such a framework, the authors developed an environmental semantics aided beam prediction and blockage prediction scheme. Note that extensive works have realized the application of multi-modal sensing in predicting the optimal beam pair from the pre-defined codebook, which is actually a classification problem in the deep learning community.} However, when the angular resolution of the utilized pre-defined codebook is low, inaccurate alignment between beam pairs will inevitably occur when the vehicle is in similar positions and the achievable communication rate will reduce. Although improving the resolution of codebook can address this problem to some extent, it also poses greater challenges to the beam prediction task as the total number of categories increases. Therefore, utilizing multi-modal sensing to aid in continuous motion parameter tracking and predictive beamforming still calls for more research efforts in the future. In summary, SoM-enhance-based predictive beamforming design is a promising candidate technique to meet the stringent communication requirements in B5G/6G networks while decreasing the training overhead and improving the system efficiency.
	
	{\color{black}UAVs are the fundamental components of the next-generation aerial networks and the key enabler for the futuristic applications such as security monitoring systems \cite{alkhateebDRONE}. Different from vehicles or robots which typically move in the azimuthal plane without position changes in the elevation plane, UAVs move in six degrees of freedom. Such a highly mobile nature requires frequent alignment of narrow beams to guarantee sufficient receive SNR, which motivates researchers to develop sensing information based solutions to address this challenge. UAVs are mostly equipped with the IMU and the GPS integrated navigation system. Zhang \textit{et al.} \cite{zhang2019position} noticed this and designed a Gaussian Process-based motion prediction scheme, involving the exchange of measured motion state among UAVs. By predicting the position and attitude of the UAV, the optimal beamforming vector can be determined. Similarly, Miao \textit{et al.} \cite{miao2020lightweight} proposed a lightweight 3-D predictive beamforming design for maintaining the backhaul links with GPS offering positioning information. Although the above works successfully demonstrate that simple side information provided by sensors like positioning information can improve the beam alignment precision, Charan \textit{et al.} \cite{alkhateebDRONE} pointed out that merely relying on position information alone might not be sufficient by designing both vision-aided and position-aided beam prediction solutions and comparing their results. In the developed vision-aided solution, Charan \textit{et al.} \cite{alkhateebDRONE} proposed to use the RGB camera mounted on the BS to help predict the best beam rather than only utilizing the positioning information collected by the UAVs. Different from \cite{alkhateebDRONE}, \cite{alkhateebDRONE2} leveraged images collected from cameras mounted on UAVs to aid in fast and accurate beam prediction. Currently, SoM-enhance-based beam prediction schemes tailored for UAV networks have not been studied systematically but the high beam prediction accuracy achieved in \cite{alkhateebDRONE} has successfully proven the rationality and potential of SoM-enhance.
		
    Table~\ref{tab:beamforming} lists typical related works of predictive beamforming design in VCN and UAV scenarios. It can be concluded that extensive works have been conducted to apply machine learning methods to predictive beamforming due to their outstanding feature learning and predictive abilities, such as RL and ANNs. Among them, SoM-enhance-based methods leverage the rich environmental features presented in multi-modal sensory data, thereby reducing signaling overhead in communication systems while ensuring high prediction accuracy. It is noteworthy that our SoM-enhance-based predictive beamforming scheme which will be introduced in the next subsection fulfills the continuous parameter prediction rather than the commonly adopted optimal beam pairs prediction.
	}

	\begin{table*}[!t]
		\renewcommand\arraystretch{1.6} 
		\centering
		\caption{Typical related works of predictive beamforming design in VCN and UAV scenarios.}
		\label{tab:beamforming}
		\resizebox{\textwidth}{!}{
			\begin{tabular}{c|c|c|c|c|c}
				\toprule[0.35mm]
				\makecell[c]{\textbf{Ref.}}	&	\makecell[c]{\textbf{Application}\\\textbf{scenarios}} & \makecell[c]{\textbf{Methods of tracking}\\\textbf{the motion parameters}}	 & 	\makecell[c]{\textbf{Whether}\\\textbf{SoM-enhance}\\\textbf{is applied}} &	\makecell[c]{\textbf{Sensory data}}& \textbf{Wireless channel data}\\
				\midrule[0.15mm]
				\cite{liu2020radar,shaham2019fast}	& \makecell[c]{VCN \\ Scenario of straight lane}  & EKF & $\times$ & $\times$ & DFRC signal echoes \\ 
				\midrule[0.15mm]
				\makecell[c]{\cite{learningbeam}}	& \makecell[c]{VCN \\ Scenario of straight lane} & ANN-based methods & $\times$ & $\times$ & Historical channels \\ 
				\midrule[0.15mm] 
				\makecell[c]{\cite{mu2021integrated}}	& \makecell[c]{VCN \\ Scenario of straight lane} & ANN-based methods & $\times$ & $\times$ & DFRC signal echoes \\ 
				\midrule[0.15mm]			\cite{9777748}	& \makecell[c]{VCN \\ Arbitrary roadway geometries} & RL  & \checkmark & RGB images & $\times$\\ 
				\midrule[0.15mm]			\cite{klautau2019lidar,dias2019position,zheng2021deep}	& \makecell[c]{VCN \\ Arbitrary roadway geometries} & \makecell[c]{ANN-based method \\ (Optimal beam pairs prediction)}  & \checkmark & LiDAR & $\times$\\ 
				\midrule[0.15mm]			\cite{alrabeiah2020millimeter,xu2023computer}	& \makecell[c]{VCN \\ Arbitrary roadway geometries} & \makecell[c]{ANN-based method \\ (Optimal beam pairs prediction)} & \checkmark & RGB images & $\times$ \\ 
				\midrule[0.15mm]			\cite{demirhan2022radar}	& \makecell[c]{VCN \\ Arbitrary roadway geometries} & \makecell[c]{ANN-based method \\ (Optimal beam pairs prediction)} & \checkmark & FMCW Radar & $\times$ \\
				\midrule[0.15mm]			{\color{black}\cite{gu2023meta}}	& {\color{black}\makecell[c]{VCN \\ Arbitrary roadway geometries}} & {\color{black}\makecell[c]{ANN-based method \\ Meta-learning\\(Optimal beam pairs prediction)}} & {\color{black}\checkmark} & {\color{black}\makecell[c]{RGB Image}} & {\color{black}$\times$} \\
				\midrule[0.15mm]		{\color{black}\cite{chowdhuryFLASH}}	& {\color{black}\makecell[c]{VCN \\ Arbitrary roadway geometries}} & {\color{black}\makecell[c]{ANN-based method \\ Federated learning\\(Optimal beam pairs prediction)}} & {\color{black}\checkmark} & {\color{black}\makecell[c]{RGB Image\\GPS\\LiDAR}} & {\color{black}$\times$} \\
				\midrule[0.15mm]			{\color{black}\cite{alkhateebDRONE}}	& {\color{black}UAV-BS communications} & {\color{black}\makecell[c]{ANN-based method \\ (Optimal beam pairs prediction)}} & {\color{black}\checkmark} & {\color{black}\makecell[c]{RGB images \\ GPS \\ from BS}} & {\color{black}$\times$} \\ 
				\midrule[0.15mm]			{\color{black}\cite{alkhateebDRONE2}}	& {\color{black}UAV-BS communications} & {\color{black}\makecell[c]{ANN-based method \\ (Optimal beam pairs prediction)}} & {\color{black}\checkmark} & {\color{black}\makecell[c]{RGB images\\ from UAVs}} & {\color{black}$\times$} \\ 
				\midrule[0.15mm]
				\textbf{Our Work \cite{MMFF}}	& \makecell[c]{VCN \\ Arbitrary roadway geometries} & \makecell[c]{ANN-based method \\ (Continuous parameter prediction)} & \checkmark & \makecell[c]{RGB images \\ Depth map} &  \makecell[c]{CSI at sub-6 GHz\\frequency band}\\ 
				\bottomrule[0.35mm]
		\end{tabular}	}
	\end{table*}

	\subsubsection{Case Study: A SoM-Enhance-Based Predictive Beamforming Design in VCN}
	
\begin{figure}[!t]
		\centering
		\includegraphics[width=1\linewidth]{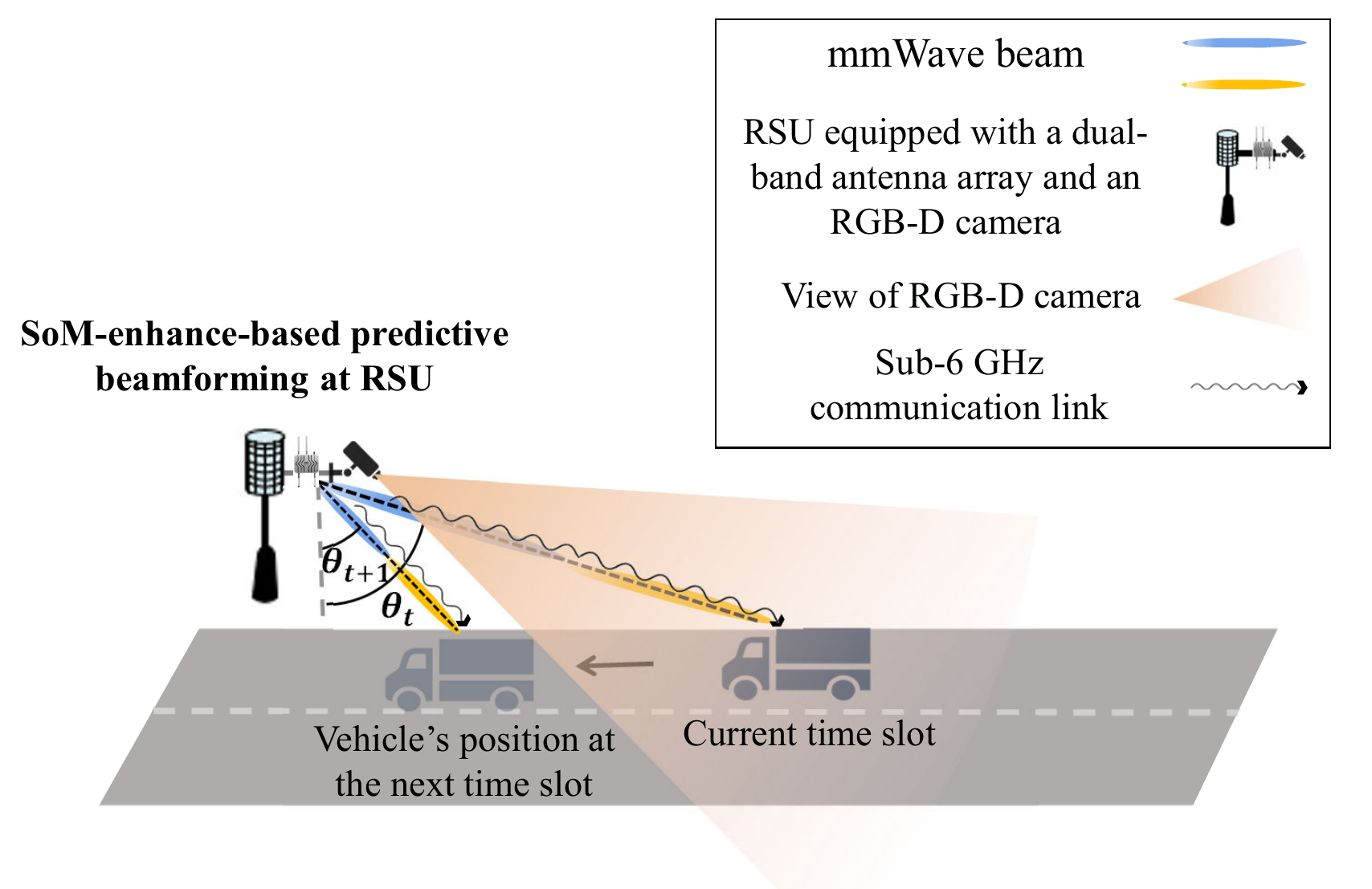}
		\caption{Illustration of the SoM-enhance-based predictive beamforming scheme in a VCN scenario.}
		\label{illus}
	\end{figure}

	In this subsection, the performance gain that SoM brings to the predictive beamforming in the VCN scenario is verified through a simulation. Fig.~\ref{illus} shows a VCN scenario where the RSU performs predictive beamforming with the aid of RGB-D cameras and communication equipment. As discussed in Section \ref{review of beamforming}, the existing predictive beamforming schemes mostly use uni-modal sensory data, such as LiDAR point clouds, RGB images, and radar echoes. Most importantly, they have only realized the prediction of the optimal beam pairs in the pre-defined codebook with low resolution, {\color{black}rather than continuously tracking the vehicle's motion parameters.} Therefore, in this subsection, we present a SoM-enhance-based predictive beamforming scheme \cite{MMFF} that is capable of continuously tracking the angle parameter of the passing vehicle and predicting its future position.
	
	In the proposed scheme, RSU periodically communicates with the passing vehicle and senses the traffic environment through multi-modal sensors. We assume that an RGB-D camera\footnote{Note that depth maps added with Gaussian noise are used for mimicking frequent radar sensing results in the simulation.} is mounted on RSU. In terms of wireless channel data, we assume that the RSU is equipped with the mMIMO uniform linear array that operates at both the mmWave band and the  sub-6 GHz bands. The proposed scheme takes that RSU utilizes the necessary control signaling transmitted at the sub-6 GHz frequency band to obtain the CSI. By this means, the electromagnetic environment changes caused by the target vehicle are obtained without the need for intentionally transmitting any extra pilot. Although the frequency bands are different, the impact of the vehicle on the signal propagation process in the mmWave frequency band and sub-6 GHz frequency band shows similarity \cite{gao2021fusionnet}. In summary, RGB images and radar ranging results are expected to provide vehicles' 2D (visual, distance) features, and the CSI of the low-frequency band is expected to provide the angular feature. We have designed an ANN to efficiently learn the feature of the vehicle's motion from the multi-modal sensory data and wireless channel data. The designed ANN comprises CNN and fully-connected networks to extract the motion feature from RGB images, radar ranging measurements, and wireless channels. In the considered simulation scenario, the target vehicle moves from one side of the RSU to the other with random drifting behavior, which is consistent with the actual situation. In this process, the RSU constantly performs angle prediction, then completes predictive beamforming periodically. 
	
\begin{figure}[!t]
		\centering
		\subfloat[]{\includegraphics[width=0.48\linewidth]{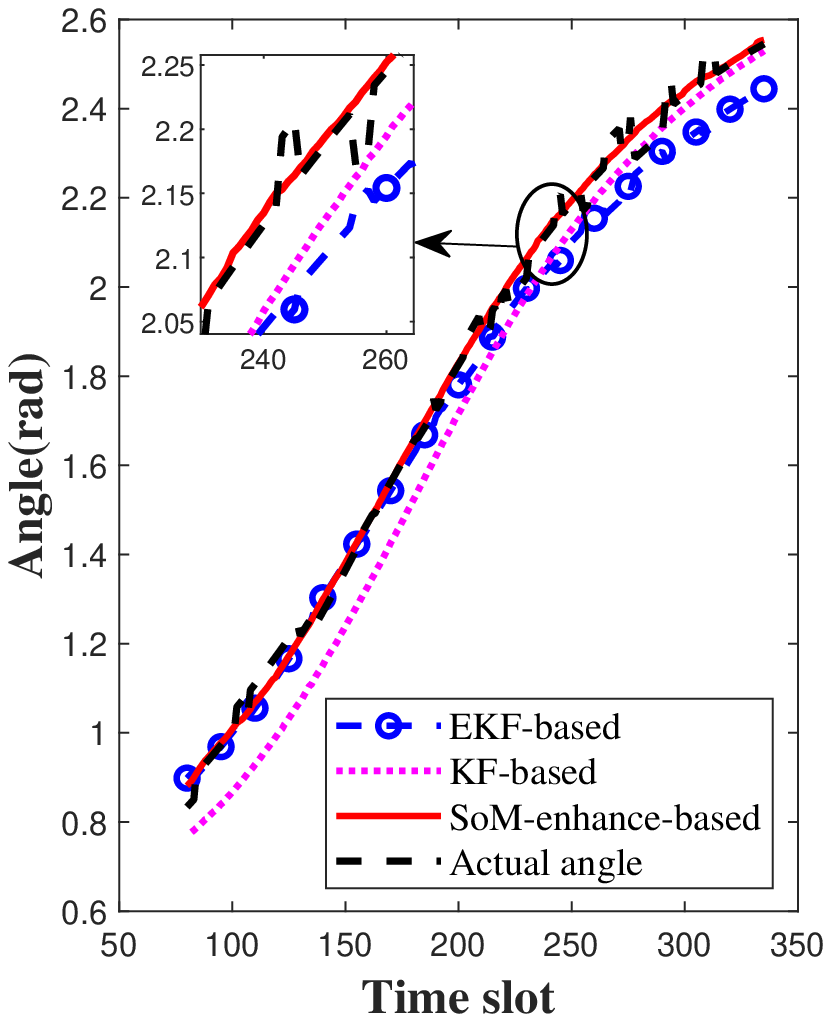}
			\label{angle}}
		\hfil
		\subfloat[]{\includegraphics[width=0.48\linewidth]{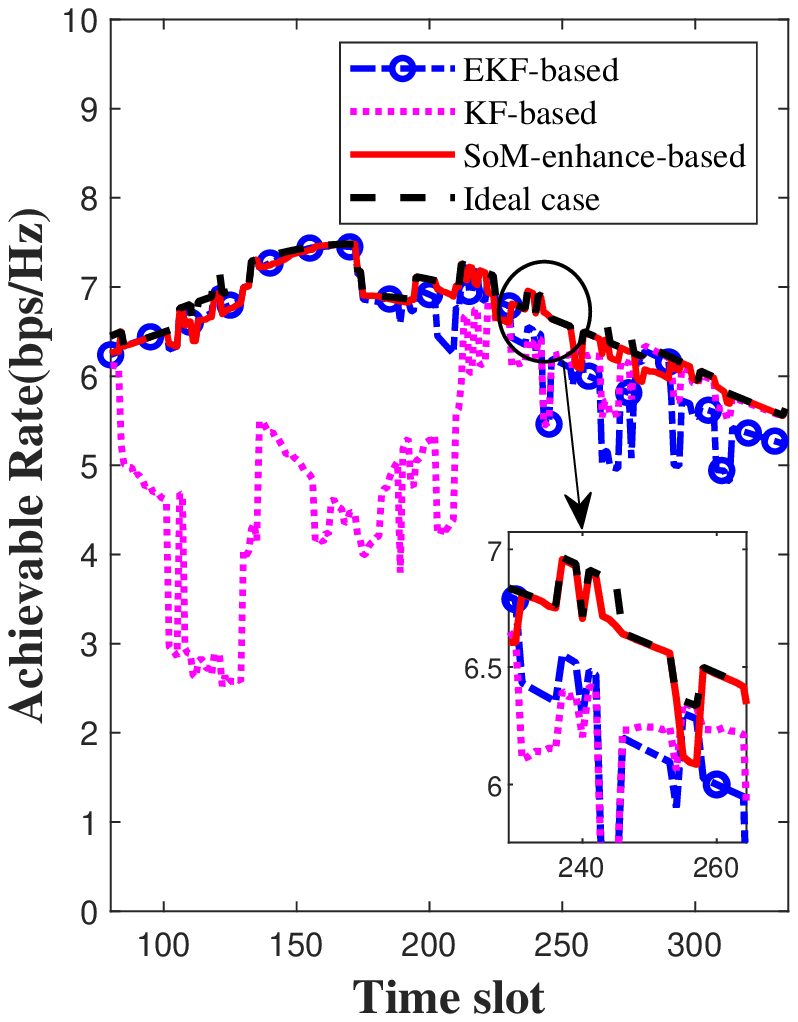}
			\label{rate}}
		\caption{Angle prediction results and achievable rates comparison of the SoM-enhance-based predictive beamforming scheme and other schemes: (a) angle prediction results; (b) achievable rates. }
		\label{beamsimu}
	\end{figure}

	The angle prediction results obtained by the designed SoM-enhance-based predictive beamforming scheme are compared with those obtained by the Kalman filter (KF) algorithm and the EKF algorithm. As shown in Fig.~\subref*{angle}, the KF algorithm cannot effectively track the movement of vehicles, because it is only applicable to linear systems. As discussed in Section \ref{review of beamforming}, most EKF-based methods assume that the vehicle moves along an ideal straight line since the nonlinear motion model brings extremely high computational complexity. Moreover, implementing the EKF algorithm requires observation values to rectify the prediction values. Consequently, RSU has to rely on extra spectrum resources to detect the target vehicle. The angle prediction results obtained by the EKF-based scheme are slightly worse than those obtained by the SoM-enhance-based scheme and the error of EKF-based scheme is gradually accumulating, as shown in Fig.~\subref*{angle}. Thanks to the accurate position feature in multi-modal sensory data and the angle evolution feature in wireless channel data, the SoM-enhance-based predictive beamforming scheme has higher prediction accuracy and stronger robustness to the vehicle's drifting behavior without a spectrum resource sacrifice. 
	
	The strong robustness of the SoM-enhance-based predictive beamforming scheme to the drifting behavior is also shown in Fig.~\subref*{rate}, where the achievable rate of the communication link based on the angle prediction results is shown. The numbers of transmit antenna and receive antenna at mmWave frequency band are set to $8$. As the vehicle passes by, its distance from the RSU decreases first and then increases, while the achievable rate behaves inversely. The achievable rate realized by the SoM-enhance-based predictive beamforming scheme is the closest to the ideal case. When the vehicle nears the RSU, the relative angle changes rapidly. As shown in Fig.~\subref*{rate}, the angle prediction error of the EKF-based scheme is gradually accumulating, leading to a rate decline. Due to the large angle prediction error via the KF-based scheme, the corresponding achieved rate is the lowest at most time slots. 
	
\begin{figure}[!t]
		\textcolor{black}{
			\centering
			\includegraphics[width=1\linewidth]{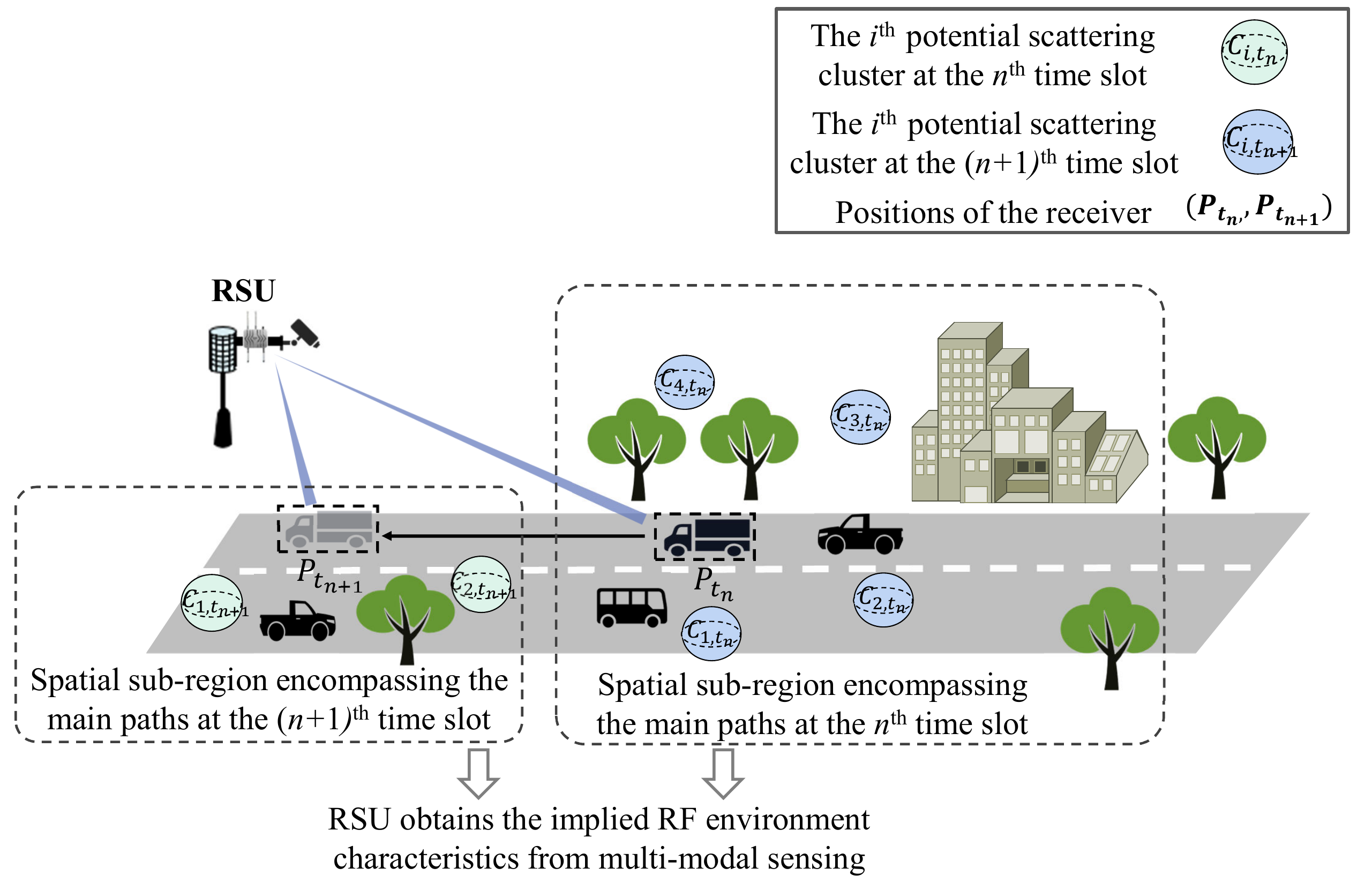}
			\caption{The potential application scenarios and possible research ideas of SoM-enhance-based channel estimation.}
			\label{estimation}}
	\end{figure}

	{\color{black}\subsection{Key Takeaways}
		In this section, we discuss the recent advances and technique challenges of SoM-enhance-based communication functionality enhancement, spanning from channel estimation, waveform design, and beamforming design. Key takeaways learned are as follows.}
	\begin{itemize}
		\item
		{\color{black}Conventional channel estimation methods that recover CSI through estimating the key parameters of a pre-assumed channel model are no longer applicable in dynamic scenarios since the network topology, channel characteristics, and corresponding channel model may vary with time. Currently, research on SoM-enhance-based channel estimation is still well underway. Since mmWave channels align better with the physical environment and are potentially associated with multi-modal sensing, relevant channel estimation schemes for mmWave communication systems may inspire the research on SoM-enhance-based channel estimation despite lack of multi-modal sensing. Benefiting from multi-modal sensing, SoM-enhance can streamline the conventional channel estimation process by utilizing the implied RF environment characteristics that are matched with the geometry of environment. Specifically, the scattering characteristics of the channel in various spatial sub-regions can be inferred from non-RF sensing. As illustrated in Fig.~\ref{estimation}, RSU can obtain the real-time environment information surrounding the receiver. Since objects near the receiver are likely to be highly correlated with the scattering clusters that cause multipath effects, RSU can transmit limited pilot signals to the sub-region encompassing the main propagation paths, thereby saving spectrum resources. Furthermore, the measurement of state information of objects by multi-modal sensing means that key parameters related to channel characteristics, such as Doppler, delay, AoA, and AoD, can be indirectly obtained. In summary, the application of SoM-enhance in channel estimation is a natural solution and deserves further study. }
	\end{itemize}
	
	\begin{itemize}
		\item
		{\color{black}Dual-function waveform, i.e., RF-ISAC, is a special case of SoM at the physical layer. Unfortunately, the existing waveform designs have weak applicability in dynamic scenarios. In this section, we have proposed a superimposed IM-OFDM to tackle time variation and boost sensing capability. To enhance the performance of dual functionalities, it is crucial to improve the utilization of data to enable a non-linear mapping from the real physical environment to the optimal waveform. However, relying solely on past wireless channel states for both sensing and communications purposes is still unreliable. An effective solution is to innovate conventional physical-layer waveform design by incorporating data-driven methods and multi-modal sensing. This approach empowers communication systems to selectively extract channel features and enhance the dual functionalities of the waveform. By acquiring key prior information about wireless channels, such as multipath and scattering, through multi-modal sensing, the signal processing algorithms and corresponding parameters can be more accurately designed, ultimately maximizing communication and sensing performances. Moreover, the sensing capability of the waveform can be further leveraged to complement existing sensors since the changes in the echo contain a wealth of physical environment information, such as directions and obstacles. The overall sensing performance can be augmented by fusing the sensing results, leading to improved machine recognition.
		}
	\end{itemize}

	\begin{itemize}
		\item
		The mobility management of mmWave narrow beams has always been a prominent issue in dynamic scenarios. Currently, many research efforts have been made to utilize uni-modal or multi-modal sensing for optimal beam pair prediction from pre-defined codebook. {\color{black} However, such schemes utilize the same beam for users in close positions, which inevitably leads to misalignment and reduces the achievable rates. SoM-enhance-based predictive beamforming promises to align the beams with low signaling overhead and high angular resolution thanks to the abundant features of the target's motion state provided by multi-modal environment information. Furthermore, SoM-enhance-based schemes also have the potential of adaptively adjusting the beamwidth and tracking targets with more complex behavior for practical considerations thanks to large-scale, multi-property, and fine-grained environmental awareness. By comprehensively monitoring and predicting the state information of the receiver and the surrounding environmental information through such environmental awareness, real-time and accurate beam management can be achieved. In addition, existing works mostly utilize uni-modal sensory data, which leads to their sensitivity to weather conditions. Therefore, leveraging multi-modal sensing to enhance the weather robustness of predictive beamforming methods deserves further research.
		}
	\end{itemize}

	\section{SoM-Concert: Reinforced Environment Sensing}	
	\label{SoMConcert} 
	As stated earlier, communication systems can be tapped for more potential under the guidance of SoM-evoke and SoM-enhance. During the operation of dynamic agents, they also need to sense the environment in real time to support complex applications and improve intelligence. This leads to a fundamental task for mobile agents, namely environment sensing. In the B5G/6G era, communications and RF-based sensing are closely coupled as the frequency band that the communication systems operate at continuously increases. \textcolor{black}{In this way, communication networks are not just a medium for information exchange, but more of a sensing unit overseeing the wireless electromagnetic environment.}  Moreover, as non-RF-based and RF-based sensors with distinct functionalities are densely deployed in the environment and on the agents, sensing functionality is envisioned to be native to future wireless communication systems \cite{Rahman2020frame}, rather than an independent function. Furthermore, just as introduced in Section~\ref{Intro of SoM}, one of SoM's objectives is to realize reinforced environment sensing. In this section, we will detail how SoM-concert fulfills this goal, with which situational awareness and intelligence can be significantly enhanced in dynamic systems.
	
	\subsection{Necessity of SoM-Concert for Environment Sensing}
	\label{neccesity}
	
	In a typical sensing system for the mobile agent, onboard sensors, including both non-RF and RF, are utilized to collect environmental information. The information takes on a multi-modal form, such as pixels for RGB-D cameras and 2D/3D points for radar and LiDAR. Based on the rich environmental information contained in sensory data, ANNs are widely used to extract environmental features and obtain sensing results. However, the redundancy of environmental information brought by a single agent through a single type of sensor is far from enough in a dynamic environment. Firstly, the limited LoS of the sensor makes it difficult to capture long-distance yet non-negligible environmental information in a dynamic environment. Secondly, ambient occlusion in complex scenarios can greatly reduce the FoV. Moreover, harsh weather degrades the quality of sensory data and thus reduces the reliability of sensing. 
	
	As a result, the mechanism of SoM-concert needs to be introduced to integrate the environmental information to improve the accuracy and robustness of sensing. The existing works on SoM-concert for environment sensing can be classified into two categories, i.e., single-agent sensing and multi-agent sensing. In single-agent sensing, multi-modal sensory data and sensing results of multiple on-board sensors can be integrated to reinforce each other. This type of approach does not rely on interaction with other agents and infrastructure, which possesses low deployment complexity but falls short in sensing range and reliability. In multi-agent sensing, communication modules are utilized as the media of information transmission to break through the limitations of single-agent sensing. Similar to single-agent sensing, the integration of environment information also exists but is manifested in multi-agent level. These two levels of SoM-concert for environment sensing can mutually reinforce each other, both driven by the same goal of improving sensing capability. According to the stage of multi-modal fusion in the sensing process, single-agent sensing can be classified into early fusion, late fusion, and hybrid fusion. Correspondingly, according to the stage of multi-agent information interaction in the sensing process, multi-agent sensing can be classified into raw-data-level fusion, feature-level fusion, and semantic-level fusion.

	\subsection{SoM-Concert for Single-Agent Sensing}
	\label{multimodal}
	For a single dynamic agent equipped with no less than two kinds of sensors, its sensing system can improve the redundancy of sensing information by utilizing complementary properties of different kinds of sensors to improve environment sensing capability by SoM-concert. For example, the camera can collect dense semantic information from the surrounding environment but suffer from a lack of depth, correspondingly, LiDAR point clouds are highly structured but relatively sparse. RGB-D images from RGB-D cameras ensure density while containing depth information, but the measurement range is very limited. Importantly, dynamic environments involve harsh scenarios such as dim light, heavy rain, and snow. RGB-D cameras and LiDARs are affected by inadequate light and scattering, respectively, which may result in insufficient environmental information collected and lead to a significant decline in sensing effects. As supplementary, RF-based sensors like radars have attracted more attention in spite of a relatively weaker ability to capture environmental information. In harsh environments, radar can maintain stronger robustness due to operating frequency and imaging principle. Some other properties like direct velocity measurement can also enhance the applicability of radar. In summary, multi-modal fusion is essential to improve the sensing ability of a single agent.
	
\begin{figure}[!t]
		\centering
		\subfloat[]{\includegraphics[width=0.5\textwidth]{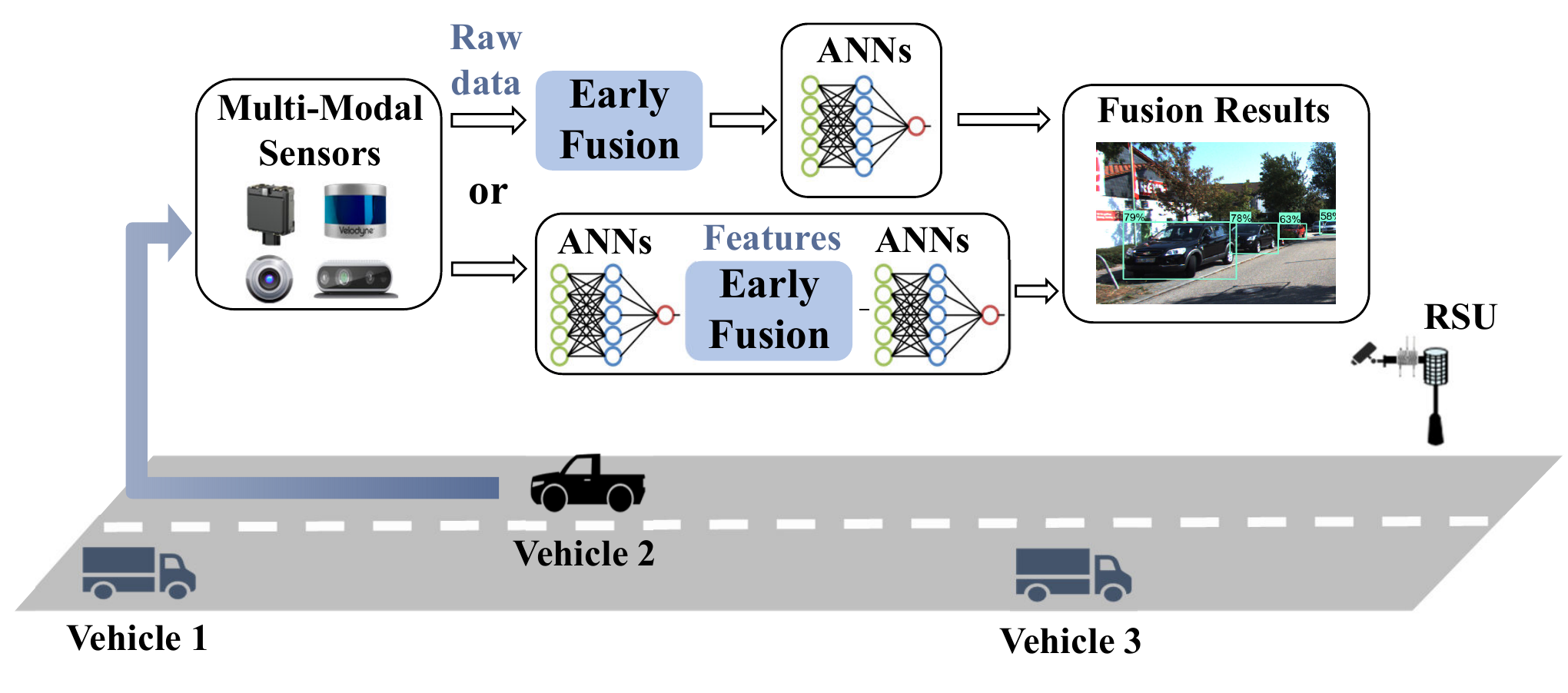}%
			\label{early}}
		\hfil
		\subfloat[]{\includegraphics[width=0.5\textwidth]{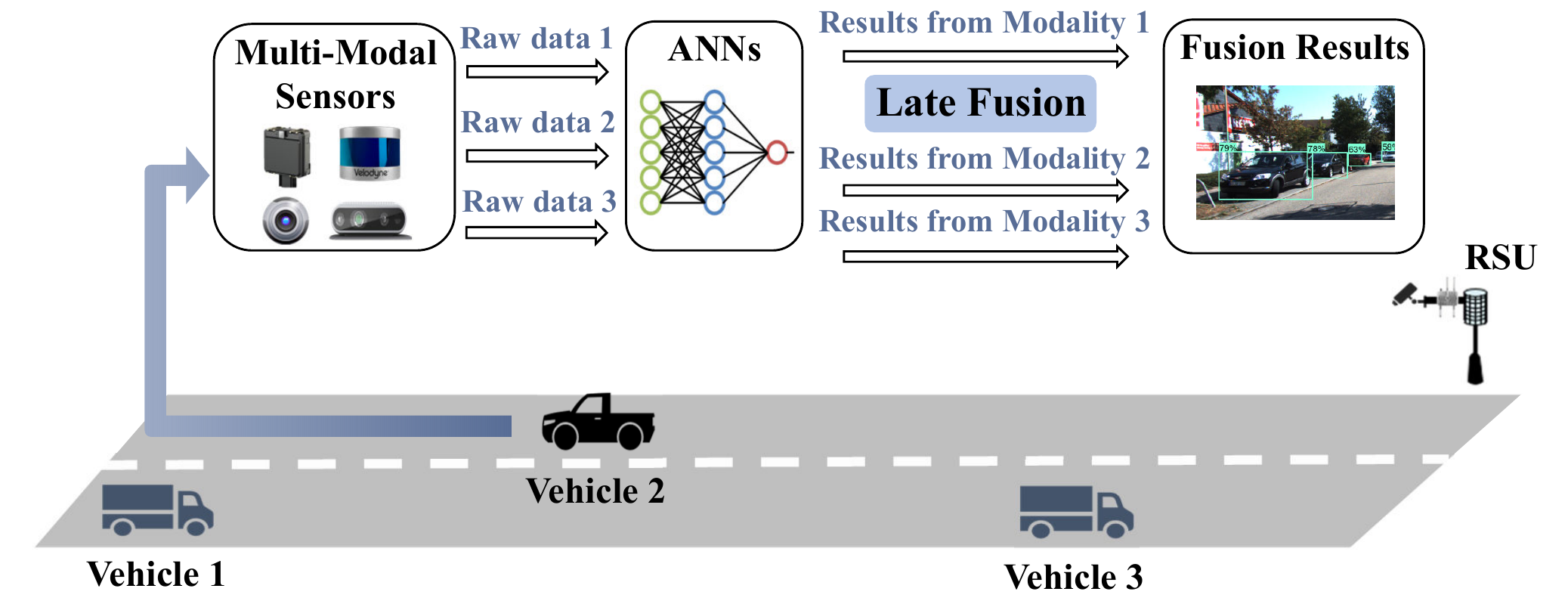}%
			\label{late}}
		\hfil
		\subfloat[]{\includegraphics[width=0.5\textwidth]{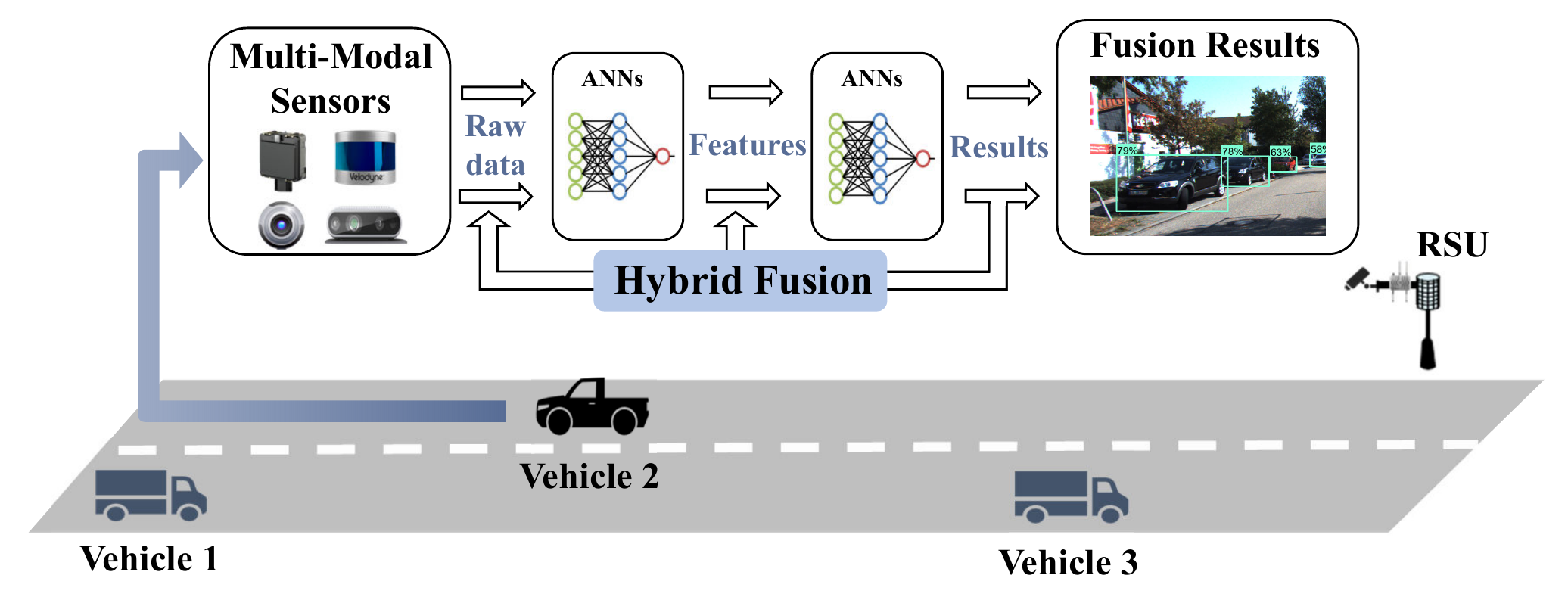}%
			\label{hybr}}
		\caption{Diagram of SoM-concert for single-agent sensing system in different levels: (a) early fusion; (b) late fusion; (c) hybrid fusion.}
		\label{SenseSym1}
	\end{figure}

	In general, multi-modal fusion strategies in a single dynamic agent can be categorized into three types: early fusion, late fusion, and hybrid fusion. The classification criteria is based on the workflow of ANNs. In sensing tasks, raw data is fed into ANNs to be processed, and environmental features are extracted. Then, sensing results are obtained. In early fusion strategies, multi-modal sensory data is fused before generating the sensing results separately. On the contrary, late fusion strategies input multi-modal sensory data into ANNs to obtain sensing results and fuse them, rather than directly fusing the raw data first and inputting them into the ANNs. In general, what early fusion schemes fuse is raw or processed raw data, while late fusion schemes fuse sensing results. As for hybrid fusion schemes, they fuse raw or processed raw data in certain modality and sensing results in others, making them flexible composite schemes. These three approaches are introduced and relevant works are reviewed in the following subsections. The system diagram of SoM-concert for single-agent sensing is shown in Fig.~\ref{SenseSym1}, and the three approaches are introduced and relevant works are reviewed in the following subsections.
	
	\subsubsection{Multi-Modal Early Fusion}
	Raw data collected by multi-modal sensors contains the complete amount of sensing information, so early fusion methods can guarantee the maximum information redundancy and therefore offer the highest theoretical performance upper bound. Many related works tend to adopt the early fusion methods. However, the challenge comes from the inconsistencies of either the data format or the network structure, leading to complicated processing. \textcolor{black}{Based on the method of integrating multi-modal heterogeneous data, multi-modal early fusion can be classified into three categories: i) grid map generation, ii) raw data projection, and iii) environment feature fusion.}

	\textbf{i) Grid map generation:}  \textcolor{black}{In the grid map generation methods, the formats of multi-modal sensory data are unified by occupancy grid maps. As shown in~\cite{early1wirges2018object}, occupancy grid maps for sensing tasks are presented in two dimensions from a bird's-eye view, which can intuitively reflect the environmental conditions around the mobile agent. Such a method is especially suitable for ground agents with low sensitivity to height.} To reflect the environmental characteristics from different levels, the grid value assignment method can be classified into two types: One is the intuitive features, such as observation point number, reflection intensity, and observation point height; the other is the advanced features, such as data distribution, occupancy probability, and velocity estimation value. In practical use, especially for multi-modal fusion, to store comprehensive environmental features more comprehensively, the value of grids can be a vector. Before generating a grid map, sensory data needs to be projected to the bird's-eye view. For those ranging sensors, such as LiDAR, radar, and depth-camera, data contains complete 3D spatial information, so the projection can be done intuitively by coordinate transformation. In~\cite{early2steyer2019grid}, LiDAR and radar point clouds are aligned before generating the grid map. But for cameras, because of the projective geometry principle, essential depth information is lost when generating 2D pixels. So in~\cite{early3zheng2022multi}, the objects are extracted by the object detection algorithm, and then the depths of objects are obtained by LiDAR. After unifying the sensor perspectives, the statistical properties of multi-modal data can be extracted by customized functions to finish the grid map generation process. To extract environmental features and sensing results from the grid map, both clustering-based~\cite{early1wirges2018object} and learning-based~\cite{early2steyer2019grid} methods are utilized. Clustering is usually based on the spatial distribution of features. The velocity distribution can also be involved if radars are available. As for the learning-based method, the data format of the grip map is equivalent to 2D pixels with customized channels rather than just RGB channels. So ANN structures which are widely used in the computer vision community can be adapted to object detection in the grid map to obtain higher performance than the clustering-based methods. However, due to the diversity of grid map assignment methods, it is necessary to adjust the network structure and retrain ANN modal parameters.
	
	\textbf{ii) Raw data projection:} \textcolor{black}{Without the bird-eye-view grid maps as intermediaries, heterogeneous 2D and 3D sensing data can also be directly unified in format through coordinate transformations, namely raw data projection. Typically, one modality of data is used as the main component, while the other modality supplements the information, which enables direct complementary between sensors.} Since the positions of multiple sensors on a single agent are relatively constant, the above transformation can be accurately obtained by off-line measurement, namely calibration. For example, Dou \textit{et al.} \cite{early4} projected LiDAR 3D point clouds to the 2D pixel plane of the camera to align multi-modal data and the aligned multi-modal data can be equivalent to data from a new sensor, which can obtain both 2D and 3D information. After that, the aligned data is fed into customized ANNs for joint processing and sensing results are output by ANNs. Such ANNs share similar base modules with ANNs for uni-modal data, however, specially designed feature fusion modules are needed to utilize multi-modal information.
	
	\textbf{iii) Environment feature fusion}: \textcolor{black}{Apart from rule-based methods like grid maps and direct projection, multi-modal data integration can also be achieved via ANNs. As environment feature extraction is an essential process in ANN-based environment sensing frameworks and CNN is the most common feature extractor. Multi-modal data can be embedded into latent feature representations and then integrated at the feature level, which grants the environment feature fusion method great flexibility.} Through the CNN-based feature extraction module, matrix data, i.e., feature maps, that contain environmental information, can be obtained from 2D or 3D raw sensory data. Different from raw sensory data, more refined semantic information is contained in the feature maps after convolution and downsampling. The data form is not differentiated by sensor category, which facilitates multi-modal fusion. Zhang \textit{et al.} \cite{early5} utilized CNNs to extract environment features collected by radar and multiple fish-eye cameras. To unify the coordinates, environment features are projected to the bird's-eye view. As feature extraction does not perturb the spatial distribution of features, the projection can be done similarly to that in raw data projection. Then the bird's-eye view is fed into a deep feature fusion module and detection head to get sensing results as the final output.

	\subsubsection{Multi-Modal Late Fusion}
	Although in early fusion methods, the key motivation is to collect multi-modal environment information with as much redundancy as possible before extracting environment features. So early fusion methods have higher performance potential. However, it suffers from drawbacks in flexibility. Especially when the number and type of sensors involved in multi-modal fusion change, the ANNs in early fusion methods are faced with retraining or even structural modification. Therefore, early fusion has been a case-by-case problem so far. On the contrary, the late fusion methods have much higher flexibility, as the multi-modal environment data is processed independently before fusion, and only the sensing results are involved in the fusion. In \cite{early6}, detection results of the camera, and LiDAR, including 2D bounding boxes and category labels of surrounding objects are fused with a designated weight. To facilitate the association of bounding boxes, LiDAR point clouds are projected to a 2D pixel plane before being fed into ANN. However, it needs to be clarified that the method in \cite{early6} does not belong to early fusion as there is no joint processing of LiDAR point clouds and images. To realize 3D detection, \cite{latenew1} leverages late fusion for refining 3D bounding boxes from LiDARs combining 2D proposals from cameras. Melotti \textit{et al}. \cite{latenew3} summarizes the scores from different modalities for the associated 3D bounding box into one final score. Moreover, \cite{latenew2} realizes road detection by combing the semantic segmentation results from LiDARs and cameras.
	
	In the late fusion methods, the feature extraction process does not involve the coupling of multi-modal data. Thus, the key issue falls on the association and weighted fusion of detection results from different sensory data. ANNs used for perception can typically output a score of the sensing result. However, the score indicates the confidence level only without quantifying the information redundancy and performance gain brought by multi-modal sensors. To maximize the multi-modal gain to realize the complementary of multi-modal sensors, weights for multi-modal sensing results need to be evaluated. Pang \textit{et al.} \cite{latenew1} utilized statistical features like confidence score, distance and IoU as the weights for the overlapping regions. Another goal-oriented approach is to train a multi-modal fusion module to determine the weight. Asvadi \textit{et al.} \cite{early6} designed a joint re-scoring module to calculate the weight of multi-modal sensing results. As the optimization objective is sensing performance, such a fusion module can lead to high performance in most cases. However, when sensory data deteriorates in a harsh environment, which is a challenge in dynamic scenarios, the relatively low interpretability of the learning-based fusion algorithms reduces the reliability of the framework. To reflect the reliability of multi-modal sensing results, and then realize complementary in harsh environments, Zheng \textit{et al.} \cite{early7} proposed a confidence evaluation scheme for learning-based sensing. The sensitivity of the ANNs to environment changes and the confidence of sensing results are quantified through the probability distribution of the data and network detection output. When the sensing failure of uni-modal sensors occurs in harsh weather, the corresponding weight will be reduced accordingly to avoid its interference with the multi-modal fusion results.
	
	\subsubsection{Multi-Modal Hybrid Fusion}
	\label{hybrid}
	Multi-modal hybrid fusion is an eclectic fusion approach that jointly considers the high sensing information redundancy and potential performance of multi-modal early fusion, and the high flexibility of multi-modal late fusion. Rather than fusing information from multiple machine senses in a particular processing stage, hybrid fusion schemes can integrate environment information from one machine sense at a certain processing stage into another machine sense's one at another processing stage. Consequently, hybrid fusion schemes involve a wide variety of specific methods and provide opportunities for trade-offs in multiple performance indicators. Although there is currently no unified classification for hybrid fusion schemes, we approximately classify them into two categories according to the correlation between pipelines that process sensory data in different modalities: i) parallel mode and ii) cascaded mode.
	
	\textbf{i) Parallel mode:} Similar to the typical fusion structure in traditional early fusion and late fusion methods, in the parallel mode, the processing and extracting of sensory data in different modalities parallel run through the whole ANN-based sensing procedure. One of the early works in hybrid fusion \cite{hybrid1} fuses environment features extracted from the bird-eye-view LiDAR point cloud, front view LiDAR point cloud, and front view RGB image. Apart from the traditional environment feature fusion pipeline in early fusion, it also utilizes proposals of 3D bounding boxes to determine the region of interest (ROI) in feature fusion, and designs a region-based fusion network. These 3D proposals consist of candidate bounding boxes that are likely to contain objects of interest. In such an approach, high-level sensing result, i.e., 3D proposals, helps feature fusion better focus on the key environment features, so it belongs to a hybrid fusion of feature and sensing results. In \cite{hybrid2}, multi-modal sensory information at different processing stages is fed into ANN to extract augmented features. Specifically, high-level results (3D proposals) and raw data (points in 3D proposals) from LiDAR, and features (semantic segmentation features) from camera are fused by ANN modules to obtain camera-augmented point-wise features. Although the processing of multi-modal sensory data is not as symmetric as one in \cite{hybrid1}, the features from RGB image and LiDAR point cloud are simultaneously processed until 3D bounding boxes are obtained by the final feature fusion module, which meets the definition of parallel. \textcolor{black}{To go beyond the LoS detection of camera and LiDAR, \cite{hybridnew1} introduces non-image modalities, namely acoustic, radar and seismic, to supplement NLoS information. Data from non-image modalities is firstly fused in both feature and sensing result levels, and then acts as a supplement to image modality for NLoS object detection tasks.}
	
	\textbf{ii) Cascaded mode:} The processing of multi-modal sensory data can also presents a cascaded architecture in the sensing pipeline. Specifically, the high-level final results of a certain modality are first obtained and then fuse with raw data or environment feature from other modalities to promote their subsequent sensing performances. Once the high-level results from a certain modality are obtained, the subsequent information processing will focus on other modalities, which can be described as cascading. Many existing works on camera and LiDAR fusion utilize cascaded mode. Thanks to the extensive research of computer vision, various ANNs for RGB image processing can be used to extract the sensing results from the image to provide ROI reference for LiDAR. In \cite{hybrid3, hybrid4, hybrid5}, 2D object bounding boxes are first obtained with RGB image and then projected to 3D LiDAR coordinates. In \cite{hybrid3}, the viewing frustum composed of 2D bounding boxes and origin in the LiDAR coordinate system is considered as the ROI of LiDAR. 3D detection results are obtained by the point cloud in ROI. This method constrains the search space of LiDAR object detection to the hot-spot areas to improve efficiency. However, the performance depends on the image processing ANNs and the redundancy of information is reduced because part of the point cloud is ignored. To further boost performance, \cite{hybrid4} selects multiple 2D proposals and \cite{hybrid5} generates a sequence of viewing frustums in different scale, where a larger amount of sensory information is utilized to improve object detection accuracy and robustness to sensing failure. Instead of object-level bounding boxes, \cite{hybrid6} selects another type of ANN to obtain pixel-level semantic segmentation results to provide a more accurate ROI reference. Based on the LiDAR points related to the projected semantic segmentation results, 3D proposals and bounding boxes can be obtained by subsequent ANN modules. By introducing dense semantic information from RGB images, desirable sensing performance can be obtained. It can be observed that mature ANN modules like image processing ANNs in \cite{hybrid3, hybrid4, hybrid5, hybrid6} are more easily utilized in cascaded mode than in parallel mode. Meanwhile, refined sensing results such as bounding boxes and semantic segmentation can help to improve the efficiency and pertinence of subsequent processing. However, the efficiency and information loss caused by single-modal processing need to be balanced.
	
	To sum up, the motivation of single-agent sensing based on SoM-concert is to maximize information redundancy of complementary sensors, including dense semantic information from RGB-D cameras, high-precision 3D structure from LiDAR, and environmental robustness and velocity measurements from radars. Early fusion makes full use of the information redundancy in raw data, but it is challenging to design a general method for different sensors and to measure the contributions of each type of sensor in sensing. Late fusion greatly improves flexibility by reducing the coupling of multiple modalities. The contribution to sensing task can be quantified by calculating the weight of uni-modal sensing results. However, when the environment features are insufficient, sensing information in a certain uni-modal processing would be lost, which may limit the overall performance. Apart from early fusion and late fusion, hybrid fusion is a composite scheme that can potentially make a balance between performance and efficiency by adjusting the pipeline correlation of sensory data in different modalities.

	\subsection{SoM-Concert for Multi-Agent Sensing}
	\label{multiagent}
	As mentioned in Section~\ref{multimodal}, through SoM-concert for single-agent sensing, a dynamic agent can obtain sensing information redundancy by multi-modal data fusion and improve sensing ability. However, in a harsh environment, sensors mounted on the intelligent agent may face simultaneous degradation in performance. For example, in a long dark tunnel, the camera imaging quality is degraded due to the dark light, the LiDAR point cloud has insufficient structural features due to the smooth inner wall, and the radar detection error is large due to multipath. Consequently, the extra cost for multi-modal sensors fusion may not lead to the expected gain. Due to the mobility of the environment, agents require a larger sensing range to avoid interference and collisions. Still, there are often more occlusions in dynamic environments, making the situation even more difficult. The above problems show that the information redundancy brought by multi-modal for a single dynamic agent is not enough to support the sensing tasks of mobile agents in dynamic environments.
	
	With the development of communication technology, sensing information can be exchanged among a large number of agents in real time thanks to large bandwidth and low latency. Therefore, a communication system can be a good medium for information sharing among dynamic agents. Since the sensory data of different agents is naturally in different perspectives and environmental states, sensory data from other agents can greatly improve the information redundancy of sensing. This is expected to be an important complement to SoM-concert for single-agent sensing in dynamic scenarios. Through multi-agent fusion, performance of sensing can be increased without increasing the cost of a single agent. Therefore, multi-agent sensing can also be augmented by SoM-concert.   
	
\begin{figure*}[!t]
		\centering
		\includegraphics[width=0.9\linewidth]{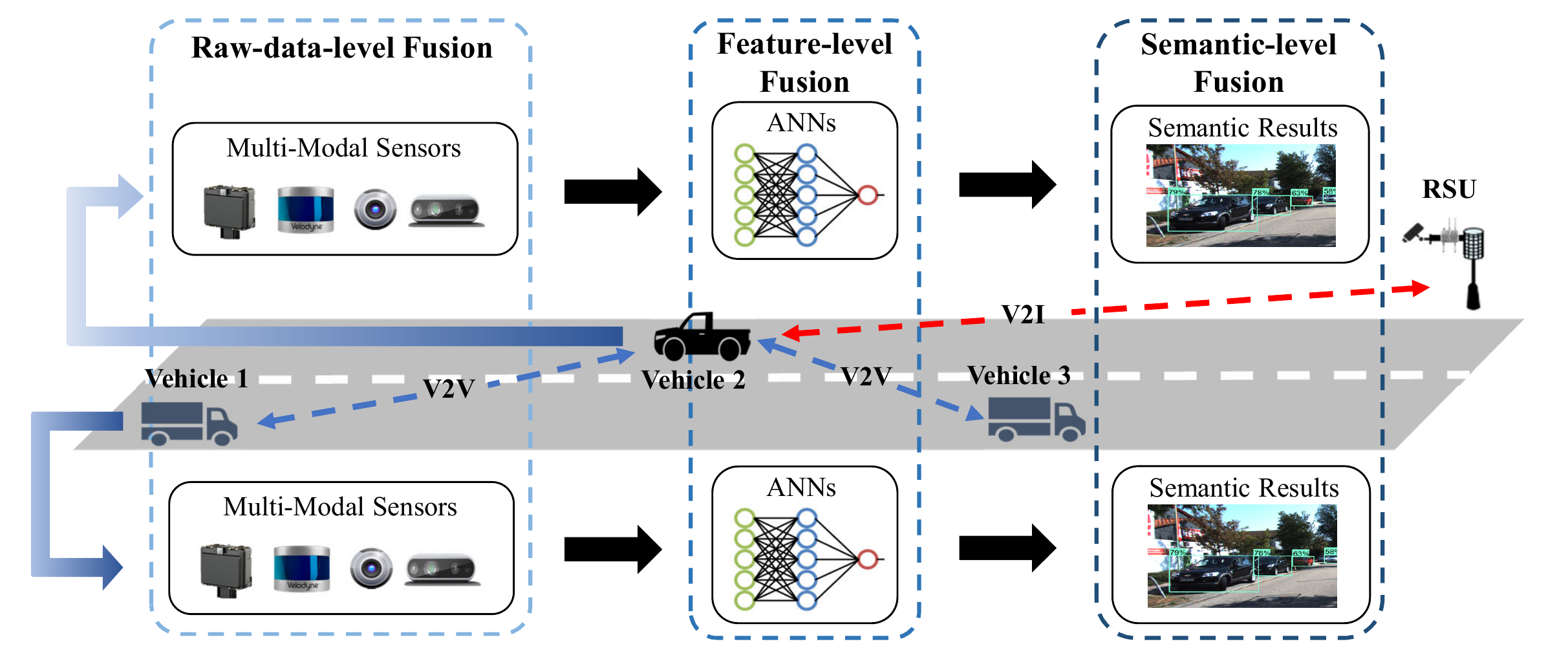}
		\caption{Diagram of SoM-concert for multi-agent sensing system.}
		\label{SenseSym2}
	\end{figure*}

	However, the non-ideality of communications is also a key factor in system design. In the sensing system, the content of the interaction between multiple agents needs to be flexibly adjusted according to the communication resources and the real-time requirements of the sensing tasks. According to the content of data interaction, existing works on SoM-concert for multi-agent sensing can be divided into the following three categories: raw-data-level systems, feature-level systems, and semantic-level systems. In the raw-data-level sensing system, agents directly share original sensory data and thus obtain the largest information redundancy from surrounding agents, but such a system has the highest requirements on the bandwidth of the communication network. In the semantic-level sensing system, sensing results are obtained locally before being shared among multiple agents, so it has the smallest communication load and the simplest data form. These make semantic-level sensing systems the most flexible. However, much effective sensing information is lost during single-agent data processing, especially in insufficiency of sensing information, restricting the performance potential of the semantic-level sensing system. In light of this issue, the feature-level sensing comes as a solution to decently balancing data volume and information redundancy by sharing environment features extracted by ANNs among multiple agents. By interacting features at different levels of the ANNs, performance and communication load can be balanced. The system diagram of SoM-concert for multi-agent sensing is shown in Fig.~\ref{SenseSym2}, and the related works of these three approaches will be introduced in subsequent subsections.
	
	\subsubsection{Raw-Data-Level Fusion}
	Similar to early fusion in single-agent multi-modal sensing, the basic form of raw-data-level sensing for multiple dynamic agents also contains grid-map sharing and raw data projection. However, different from the wired connection of multiple sensors on a single agent, wireless network resources among multiple dynamic agents are limited. Therefore, in sensing system design for multiple dynamic agents, the amount of communication data is one of the key metrics. In \cite{raw4}, camera raw data is shared to realize augmented reality, and the occluded environmental information is obtained in advance. Kim \textit{et al.} \cite{fakeautocast} proposed to share single-line LiDAR point cloud and low-resolution images to support tasks like lane detection and map match. In \cite{raw3}, LiDAR raw data from multiple vehicles is fused by data sharing and coordinate transformation. The performance of ANN-based object detection is significantly improved, as directly shown by the increase in the number of detected objects, especially in environments with more occlusion. In such a fusion scheme, the fused raw data can expand the sensing range and improve sensing accuracy. However, the problem of data volume arises because a frame of multi-line LiDAR point cloud in \cite{raw3} can be MB level or even larger. Although in \cite{fakeautocast}, the data volume can be at the 500 KB level. This poses a huge challenge to the applications in dynamic scenarios.
	
	Besides, sensory data can be processed to generate a local map before transmission and fusion. When the above processing does not involve understanding and abstracting the environment, such an approach can still be regarded as raw-data-level fusion. In \cite{early3zheng2022multi, raw5}, local grid maps created by LiDAR and cameras are shared and fused. In \cite{early3zheng2022multi}, the sensing field can be expanded and the obstacle contour information can be completed. In \cite{raw5}, the fused grid map is used in upper applications such as driving assistance. However, as the data volume of a single local grid map can also be up to MB level, such approaches also pose a great challenge to the communication systems.
	
	To bypass the data volume challenge, attempts also have been made to selectively send part of the raw data. In \cite{autocast}, based on the assumption that agents can share their trajectory planning results, the visibility and relevance of the region are evaluated. An agent first requests data corresponding to the regions that cannot be sensed by itself or are more relevant to its trajectory. Priority scheduling is calculated before data exchange to deal with unstable communication conditions. The size of a single packet can be reduced to a hundred KB level. However, the cost is an additional load caused by a large number of real-time computations.
	
	To sum up, SoM-concert for multi-agent sensing can be classified as a raw-data-level system when the sensory data interacted and fused among multiple agents is the raw data or the reformatted one. Since the sensing information contained in multi-agent data is preserved as much as possible before joint processing, raw-data-level sensing has the highest performance potential. However, the main drawback of its application in the dynamic environment is the high communication load, which makes it difficult to be real-time applied to a sensing system. As a result, most of these approaches are more applicable for offline tasks (such as static environment mapping) rather than sensing in dynamic scenarios.

	\subsubsection{Feature-Level Fusion}
	In the typical sensing process of ANNs, as shown in \cite{ann1voxel,ann2point,ann3rcnn}, the environment features are extracted and abstracted by different modules of the network, and finally, the semantic understanding of the environment is obtained through the detection head. The environment features, namely feature maps, have a more refined form compared with the raw data and have more information redundancy compared with the sensing results. Meanwhile, due to the abstraction of features by ANNs, feature maps are usually sparse and thus easy to be compressed. Therefore, sharing the feature maps can help to obtain as much information redundancy of multiple agents as possible under the condition of limited communication bandwidth. Several works \cite{fcooper, littlecooper} fused LiDAR feature maps from multiple agents to increase the number of detected objects. The experimental results have proven that by feature complementation, the multi-agent sensing results can be greater than the sum of the single-agent results. Qualitatively speaking, similar to raw-data-level fusion, sensing information contained in multi-agent feature maps can be gathered to enhance the agent's understanding of the environment, even if they are simply concatenated such as the max-out function in \cite{fcooper} and the max-norm function in \cite{littlecooper}. As it is difficult to find the direct physical meaning of the value of the feature map, \cite{coff, slimfcp} attempted to use more refined metrics to assign weights in the fusion of multi-agent feature maps. In general, feature maps from surrounding agents are weighted according to information redundancy. Feature maps that have higher information entropy, come from distant agents or have a large Euclidean distance from the local feature maps are considered to be more informative. \cite{coff} proposed the weighting equations based on the above assumptions to increase the interpretability of the feature fusion. In \cite{slimfcp}, the weight of the feature information amount also serves to reduce the communication load. At the transmitter, a feature encoding module followed by an irrelevant feature removal module has been utilized to reduce the size of the feature map and increase sparsity. Then a channel selection module inspired by \cite{senet} is used to sort the channels of the feature map by importance for sensing. In the trade-off between performance and data volume, the size of a feature map frame can float from 50KB to 1MB, which is applicable in communication networks with different bandwidths.  
	
	\subsubsection{Semantic-Level Fusion}
	As described in the previous subsections, both the raw-data-level fusion and the feature-level fusion can exploit the information redundancy brought by the multi-agent sensing information, similar to SoM-concert for single-agent. However, in practical scenarios, sometimes more loosely coupled frameworks are needed to cope with heterogeneous sensor configurations and low communication bandwidth among different agents. Inspired by single-agent late fusion, the specific single-agent sensing information extraction process can be ``transparent" to multi-agent information fusion, that is, semantic-level fusion. In the semantic-level fusion approaches, the information shared via communications is single-agent semantic comprehension of the environment, specifically bounding boxes and labels of surrounding objects or semantic segmentation of the environment. Therefore, the fusion architecture does not limit the types of sensors for different agents. Meanwhile, because the redundant environmental information contained by the sensors has been abstracted by ANNs, the output data of ANNs takes extremely small communication resources of KB-level. In \cite{semanticseg}, the semantic segmentation results of road elements such as lane lines and lamps are shared and fused in the bird's-eye view. In \cite{semanticob1, semanticob2}, multiple agents and multiple sensors are used to locate and classify the surrounding dynamic objects, to support real-time tracking of environmental targets. Such a semantic-level fusion problem can be modeled as the association and fusion generally. Sensing results from any agent and any sensor are first mapped to a unified coordinate, then sensing results belonging to the same environmental element are associated and finally fused. Referring to single-agent multi-modal late fusion, to maximize performance gain, the weight of semantic information from different agents needs to be quantified. Especially for dynamic scenarios, the sensing of a single agent is highly likely to be affected by the complexity of the environment, special weather, and lighting. To enhance the robustness of the system, the weight of abnormal data in fusion needs to be reduced respectively. As discussed in \cite{early7}, the environmental sensitivity of ANNs can be evaluated quantitatively, which also serves as the input of the weight function. The simulations in \cite{early7} show that the uncertainty of sensing can be reduced through weighted fusion.   
	
	\begin{table*}[!t]
		\renewcommand\arraystretch{2.7} 
		\renewcommand\cellalign{cc}
		\centering
		\caption{Typical related works of SoM-concert for environment sensing}
		\label{tab:tablesensing}
		\resizebox{\textwidth}{!}{
			\begin{tabular}{c|c|c|c|c|c|c|c|c}
				\toprule[0.35mm]
				\makecell[c]{\textbf{Sensing} \\ \textbf{entities}} & \textbf{Level of fusion} & \textbf{Ref.} & \makecell[c]{\textbf{Communication} \\ \textbf{load}} & \makecell[c]{\textbf{Information} \\ \textbf{redundancy}} & \textbf{Performance} & \textbf{Scalability} & \textbf{Interpretability} &\textbf{\textcolor{black}{Use cases}}\\ \midrule[0.15mm]
				\multirow{3}{*}{Single-Agent} & Early fusion &\makecell[c]{\cite{early1wirges2018object,early4, early5 } \\ \cite{early2steyer2019grid,early3zheng2022multi}}& -- & High & High & Low & Low & \textcolor{black}{Plug-and-play}\\ 
				\cline{2-9} & Late fusion & \makecell[c]{\cite{early6, latenew2,early7},\\ \cite{latenew1,latenew3}}& -- & Low & Limited & High  & High & \textcolor{black}{Edge computing}\\ 
				\cline{2-9} & Hybrid fusion & \cite{hybrid1, hybrid2, hybridnew1, hybrid3, hybrid4, hybrid5, hybrid6}& -- & Medium & Medium & Medium  & Medium & \textcolor{black}{Mobile computing}\\\midrule[0.15mm]
				\multirow{3}{*}{Multi-Agent}  & \makecell[c]{Raw-data-level \\ fusion} & \cite{raw4, fakeautocast, raw3, early3zheng2022multi, raw5, autocast}& High & High & High & Medium & Medium & \makecell[c]{\textcolor{black}{Abundant communication} \\ \textcolor{black}{resource between}\\ \textcolor{black}{multiple RSUs}}\\ \cline{2-9} & \makecell[c]{Feature-level \\ fusion}  & \cite{fcooper, littlecooper, coff, slimfcp, v2vnet, privacy}& Flexible & Medium & Medium  & Low    & Low & \makecell[c]{\textcolor{black}{Moderate communication} \\ \textcolor{black}{resource between} \\ \textcolor{black}{homogeneous agents}}   \\ 
				\cline{2-9} & \makecell[c]{Semantic-level \\ fusion} & \cite{semanticseg, semanticob1, semanticob2, early7, malicious1}& Low      & Low    & Limited & High   & High  & \makecell[c]{\textcolor{black}{Limited communication} \\ \textcolor{black}{resource between}\\ \textcolor{black}{heterogeneous agents}} \\ 
				\bottomrule[0.35mm]
		\end{tabular}}
	\end{table*}

	{\color{black}\subsection{Key Takeaways}
	\label{sensesum}
	In this section, we discuss the related works and technical features of SoM-concert-based reinforced environment sensing, including single-agent sensing and multi-agent sensing at various levels. For clarity, the related works of SoM-concert for environment sensing are summarized in Table.~\ref{tab:tablesensing}. Key takeaways learned are as follows.}
	\begin{itemize}
		\item
			{\color{black}The motivation of single-agent sensing based on SoM-concert is to maximize information redundancy of complementary sensors, including dense semantic information from RGB-D cameras, high-precision 3D structure from LiDAR, and environmental robustness and velocity measurements from radars. 
            Among the fusion methods, late fusion is the most practical one due to the decoupling of  multi-modal sensory data processing, which allows a plug-and-play deployment on off-the-shelf multi-modal sensing results from existing algorithms as demonstrated in \cite{latenew1}. However, in situations where the environmental features are insufficient, such as long dark tunnels or harsh weather conditions, individual processing of each sensor would cause significant information loss and performance degradation. Moreover, optimal fusion weight for different sensors in complex scenarios requires complicated hand-made rules. To address such challenges, one approach is to couple sensory data through ANNs during feature extraction process, which is adopted in early fusion and hybrid fusion. This can lead to performance gains since ANNs can learn the complex complementary relationships among the sensors. Early fusion makes full use of the raw data to maximize information redundancy but results in complicated joint processing, which is suitable for RSUs equipped with edge servers as demonstrated in \cite{usecase1, usecase2} rather than mobile agents with limited computation resources to ensure real-time sensing. Hybrid fusion balances efficiency and performance by utilizing efficient sensors (e.g., cameras) to guide the ROI of sensors with large data volumes (e.g., LiDARs) as demonstrated in~\cite{hybrid3, hybrid4, hybrid5}, which can be a composite scheme for resource-limited mobile agents. Leveraging existing single-modal sensing frameworks also aids in transitioning to multi-modal sensing.}
        \end{itemize}
            
        \begin{itemize}
		\item
		{\color{black}To overcome the limitation of LoS and FoV, SoM-concert for multi-agent sensing is also indispensable. Although sensor types and tasks are similar to those in single-agent sensing, communication would greatly impact fusion scheme design in multi-agent case. Like single-agent early fusion, both raw-data-level fusion and feature-level fusion merge raw data or environmental features from different agents. Raw-data-level fusion offers the highest redundancy and performance but struggles with real-time deployment on mobile agents due to communication load concerns. It is suitable for cooperative sensing among multiple RSUs with stable networks, providing powerful sensing support for mobile agents~\cite{usecase3}. 
        Feature-level fusion balances data volume and performance via ANNs before transmission. But it highly depends on the homogeneous sensor configurations and neural network settings among multiple agents, due to the limited generalization of ANNs. For example, \cite{fcooper, littlecooper, coff, slimfcp} assume that multiple agents have identical sensor and algorithm configurations. Semantic-level fusion extends single-agent late fusion to multiple agents, practical for heterogeneous mobile agents with limited communications. However, the effectiveness of the semantic-level fusion may significantly reduce in harsh weather and obstructed environments. In dynamic scenarios, ROI for sensing and CSI are both time-varying. Mobile agents can utilize the CSI between neighboring intelligent agents to adaptively adjust the level of fusion and optimally utilize time-varying communication conditions. Under favorable communication conditions, transmitting richer redundant data can improve sensing performance. Conversely, in adverse communication conditions, refining information in semantic fusion reduces the communication load. Furthermore, environmental feature extraction can be jointly designed with feature transmission, that is, allocating more communication resources strategically to ensure the transmission of crucial information to meet the sensing requirement in dynamic scenarios.}
	\end{itemize}

	\section{Future Research for SoM}
	\label{F}
\begin{figure}[!t]
		\centering
		\includegraphics[width=1\linewidth]{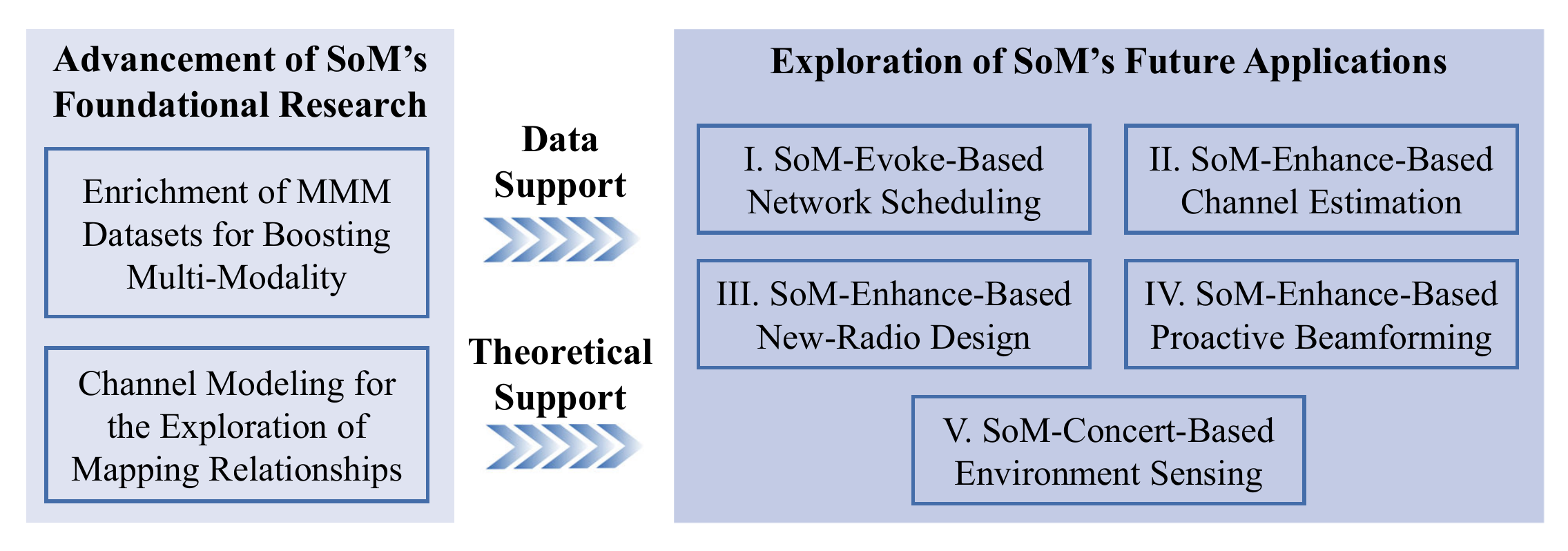}
		\caption{Illustration of the two aspects of future research for SoM.}
		\label{future}
	\end{figure}

	In this section, we highlight open issues and potential directions to stimulate and guide future research in the field of SoM, with a special focus on foundational research and future applications of SoM. The relationship between these two aspects is shown in Fig.~\ref{future}. The MMM datasets and mapping relationships form the essential data foundation and theoretical basis for the subsequent SoM research. The broader applications of SoM can only be explored based on more available MMM datasets as well as the in-depth knowledge of the mapping relationships and are well worthy for further research.
	
	\subsection{Advancement of SoM's Foundational Research}
	\subsubsection{Enrichment of MMM Datasets for Boosting Multi-Modality}
	ANNs are the core of realizing the intelligent multi-modal sensing-communication integration. The capability of ANNs relies on high-quality training data obtained from various dynamic scenarios. Nowadays, the research of desirable MMM datasets is scarce. As introduced in Section \ref{ammm}, the M$^3$SC dataset is constructed under a dynamic VCN crossroad scenario to bridge this gap. However, the sensory data can be further enriched and the simulated scenarios can also be  diversified by taking more scenarios, e.g., UAV networks and mountainous highways, into account. Furthermore, the detailed information of the communication links, such as latency, signal strength, and DoD/DoA, will be a more solid basis for the research of SoM-based communication system design. In the future, MMM datasets containing a wider variety of multi-modal sensory data and wireless channel data should be constructed under various emerging application scenarios in the B5G/6G era.

	\subsubsection{Channel Modeling for the Exploration of Mapping Relationships}
	Research on the complex mapping relationships between the tightly interwoven multi-modal sensing and communication functionality is indispensable for SoM research. To fulfill this requirement, proper channel modeling methods for SoM need to be accordingly customized. For the capturing of large-scale fading characteristics in channel modeling for SoM, the complex mapping relationship between multi-modal sensing and path loss/shadowing needs to be investigated urgently. For the capturing of small-scale fading characteristics in channel modeling for SoM, the complex mapping relationship between objects in the physical environment and scattering clusters in the electromagnetic environment needs to be revealed. Moreover, the relationship between key scatterer-related parameters, e.g., angle, power, and delay, and key object-related parameters, e.g., texture, material, and pose (position and orientation), also needs to be investigated. In the future, with the help of channel modeling customized for SoM, the complex mapping relationships between multi-modal sensing and communications need to be adequately explored to support the subsequent SoM research.

	\subsection{Exploration of SoM's Future Applications}
	\subsubsection{SoM-Evoke-Based Network Scheduling}    
	One key challenge in wireless communication network design is to properly schedule transmissions so that nodes in networks are able to share the common spectrum efficiently. Based on more available MMM datasets, research on SoM-evoke-based wireless network scheduling can be carried out. With the help of multi-modal sensory data, link state prediction and transmission scheduling can be conducted to improve the throughput of wireless networks. To be specific, multi-modal sensing is expected to be utilized to obtain CSI and channel statistics. This information enables the determination of channel quality and prediction of link state \cite{2019evoke}. As a consequence, links with a decent state can be exploited to transmit the data stream, thereby enhancing the throughput of the wireless network. In the future, more SoM-evoke-based wireless network scheduling works can be conducted to achieve the goal of directly enhancing the performance of wireless network with the help of multi-modal sensing.

	\subsubsection{SoM-Enhance-Based Channel Estimation}
	
	Currently, a few studies have adopted ANNs to estimate mmWave channels, utilizing the rich geometry characteristics but leaving out multi-modal sensing. Moreover, ANN-based schemes are prone to experience performance degradation when the actual data statistics deviate from {\color{black}those} of the training data \cite{hu2020deep}. Therefore, enhancing the scalability and interpretability of SoM-enhance-based channel estimation schemes is necessary. For this purpose, the mapping relationships between multi-modal sensing and electromagnetic propagation characteristics of different frequency bands are the prerequisite. Moreover, diversified training data collected from various practical scenarios and corresponding feature extraction mechanisms are needed to improve the scalability of SoM-enhance-based channel estimation. It is envisioned that the SoM-enhance-based channel estimation methods can flexibly adjust deployment details to adapt to different environments and communication systems in the future.

	\subsubsection{SoM-Enhance-Based New-Radio Design}
	
	Dual-function waveform design in higher frequency band (THz and beyond)  represents an important trend in the development of RF-ISAC in academia and industry due to more abundant bandwidth resources. However, there are many challenges that waveform design in high frequency band needs to address. For example, due to the hardware limitations of the transceivers, complex phenomena such as beam squint and beam split will occur, which will result in a huge performance loss of the communication and sensing functionality.  Fortunately, with the stronger correlation between channels in high frequency band and physical geometry, precoding parameters such as phase shifts and time delays can be more appropriately designed with richer environmental characteristics reflected by multi-modal sensing, thereby facilitating the design of dual-function waveform. The relative locations of users and the reflection properties of scatters can be captured implicitly from multi-modal sensing through DL-based methods, which help to compensate for the delay between antennas and proactively control the beam squint and beam split effects in high frequency band.

	\subsubsection{SoM-Enhance-Based Proactive Beamforming}
	
	Accurate real-time beam alignment acts as a premise for ultra-reliable communications. Current predictive beamforming schemes fall short in tracking the motion parameters of mobile users with complex trajectories and behavior, such as vehicle's emergency braking and sudden lane change. With a large-scale MIMO, the communication links established based on the existing predictive beamforming schemes may face instability or even experience frequent outages since the available motion state characteristics of mobile users are insufficient. For instance, in VCNs, vehicles' complex behavior and trajectory are often closely related to their surroundings, such as pedestrians, nearby vehicles, and obstacles. To accurately track the complex movement parameters of the target vehicle, it is crucial to acquire comprehensive information concerning the complex behavioral interactions and environmental information surrounding it. In the future, the SoM-enhance-based beamforming design, capable of analyzing the complex interactions between users in complex scenarios based on multi-modal environment information, deserves further study.

	\subsubsection{SoM-Concert-Based Environment Sensing}
	In most existing works on SoM-concert, communications mainly act as a medium for the interaction of environmental information. Building upon MMM datasets, future research can be conducted in two areas to further explore the potential of communications in SoM-concert for environment sensing. Firstly, viewing communications as a way of transmitting information, communication resource planning can be performed in insufficient and insecure communication networks. By quantifying the information redundancy of sensory data, sensing performance and communication load can be balanced. Secondly, communication functionality can be viewed as an additional source of sensing information. In situations where sensors may not work properly, such as in dark tunnels with blurry features or during harsh weather conditions, communication functionality can be used to serve as an off-the-shelf sensing apparatus. Through channel parameter estimation and physical layer sensing, communication functionality is expected to introduce a new information domain. 
	
	Moreover, as 
	power, spectrum, storage, and computational resources are coupled in environment sensing tasks, the multi-resource from heterogeneous agents needs to be dealt with in a holistic manner. Currently, existing resource management methods of dynamic systems are limited to communication resources \cite{letaief2006dynamic}. To maximize resource utility and achieve the upper bound of sensing performance, there is an urgent need for research on multi-resource management methods, including routing, scheduling, and task offloading, which are still challenging. The spatiotemporal distribution of multi-resource is highly related to the user locations and the surroundings. In this case, by employing abundant environment awareness collected by widely deployed sensors, an approach for obtaining global knowledge of the dynamic system can be provided to guide the multi-resource management optimization through user action selection and multi-user collaborative design.

	\section{Conclusion}
	\label{C}

    In this survey, we propose the innovative concept of SoM as a unified framework of the intelligent multi-modal sensing-communication integration. Inspired by the human brain, we emphasize that such integration is intelligently achieved through SoM processing, with ANNs at its core, employed in a model-enhanced, data-driven manner. SoM offers three operational modes for holistic design, excelling particularly in dynamic scenarios, distinguishing it from RF-ISAC's exclusive focus on RF-based radar sensing. We initially trace the historical evolution and deficiencies of RF-ISAC, underscoring the vital role of multi-modal sensing in dynamic scenarios. We also introduce the M$^3$SC dataset as a cornerstone for SoM research. Subsequently, we delve into the complex mapping relationships between multi-modal sensing and communications, forming the theoretical foundation of SoM research. We highlight that such intricate mapping relationships can be explored with the aid of channel modeling customized for SoM. Additionally, we present our relevant research to illuminate future directions. Subsequently, our attention turns to reviewing the specific research pertaining to the enhancement of communication and sensing functionalities through SoM. We comprehensively assess advancements in SoM-enhanced transceiver design, covering critical aspects such as channel estimation, dual-function waveform, and predictive beamforming. Furthermore, we underscore the significance of SoM-concert in dynamic environment sensing, classifying related works into single-agent and multi-agent categories. Finally, we identify key research gaps and outline future directions. In summary, this survey pioneers the intelligent multi-modal sensing-communication field by proposing the SoM framework and thoroughly reviewing the current research status. We anticipate that SoM serves as a pivotal catalyst for advancing sensing and communication capabilities, ultimately contributing to the realization of pervasive connectivity and networked intelligence in the B5G/6G era.


\begin{thebibliography}{10}
		
		\bibitem{ntt2020white}
		{\em 5G Evolution and 6G}, NTT~Docomo, Inc., Tokyo, Japan, White Paper, Jan. 2020. [Online]. Available: \url{https://www.nttdocomo.co.jp/english/binary/pdf/corporate/technology/whitepaper_6g/DOCOMO_6G_White_PaperEN_20200124.pdf}
		
		\bibitem{zhang20196g}
		Z. Zhang \emph{et al.}, ``6G Wireless Networks: Vision, Requirements, Architecture, and Key Technologies,'' \emph{IEEE Veh. Technol. Mag.}, vol.~14, no.~3, pp.~28--41, Sep.~2019.
		
		\bibitem{chowdhury20206g}
		M.~Z. Chowdhury, M.~Shahjalal, S.~Ahmed, and Y.~M. Jang, ``6G Wireless Communication Systems: Applications, Requirements, Technologies, Challenges, and Research Directions,'' {\em IEEE Open J. Commun. Society}, vol.~1, pp.~957--975, 2020.
		
		{\color{black}\bibitem{dang2020should}
		S.~Dang, O.~Amin, B.~Shihada, and M.-S. Alouini, ``What should 6G be?,'' {\em Nat. Electron.}, vol.~3, no.~1, pp.~20--29, Jan.~2020.}
		
		
		{\color{black}\bibitem{liu2022integrated}
		F.~Liu {\em et~al.}, ``Integrated Sensing and Communications: Towards Dual-functional Wireless Networks for 6G and Beyond,'' {\em IEEE J. Sel. Areas Commun.}, vol.~40, no.~6, pp.~1728--1767, Jun.~2022.}
		
		\bibitem{Paul2017}
		B.~Paul, A.~R.~Chiriyath, D.~W.~Bliss, ``Survey of RF Communications and Sensing Convergence Research,'' {\em IEEE Access}, vol.~5, pp.~252--270, 2017.
		
		
		
		\bibitem{roberton2003integrated}
		M.~Roberton and E.~Brown, ``Integrated radar and communications based on chirped spread-spectrum techniques,'' in {\em IEEE MTT-S Int. Microw. Symp. Dig.}, Philadelphia, PA, USA, Oct.~2003, pp.~611--614.
		
		\bibitem{saddik2007ultra}
		G.~N. Saddik, R.~S. Singh, and E.~R. Brown, ``Ultra-Wideband Multifunctional Communications/Radar System,'' {\em IEEE Trans. Microwave Theory Tech. Techniques}, vol.~55, no.~7, pp.~1431--1437, Jul.~2007.
		
		\bibitem{jamil2008integrated}
		M.~Jamil, H.-J. Zepernick, and M.~I. Pettersson, ``On integrated radar and communication systems using Oppermann sequences,'' in {\em Proc. IEEE Military Commun. Conf.}, San Diego, CA, USA, Nov.~2008, pp.~1--6.
		
		\bibitem{tigrek2012ofdm}
		R.~F. Tigrek, W.~J. De~Heij, and P.~Van~Genderen, ``OFDM Signals as the Radar Waveform to Solve Doppler Ambiguity,'' {\em IEEE Trans. Aerosp. Electron. Syst.}, vol.~48, no.~1, pp.~130--143, Jan.~2012.
		
		\bibitem{hu2014radar}
		L.~Hu, Z.~Du, and G.~Xue, ``Radar-communication integration based on OFDM signal,'' in {\em Proc. IEEE Int. Conf. Signal Process. Commun. Comput. (ICSPCC)},  Guilin, China, Aug.~2014, pp.~442--445.
		
		\bibitem{marzetta2010noncooperative}
		T.~L. Marzetta, ``Noncooperative Cellular Wireless with Unlimited Numbers of Base Station Antennas,'' {\em IEEE Trans. Wireless Commun.}, vol.~9, no.~11, pp.~3590--3600, Nov.~2010.
		
		\bibitem{bliss2003multiple}
		D.~Bliss and K.~Forsythe, ``Multiple-input multiple-output (MIMO) radar and imaging: degrees of freedom and resolution,'' in {\em Proc. IEEE Conf. Rec. 37th Asilomar Conf. Signals, Syst. Comput.}, Pacific Grove, CA, USA, Nov.~2003, pp.~54--59.
		
		\bibitem{commu1} C.~Sturm and W.~Wiesbeck, ``Waveform Design and Signal Processing Aspects for Fusion of Wireless Communications and Radar Sensing,"  \emph{Proc. IEEE}, vol.~99, no.~7, pp.~1236--1259, Jul.~2011.
		
		\bibitem{liu2020joint}
		F.~Liu, C.~Masouros, A.~P. Petropulu, H.~Griffiths, and L.~Hanzo, ``Joint Radar and Communication Design: Applications, State-of-the-Art, and the Road Ahead,'' {\em IEEE Trans. Commun.}, vol.~68, no.~6, pp.~3834--3862, Jun.~2020.
		
		\bibitem{zhang2020perceptive}
		A.~Zhang, M.~L. Rahman, X.~Huang, Y.~J. Guo, S.~Chen, and R.~W. Heath, ``Perceptive Mobile Networks: Cellular Networks With Radio Vision via Joint Communication and Radar Sensing,'' {\em IEEE Vehic. Tech. Mag.}, vol.~16, no.~2, pp.~20--30, Jun.~2021.

           \bibitem{chengbookVCN}
           X.~Cheng, S.~Gao, and L.~Yang, {\em mmWave Massive MIMO Vehicular Communications.} Switzerland: Springer Nature, 2022.
		\bibitem{klautau2019lidar}
		A.~Klautau, N.~Gonz{\'a}lez-Prelcic, and R.~W. Heath, ``LIDAR Data for Deep Learning-Based mmWave Beam-Selection,'' {\em IEEE Wireless Commun. Lett.}, vol.~8, no.~3, pp.~909--912, Jun.~2019.
		
		
			
		\bibitem{alkhateebDRONE}
		G.~Charan {\em et al.}, ``Towards Real-World 6G Drone Communication: Position and Camera Aided Beam Prediction,'' in {\em Proc. IEEE Global Commun. Conf. (GLOBECOM)}, Rio de Janeiro, Brazil, Dec.~2022, pp.~2951-–2956.
			
		\bibitem{alkhateebDRONE2}
		G.~Charan, A.~Hredzak, and A.~Alkhateeb, ``Millimeter Wave Drones with Cameras: Computer Vision Aided Wireless Beam Prediction'' Nov.~2022, {\em arxiv:2211.07569}.
				
		\bibitem{alrabeiah2020millimeter}
		M.~Alrabeiah, A.~Hredzak, and A.~Alkhateeb, ``Millimeter Wave Base Stations with Cameras: Vision-Aided Beam and Blockage Prediction,'' in {\em Proc. IEEE 91st Vehicular Technol. Conf. (VTC2020-Spring)}, Antwerp, Belgium, May.~2020, pp.~1--5.
	
		
		
		\bibitem{zheng2019radar}
		L.~Zheng, M.~Lops, Y.~C. Eldar, and X.~Wang, ``Radar and Communication Coexistence: An Overview: A Review of Recent Methods,'' {\em IEEE Signal Process. Mag.}, vol.~36, no.~5, pp.~85--99, 2019.
		
		\bibitem{ma2020joint}
		D.~Ma, N.~Shlezinger, T.~Huang, Y.~Liu, and Y.~C. Eldar, ``Joint Radar-Communication Strategies for Autonomous Vehicles: Combining Two Key Automotive Technologies,'' {\em IEEE Signal Process. Mag.}, vol.~37, no.~4, pp.~85--97, Jul.~2020.
		
		\bibitem{feng2020JRC}
		Z.~Feng, Z.~Fang, Z.~Wei, X.~Chen, Z.~Quan, and D.~Ji, ``Joint Radar and Communication: A survey'' {\em China Commun.}, vol.~17, no.~1, pp.~1--27, Jan.~2020.
		
		\bibitem{cui2021integrating}
		Y.~Cui, F.~Liu, X.~Jing, and J.~Mu, ``Integrating Sensing and Communications for Ubiquitous IoT: Applications, Trends, and Challenges,'' {\em IEEE Netw.}, vol.~35, no.~5, pp.~158--167, Sep./Oct.~2021.
		
		
		\bibitem{mishra2019toward}
		K.~V. Mishra, M.~B. Shankar, V.~Koivunen, B.~Ottersten, and S.~A. Vorobyov, ``Toward Millimeter-Wave Joint Radar Communications: A Signal Processing Perspective,'' {\em IEEE Signal Process. Mag.}, vol.~36, no.~5, pp.~100--114, Sep.~2019.
		
		{\color{black}
		\bibitem{zhang2021overview}
		J.~A. Zhang {\em et~al.}, ``An Overview of Signal Processing Techniques for Joint Communication and Radar Sensing,'' {\em IEEE J. Sel. Topics Signal Process.}, vol.~15, no.~6, pp.~1295--1315, Nov.~2021.}
		
		\bibitem{wei2023} 
		Z.~Wei {\em et~al.}, ``Integrated Sensing and Communication Signals Towards 5G-A and 6G: A Survey,'' {\em IEEE Internet Things J.}, vol.~10, no.~13, pp.~11068--11092, Jul.~2023.
		
		\bibitem{liu2022survey}
		A.~Liu {\em et~al.}, ``A Survey on Fundamental Limits of Integrated Sensing and Communication,'' {\em IEEE Commun. Surveys Tuts.}, vol.~24, no.~2, pp.~994--1034, 2nd Quart., 2022.
		
		
		\bibitem{zhang2021enabling}
		J.~A. Zhang {\em et~al.}, ``Enabling Joint Communication and Radar Sensing in Mobile Networks—A Survey,'' {\em IEEE Commun. Surveys Tuts.}, vol.~24, no.~1, pp.~306--345, 1st Quart., 2021.
		
		
		\bibitem{CXISAC}
		X.~Cheng, H.~Zhang, Z.~Yang, Z.~Huang, S.~Li, and A.~Yu, ``Integrated sensing and communications for internet of vehicles: current status and development trend,'' {\em Journal on Communications}, vol.~43, no.~8, pp.~188--202, Aug.~2022.
		
		\bibitem{Chengisac}
		X.~Cheng, D.~Duan, S.~Gao, and L.~Yang, ``Integrated Sensing and Communications (ISAC) for Vehicular Communication Networks (VCN),'' \emph{IEEE Internet Things J.}, vol.~9, no.~23, pp.~23441--23451, Dec.~2022.
		\bibitem{mahmood2020white}
		N.~H.~Mahmood {\em et~al.}, ``White Paper on Critical and Massive Machine Type Communication Towards 6G,'' Apr.~2020, {\em arxiv:2004.14146}.
		
		\bibitem{chengbook}
		X.~Cheng, R.~Zhang, and L.~Yang, {\em 5G-Enabled Vehicular Communications and Networking}. 1nd ed. Cham, Switzerland: Springer, 2019.
		
		
		\bibitem{smartcity1}
		H.~Kim and J.~Ben-Othman, ``Toward Integrated Virtual Emotion System with AI Applicability for Secure CPS-Enabled Smart Cities: AI-Based Research Challenges and Security Issues," {\em IEEE Netw.}, vol.~34, no.~3, pp.~30--36,~May/Jun.~2020.
		
		
		\bibitem{industry1}
		H.~Yang, A.~Asheralieva, J.~Zhang, M.~M. Karim, D.~Niyato, and K.~A. Raza,
		``User-centric blockchain for industry 5.0 applications,'' in {\em Proc. IEEE Int. Conf. Commun. Workshops (ICC Workshops)}, Seoul, South Korea, May.~2022, pp.~25--30.
		
		
		\bibitem{industry2}
		Q.~Tang, F.~R.~Yu, R.~Xie, A.~Boukerche, T.~Huang, and Y.~Liu, ``Internet of Intelligence: A Survey on the Enabling Technologies, Applications, and Challenges,” {\em IEEE Commun. Surveys Tuts.}, vol.~24, no.~3, pp.~1394--1434, 3rd Quart., 2022.
		
		\bibitem{fan2022radar}
		Y.~Fan, S.~Gao, D.~Duan, X.~Cheng, and L.~Yang, ``Radar Integrated MIMO Communications for Multi-Hop V2V Networking,'' {\em IEEE Wireless Commun. Lett.}, vol.~12, no.~2, pp.~307--311, Feb.~2023.
		
		\bibitem{ali2020passive}
		A.~Ali, N.~Gonz{\'a}lez-Prelcic, and A.~Ghosh, ``Passive Radar at the Roadside Unit to Configure Millimeter Wave Vehicle-to-Infrastructure Links,'' {\em IEEE Trans. Veh. Technol.}, vol.~69, no.~12, pp.~14903--14917, Sep.~2020.
		
		\bibitem{xu2023computer}
		W.~Xu, F.~Gao, X.~Tao, J.~Zhang and A.~Alkhateeb, ``Computer Vision Aided mmWave Beam Alignment in V2X Communications," {\em IEEE Trans. Wireless Commun.}, vol.~22, no.~4, pp.~2699--2714, Apr.~2023.
		
		\bibitem{alouini21CVbeam}
			Y.~Tian, G.~Pan, and M.-S.~Alouini, ``Applying Deep-Learning-Based Computer Vision to Wireless Communications: Methodologies, Opportunities, and Challenges,” {\em IEEE Open J. Commun. Soc.}, vol.~2, pp.~132--143, Jan.~2021.
		
        \bibitem{xu2023multi}
			W.~Xu, F.~Gao, Y.~Zhang, C.~Pan, and G.~Liu, ``Multi-User Matching and Resource Allocation in Vision Aided Communications,'' {\em IEEE Trans. Commun.}, vol.~71, no.~8, pp.~4528--4543, Aug.~2023.
				
		\bibitem{Rahman2020frame}
		M. L. Rahman, J. A. Zhang, X. Huang, Y. J. Guo, and R. W. Heath, ``Framework for a Perceptive Mobile Network Using Joint Communication and Radar Sensing,'' {\em IEEE Trans. Aerosp. Electron. Syst.}, vol.~56, no.~3, pp.~1926--1941, Jun.~2020.
		
		\bibitem{myCOMST}
		X.~Cheng, Z.~Huang, and L.~Bai, ``Channel Nonstationarity and Consistency for Beyond 5G and 6G: A Survey,'' \emph{IEEE Commun. Surveys Tutor.}, vol.~24, no.~3, pp.~1634--1669, 3rd Quart., 2022.
		
		\bibitem{LS}
		M.~Henkel, C.~Schilling, and W.~Schroer, ``Comparison of Channel Estimation Methods for Pilot Aided OFDM Systems,'' in {\em Proc. IEEE VTC Spring}, Dublin, Ireland, Apr.~2007, pp.~1435--1439.
		
		{\color{black}
		\bibitem{mmse}
		J.-J.~Van De Beek, O.~Edfors, M.~Sandell, S.~K.~Wilson, and P.~O. Borjesson, ``On channel estimation in OFDM systems,'' in {\em Proc. IEEE 1031 45th Veh. Technol. Conf.}, Chicago, IL, USA, vol.~2, Jul.~1995, pp.~815–-819.}
		
	
			\bibitem{sensingaidedCE2022}
			S.~Jiang and A.~Alkhateeb, ``Sensing aided OTFS channel Estimation for Massive MIMO Systems,'' Sep.~2022, {\em arxiv:2209.11321}.
			
			\bibitem{sensingaidedCE2023}
			R.~Mundlamuri, R.~Gangula, C.~K. Thomas, F.~Kaltenberger, and W.~Saad, ``Sensing aided Channel Estimation in Wideband Millimeter-Wave MIMO Systems,'' Feb.~2023, {\em arxiv:2302.02065}.
			
			\bibitem{gao2021prediciton}
			Y.~Yang, F.~Gao, C.~Xing, J.~An, and A.~Alkhateeb, ``Deep Multimodal Learning: Merging Sensory Data for Massive MIMO Channel Prediction,'' {\em IEEE J. Sel. Areas Commun.}, vol.~39, no.~7, pp.~1885--1898, Jul.~2021.
			
			\bibitem{xu2021deep}
			W.~Xu, F.~Gao, J.~Zhang, X.~Tao, and A.~Alkhateeb, ``Deep Learning Based Channel Covariance Matrix Estimation With User Location and Scene Images,'' \emph{IEEE Trans. Commun.}, vol.~69, no.~12, pp.~8145--8158, Dec.~2021.
			
			{\color{black}
			\bibitem{alkhateebCCP}
			A.~Taha and A.~Alkhateeb, ``Situation-Aware Channel Covariance Prediction for Deep Learning Aided Massive MIMO Systems," in {\em Proc. 54th Asilomar Conf. Signals, Syst., Comput.}, Pacific Grove, CA, USA, Nov.~2020, pp.~1342–-1346.}
			
			\bibitem{alkhateebLiDARBeam}
			S.~Jiang, G.~Charan and A.~Alkhateeb, ``LiDAR Aided Future Beam Prediction in Real-World Millimeter Wave V2I Communications,"{\em IEEE Wireless Commun. Lett.}, vol.~12, no.~2, pp.~212--216, Feb.~2023.
			
			\bibitem{feifeibeam}
			F.~Wen, W.~Xu, F.~Gao, C.~Pan and G.~Liu, ``Vision Aided Environment Semantics Extraction and Its Application in mmWave Beam Selection," \emph{IEEE Commun. Lett.}, vol.~27, no.~7, pp.~1894--1898, Jul.~2023.	
			
			
			\bibitem{chowdhuryV2I}
			G.~Reus-Muns {\em et al.}, ``Deep Learning on Visual and Location Data for V2I mmWave Beamforming,” in {\em Proc. Int. Conf. Mobility, Sens. Netw. (MSN)}, Exeter, U.K., Dec.~2021, pp. 559--566.  

   		\bibitem{gonzalez2016radar}
		N.~Gonzalez-Prelcic, R.~M{\'e}ndez-Rial, and R.~W.~Heath, ``Radar aided beam alignment in MmWave V2I communications supporting antenna diversity,'' in {\em Proc. UCSD Inf. Theory Appl. Workshop}, La Jolla, CA, USA, Jan.~2016, pp.~1--7.
  
			\bibitem{charan2021vision}
			G.~Charan, M.~Alrabeiah, and A.~Alkhateeb, ``Vision-Aided 6G Wireless Communications: Blockage Prediction and Proactive Handoff,'' {\em IEEE Trans. Veh. Technol.}, vol.~70, no.~10, pp.~10193--10208, Oct.~2021.
			
			
			\bibitem{chen2022computer}
			J.~Chen, F.~Gao, X.~Tao, G.~Liu, C.~Pan, and A.~Alkhateeb, ``Computer Vision Aided Codebook Design for MIMO Communications Systems,'' {\em IEEE Trans. Wireless Commun.}, vol.~22, no.~5, pp.~3341--3354, May.~2023.
		
		\bibitem{early1wirges2018object}
		S.~Wirges, T.~Fischer, C.~Stiller, and J.~B.~Frias, ``Object Detection and Classification in Occupancy Grid Maps Using Deep Convolutional Networks," in {\em Proc. 21st Int. Conf. Intell. Transp. Syst. (ITSC)}, Maui, HI, USA, Nov.~2018, pp.~3530--3535.
		
		\bibitem{early5}
		J.~Zhang, M.~Zhang, Z.~Fang, Y.~Wang, X.~Zhao, and S.~Pu, ``RVDet: Feature-level Fusion of Radar and Camera for Object Detection," in {\em Proc. IEEE Int. Intell. Transp. Syst. Conf. (ITSC)}, Indianapolis, IN, USA, Oct.~2021, pp.~2822--2828.
		
		\bibitem{early4}
		J.~Dou, J.~Xue, and J.~Fang, ``SEG-VoxelNet for 3D Vehicle Detection from RGB and LiDAR Data," in {\em Proc. Int. Conf. Robot. Automat. (ICRA)}, Montreal, QC, Canada, May.~2019, pp.~4362--4368.
		
		\bibitem{early6}
		A.~Asvadi, L.~Garrote, C.~Premebida, P.~Peixoto, and U.~J.~Nunes, ``Multimodal vehicle detection: Fusing 3D-LIDAR and color camera data," {\em Pattern Recognit. Lett.}, vol.~115, pp.~20--29, Nov.~2018.
		
		\bibitem{early2steyer2019grid}
		S.~Steyer, C.~Lenk, D.~Kellner, G.~Tanzmeister, and D.~Wollherr, ``Grid-Based Object Tracking With Nonlinear Dynamic State and Shape Estimation," {\em{IEEE} Trans. Intell. Transport. Syst.}, vol.~21, no.~7, pp.~2874--2893, Jul.~2020.
		
		\bibitem{latenew2}
		S.~Gu, Y.~Zhang, J.~Tang, J.~Yang, J. M.~Alvarez and H.~Kong, ``Integrating Dense LiDAR-Camera Road Detection Maps by a Multi-Modal CRF Model," in \emph{IEEE Trans. Veh. Technol.}, vol.~68, no.~12, pp.~11635--11645, Dec.~2019.
		\bibitem{frangeul2016cross}
		L.~Frangeul, G.~Pouchelon, L.~Telley, S.~Lefort, C.~Luscher, and D.~Jabaudon, ``A cross-modal genetic framework for the development and plasticity of sensory pathways,'' \emph{Nature}, vol.~538, no.~7623, pp.~96--98, Sep.~2016.
		
		\bibitem{mattingley2001unconscious}
		J.~B.~Mattingley, A.~N.~Rich, G.~Yelland, and J.~L.~Bradshaw, ``Unconscious priming eliminates automatic binding of colour and alphanumeric form in synaesthesia,'' \emph{Nature}, vol.~410, no.~6828, pp.~580--582, Mar.~2001.
		
		\bibitem{beeli2005coloured}
		G.~Beeli, M.~Esslen, and L.~J{\"a}ncke, ``When coloured sounds taste sweet,'' \emph{Nature}, vol.~434, no.~7029, pp.~38--38, Mar.~2005.
		
		\bibitem{niyatoIoT}
		D.~C.~Nguyen {\em et al.}, ``6G Internet of Things: A Comprehensive Survey," {\em IEEE Internet Things J.}, vol.~9, no.~1, pp.~359--383, Jan.~2022.
		
		
			\bibitem{jsac_DRL_mimo}
			M.~Fozi, A.~R.~Sharafat, and M.~Bennis, ``Fast MIMO beamforming via deep reinforcement learning for high mobility mmWave connectivity,'' {\em IEEE J. Sel. Areas Commun.}, vol.~40, no.~1, pp.~127--142, Jan.~2022.
			
			\bibitem{gu2023meta}
			J.~Gu, L.~Collins, D.~Roy, A.~Mokhtari, S.~Shakkottai, and K.~R. Chowdhury, ``Meta-learning for image-guided millimeter-wave beam selection in unseen environments,'' in {\em Proc. IEEE Int. Conf. Acoust., Speech Signal Process. (ICASSP)}, Greek, Jun.~2023, pp.~1--5.
			
			\bibitem{chowdhuryFLASH}
			B.~Salehi, J.~Gu, D.~Roy, and K.~Chowdhury, ``FLASH: Federated learning for automated selection of high-band mmWave sectors,” in {\em Proc. IEEE Conf. Comput. Commun.}, London, U.K., May.~2022.
		
			\bibitem{chowdhurymultimodal}
			B.~Salehihikouei {\em et al.}, ``Deep Learning on Multimodal Sensor Data at the Wireless Edge for Vehicular Network,” {\em IEEE Trans. Veh. Technol.}, vol.71, no.~7, pp.~7639--7655, Apr.~2022.

		
		\bibitem{kumari2017ieee}
		P.~Kumari, J.~Choi, N.~Gonz{\'a}lez-Prelcic, and R.~W. Heath, ``IEEE 802.11ad-Based Radar: An Approach to Joint Vehicular Communication-Radar System,'' {\em IEEE Trans. Veh. Technol.}, vol.~67, no.~4, pp.~3012--3027, Apr.~2017.
		
		\bibitem{grossi2017opportunistic}
		E.~Grossi, M.~Lops, L.~Venturino, and A.~Zappone, ``Opportunistic automotive radar using the IEEE 802.11ad standard,'' in {\em Proc. IEEE Radar Conf. (RadarConf)}, Seattle, WA, USA, May.~2017, pp.~1196--1200.
		
		\bibitem{liu2017multiobjective}
		Y.~Liu, G.~Liao, Z.~Yang, and J.~Xu, ``Multiobjective optimal waveform design for OFDM integrated radar and communication systems,'' {\em Signal Process.}, vol.~141, pp.~331--342, Dec.~2017.
		
		\bibitem{li2017joint}
		B.~Li and A.~P.~Petropulu, ``Joint Transmit Designs for Coexistence of MIMO Wireless Communications and Sparse Sensing Radars in Clutter,'' {\em IEEE Trans. Aerosp. Electron. Syst.}, vol.~53, no.~6, pp.~2846--2864, Dec.~2017.
		
		{\color{black}
		\bibitem{zheng2017adaptive}
		L.~Zheng, M.~Lops, and X.~Wang, ``Adaptive Interference Removal for Uncoordinated Radar/Communication Coexistence,'' {\em IEEE J. Sel. Topics Signal Process.}, vol.~12, no.~1, pp.~45--60, Feb.~2017.
	}
		
		\bibitem{nartasilpa2018communications}
		N.~Nartasilpa, A.~Salim, D.~Tuninetti, and N.~Devroye, ``Communications System Performance and Design in the Presence of Radar Interference,'' {\em IEEE Trans. Commun.}, vol.~66, no.~9, pp.~4170--4185, Apr.~2018.
		
		\bibitem{hass2016dual}
		A. Hassanien, M. G. Amin, Y. D. Zhang, and F. Ahmad, ``Dual-function radar-communications: Information embedding using sidelobe control and waveform diversity,'' {\em IEEE Trans. Signal Process.}, vol.~64, no.~8, pp.~2168--2181, Apr.~2016.
		
		{\color{black}
		\bibitem{ma2021frac}
		D. Ma, N. Shlezinger, T. Huang, Y. Liu, and Y. C. Eldar, ``FRaC: FMCW-based joint radar-communication system via index modulation,'' {\em IEEE J. Sel. Topics Signal Process.}, vol.~15, no.~6, pp.~1348--1364, Nov.~2021
	}
		\bibitem{ruan2019evoke}
		Y.~Ruan, Y.~Li, C.-X.~Wang, R.~Zhang, and H.~Zhang,	``Power Allocation in Cognitive Satellite-Vehicular Networks From Energy-Spectral Efficiency Tradeoff Perspective'' {\em IEEE Trans. Cogn. Commun. Netw.}, vol.~5, no.~2, pp.~318–329, Jun.~2019.
		
		\bibitem{early3zheng2022multi}
		X.~Zheng, Y.~Li, D.~Duan, L.~Yang, C.~Chen, and X.~Cheng, ``Multivehicle Multisensor Occupancy Grid Maps (MVMS-OGM) for Autonomous Driving," {\em IEEE Internet Things J.}, vol.~9, no.~22, pp.~22944--22957, Nov.~2022.
		
		\bibitem{early7}
		X.~Zheng, S.~Li, Y.~Li, D.~Duan, L.~Yang and X.~Cheng, ``Confidence Evaluation for Machine Learning Schemes in Vehicular Sensor Networks," {\em {IEEE} Trans. Wireless Commun.}, vol.~22, no.~4, pp.~2833--2846, Apr.~2023.
		
		
			\bibitem{sss-h1}
			S.~Wu, S.~Hur, K.~Whang, and M.~Nekovee, ``Intra-Cluster Characteristics of 28 GHz Wireless Channel in Urban Micro Street Canyon,'' in \emph{Proc. IEEE Glob. Commun. Conf. (GLOBECOM)}, Washington, DC, USA, Dec.~2016, pp.~1--6.
			
		{\color{black}	
			\bibitem{sss-h2}
			S.~Hur \emph{et al.}, ``Proposal on Millimeter-Wave Channel Modeling for 5G Cellular System,'' \emph{IEEE J. Sel. Topics Signal Process.}, vol.~10, no.~3, pp.~454--469, Apr.~2016.
		}
			\bibitem{sss-h3}
			J.~Saunders, S.~Saeedi, and W.~Lil, ``Parallel Reinforcement Learning Simulation for Visual Quadrotor Navigation," in \emph{Proc. 2023 IEEE International Conference on Robotics and Automation (ICRA)}, London, U.K., Jun.~2023, pp.~1357--1363.
		
		
		
		
		\bibitem{alkhateeb2019deepmimo}
		A.~Alkhateeb, ``DeepMIMO: A Generic Deep Learning Dataset for Millimeter Wave and Massive MIMO Applications,'' Feb.~2019, {\em arxiv:1902.06435}.

		
		\bibitem{klautau20185g}
		A.~Klautau, P.~Batista, N.~Gonz{\'a}lez-Prelcic, Y.~Wang, and R.~W. Heath, ``5G MIMO Data for Machine Learning: Application to Beam-Selection Using Deep Learning,'' in {\em Proc. Inf. Theory Appl. Workshop (ITA)}, San Diego, CA, USA, Feb.~2018, pp.~1--9.
		
		\bibitem{manivasagam2020lidarsim}
		S.~Manivasagam {\em et~al.}, ``LiDARsim: Realistic LiDAR Simulation by Leveraging the Real World,'' in {\em Proc. IEEE Conf. Comput. Vis. Pattern Recognit. (CVPR)}, Seattle, WA, USA, Jun.~2020, pp.~11167--11176.
		
		\bibitem{li2022v2x}
		Y.~Li {\em et~al.}, ``V2X-Sim: Multi-Agent Collaborative Perception Dataset and Benchmark for Autonomous Driving,'' {\em IEEE Robot. Autom. Lett.}, vol.~7, no.~4, pp.~10914--10921, Oct.~2022.
		
		\bibitem{xu2022opv2v}
		R.~Xu, H.~Xiang, X.~Xia, X.~Han, J.~Li, and J.~Ma, ``OPV2V: An Open Benchmark Dataset and Fusion Pipeline for Perception with Vehicle-to-Vehicle Communication,'' in {\em Proc. Int. Conf. Robot. Automat.}, Philadelphia, PA, USA, May.~2022, pp.~2583--2589.
		
		\bibitem{sun2022shift}
		T.~Sun {\em et~al.}, ``SHIFT: A Synthetic Driving Dataset for Continuous Multi-Task Domain Adaptation,'' in {\em Proc. IEEE Conf. Comput. Vis. Pattern Recognit. (CVPR)}, New Orleans, LA, USA, Jun.~2022, pp.~21371--21382.
		
		\bibitem{waldo}
		National Institute of Standards and Technology (NIST). PS-002 WALDO. [Online]. Available: {\em \url{https://github.com/usnistgov/PS-002-WALDO}}
		
		\bibitem{alrabeiah2020viwi}
		M.~Alrabeiah, A.~Hredzak, Z.~Liu, and A.~Alkhateeb, ``ViWi: A Deep Learning Dataset Framework for Vision-Aided Wireless Communications,'' in {\em Proc. IEEE 91st Vehicular Technol. Conf. (VTC2020-Spring)}, Antwerp, Belgium, May.~2020, pp.~1--5.

			\bibitem{gu2022multimodality}
			J.~Gu, B.~Salehi, D.~Roy, and K.~R.~Chowdhury, ``Multimodality in mmWave MIMO Beam Selection Using Deep Learning: Datasets and Challenges,'' {\em IEEE Commun. Mag.}, vol.~60, no.~11, pp.~36--41, Nov.~2022.	
		
	
		\bibitem{DeepSense}
		A.~Alkhateeb {\em et~al.}, ``DeepSense 6G: A Large-Scale Real-World Multi-Modal Sensing and Communication Dataset,'' \emph{IEEE Commun. Mag.}, vol.~61, no.~9, pp.~122--128, Sep.~2023.
		
		\bibitem{M3C}
		X.~Cheng \emph{et al.}, ``M$^3$SC: A Generic Dataset for Mixed Multi-Modal (MMM) Sensing and Communication Integration,'' \emph{China Commun.}, vol.~20, no.~11, pp.~13--29, Nov.~2023.
		

		\bibitem{WI}
		Remcom. \emph{Wireless InSite}. Accessed: Mar.~2022. [Online]. Available: https://www.remcom.com/wireless-insite-em-propagation-software
	

		\bibitem{CARLA}
		A.~Dosovitskiy, G.~Ros, F.~Codevilla, A.~Lopez, and V.~Koltun, ``CARLA: An open urban driving simulator,'' in \emph{Proc. 1st Annu. Conf. Robot Learn. (CoRL)}, Mountain View, CA, USA, Nov.~2017, pp. 1--16.
		
		
		\bibitem{AirSim}
		S.~Shah, D.~Dey, C.~Lovett, and A.~Kapoor, ``AirSim: High-Fidelity Visual and Physical Simulation for Autonomous Vehicles,'' in \emph{Field and Service Robotics}, M. Hutter and R. Siegwart, Eds. Cham, Switzerland: Springer, 2018, pp.~621--635.
		

		\bibitem{WaveFarer}
		Remcom, \emph{WaveFarer}. Accessed: Mar. 2022. [Online]. Available: https://www.remcom.com/wavefarer-automotive-radar-software

		\bibitem{SUMO}
		P.~A.~Lopez \emph{et al.}, ``Microscopic traffic simulation using SUMO," in \emph{Proc. 2018 21st International Conference on Intelligent Transportation Systems (ITSC)}, Maui, HI, USA, Nov. 2018, pp. 2575--2582.
		\bibitem{CloudRT} 
		D.~He, B.~Ai, K.~Guan, L.~Wang, Z.~Zhong and T.~Kürner, ``The Design and Applications of High-Performance Ray-Tracing Simulation Platform for 5G and Beyond Wireless Communications: A Tutorial,'' \emph{IEEE Commun. Surveys Tutor.}, vol.~21, no.~1, pp.~10--27, 1st Quart., 2019.
		
		\bibitem{CST} 
		\emph{Electromagnetic Simulation Solvers|CST Studio Suite}. Accessed:
			Nov. 17, 2020. [Online]. Available: https://www.3ds.com/productsservices/simulia/products/cst-studio-suite/solvers/
		
		
		\bibitem{TITS_my}	
		Z.~Huang and X.~Cheng, ``A 3-D Non-Stationary Model for Beyond 5G and 6G Vehicle-to-Vehicle mmWave Massive MIMO Channels,'' \emph{IEEE Trans. Intell. Transp. Syst.}, vol.~23, no.~7, pp.~8260--8276, Jul.~2022.
		
		\bibitem{fsfef}
		X.~Cheng, Z.~Huang, and S.~Chen, ``Vehicular communication channel measurement, modelling, and application for beyond 5G and 6G,'' \emph{IET Commun.}, vol.~14, no.~19, pp.~3303--3311, Nov.~2020.
		
		\bibitem{TWCMY}
		Z.~Huang and X.~Cheng, ``A General 3D Space-Time-Frequency Non-Stationary Model for 6G Channels,'' \emph{IEEE Trans. Wireless Commun.}, vol.~20, no.~1, pp.~535--548, Jan.~2021.
		
		\bibitem{sensor-4}
		Q.~Zhang, H.~Sun, Z.~Wei, and Z.~Feng, ``Sensing and Communication Integrated System for Autonomous Driving Vehicles,'' in \emph{Proc. IEEE Int. Conf. Comput. Commun. (INFOCOM) Workshops}, Toronto, ON, Canada, Jul.~2020, pp.~1278--1279.
		
		
		\bibitem{Rappaport1-4}
		T.~S.~Rappaport \emph{et al.}, ``Wireless Communications and Applications Above 100 GHz: Opportunities and Challenges for 6G and Beyond,'' \emph{IEEE Access}, vol.~7, pp.~78729--78757, 2019.
		
		\bibitem{newpaper}
		Z.~Huang, X.~Cheng, and X.~Yin, ``A General 3D Non-Stationary 6G Channel Model With Time-Space Consistency,'' \emph{IEEE Trans. Commun.}, vol.~70, no.~5, pp.~3436--3450, May.~2022.
		
		\bibitem{huangj1}
		J.~Huang, C.-X.~Wang, H.~Chang, J.~Sun, and X.~Gao, ``Multi-Frequency Multi-Scenario Millimeter Wave MIMO Channel Measurements and Modeling for B5G Wireless Communication Systems,'' \emph{IEEE J. Sel. Areas Commun.}, vol.~38, no.~9, pp.~2010--2025, Sep.~2020.
		
		
		\bibitem{3GPP}
		\emph{Technical Specification Group Radio Access Network; Study on Channel Model for Frequencies From 0.5 to 100 GHz (Release 14), Version 14.2.0}, document TR 38.901, 3GPP, Sophia Antipolis, France, Sep. 2017. [Online]. Available: http://www.3gpp.org/DynaReport/38901.htm
		
		
		
		\bibitem{yolo}
		A.~Bochkovskiy, C.-Y.~Wang, and H.-Y.~Mark Liao, ``YOLOv4: Optimal speed and accuracy of object detection" Apr.~2020, {\em arxiv:2004.10934}.
		
		
		
		\bibitem{Gao1}
		X.~Gao, F.~Tufvesson, and O.~Edfors, ``Massive MIMO channels--Measurements and models,'' in \emph{Proc. Rec. Annu. Asilomar Conf. Signals Syst. Comput. (ASILOMAR)}, Pacific Grove, CA, USA, May.~2013, pp.~280--284.
		
		\bibitem{Gaonew}
		X.~Gao, J.~Flordelis, G.~Dahman, F.~Tufvesso, and O.~Edfors, ``Massive MIMO channel modeling--Extension of the COST 2100 model,'' in \emph{Proc. Joint NEWCOM/COST Workshop Wireless Commun. (JNCW)}, Barcelona, Spain, Oct.~2015, pp.~1--8.
		
		\bibitem{liuy-4}
		L.~Liu \emph{et al.}, ``The COST 2100 MIMO channel model,'' \emph{IEEE Wireless	Commun.}, vol.~19, no.~6, pp.~92--99, Dec.~2012.
		

			\bibitem{xiong}
			B.~Xiong \emph{et al.}, ``Novel Multi-Mobility V2X Channel Model in the Presence of Randomly Moving Clusters,” \emph{IEEE Trans. Wireless Commun.}, vol.~20, no.~5, pp.~3180--3195, May.~2021. 
			
			\bibitem{myTVT}
			Z.~Huang, L.~Bai, X.~Cheng, X.~Yin, P.~Mogensen, and X.~Cai, ``A Non-Stationary 6G V2V Channel Model With Continuously Arbitrary Trajectory,'' \emph{IEEE Trans. Veh. Technol.}, vol.~72, no.~1, pp.~4--19, Jan.~2023.
		
		\bibitem{5393069}
		O.~Renaudin, V.~Kolmonen, P. Vainikainen, and C.~Oestges, ``Non-Stationary Narrowband MIMO Inter-Vehicle Channel Characterization in the 5-GHz Band,'' \emph{{IEEE} Trans. Veh. Technol.}, vol.~59, no.~4, pp.~2007--2015, May.~2010.
		
		{\color{black}
		\bibitem{dewd}
		S.~Knörzer, M.~A.~Baldauf, T.~Fugen, and W.~Wiesbeck, ``Channel Analysis for an OFDM-MISO Train Communications System Using Different Antennas,'' in \emph{Proc. IEEE Veh. Technol. Conf. (VTC-Fall)}, Baltimore, MD, USA, Oct.~2007, pp.~809--813.}
		
		\bibitem{ZTECom}
		Z.~Huang, X.~Cheng, and N.~Zhang, ``An Improved Non-Geometrical Stochastic Model for Non-WSSUS Vehicle-to-Vehicle Channels,'' \emph{{ZTE} Commun.}, vol.~17, no.~4, pp.~62--71, Dec.~2019.
		
		\bibitem{NGSMDavid}
		D.~W.~Matolak and Q. Wu, ``Markov models for vehicle-to-vehicle channel multipath persistence processes,'' in \emph{Proc. IEEE Veh. Tech. Society Wireless Access in Veh. Env. (WAVE)}, Dearborn, MI, USA, Dec.~2008, pp.~8--9.
		
		
		\bibitem{7473100}
		Z.~Huang, X.~Zhang, and X.~Cheng, ``Non-geometrical stochastic model for non-stationary wideband vehicular communication channels,'' \emph{{IET} Commun.}, vol.~14, no.~1, pp.~54--62, Jan.~2020.
		
		\bibitem{8108211}
		L.~Bai, C.-X.~Wang, S.~Wu, J.~Sun, and W.~Zhang, ``A 3-D Wideband Multi-Confocal Ellipsoid Model for Wireless Massive MIMO Communication Channels with Uniform Planar Antenna Array,'' in \emph{Proc. IEEE Veh. Technol. Conf. (VTC-Spring)}, Sydney, NSW, Australia, Jun.~2017, pp.~1--6.		
		
		\bibitem{e3e3}
		A.~G.~Zajic, G.~L.~Stuber, T.~G.~Pratt, and S.~T.~Nguyen, ``Wideband MIMO Mobile-to-Mobile Channels: Geometry-Based Statistical Modeling With Experimental Verification,'' \emph{IEEE Trans. Veh. Technol.}, vol.~58, no.~2, pp.~517--534, Feb.~2009.
		
		\bibitem{xilaren}
		E.~Michailidis, N.~Nomikos, P.~Trakadas, and A.~G.~Kanatas, ``Three-Dimensional Modeling of mmWave Doubly Massive MIMO Aerial Fading Channels," \emph{IEEE Trans. Veh. Technol.}, vol.~69, no.~2, pp.~1190--1202, Feb.~2020.		
		
		\bibitem{5GCM} 
		Aalto University, AT\&T, BUPT, CMCC, Ericsson, Huawei, Intel, KT Corporation, Nokia, NTT DOCOMO, New York University, Qualcomm, Samsung, University of Bristol, and University of Southern California. ``5G Channel Model for Bands Up to 100 GHz." Oct. 2016. [Online]. Available: http://www.5gworkshops.com/5GCM.html
		
		\bibitem{8647188}
		S.~Ju and T.~S.~Rappaport, ``Millimeter-Wave Extended NYUSIM Channel Model for Spatial Consistency,'' in \emph{Proc. IEEE Glob. Commun. Conf. (GLOBECOM)}, Abu Dhabi, United Arab Emirates, Dec.~2018, pp.~1--6.
		
		\bibitem{XUESONG1}
		X.~Cai, G.~Zhang, C.~Zhang, W.~Fan, J.~Li, and G.~F.~Pedersen, ``Dynamic Channel Modeling for Indoor Millimeter-Wave Propagation Channels Based on Measurements," \emph{{IEEE} Trans. Commun.}, vol.~68, no.~9, pp.~5878--5891, Sep.~2020.
		
		\bibitem{METIS}
		V. Nurmela \emph{et al.}, \emph{METIS Channel Models}, document ICT-317669-METIS/D1.4, METIS, New York, NY, USA, Jul.~2015.
		
		\bibitem{IMT}
		\emph{Preliminary Draft New Report ITU-R M.[IMT-2020.EVAL]}, document R15-WP5D-170613-TD-0332, Int. Telecommun. Union, Niagara Falls, ON, Canada, Jun.~2017.
		
		\bibitem{COST2100f}
		N. Cardona, \emph{Cooperative Radio Communications for Green Smart Environments}. Gistrup, Denmark: River, 2016.
		
		\bibitem{science}
		M.~I.~Jordan and T.~M.~Mitchell, ``Machine learning: Trends, perspectives, and prospects,'' \emph{Science}, vol.~349, no.~6245, pp.~255--260, Jul.~2015.
		
		{\color{black}
		\bibitem{xietongUGV} 
		Z. Huang, L. Bai, M. Sun, and X. Cheng, ``A 3D Non-Stationarity and Consistency Model for Cooperative Multi-Vehicle Channels,'' \emph{IEEE Trans. Veh. Technol.}, vol.~72, no.~9, pp.~11095--11110, Sep.~2023.
		
		\bibitem{alkhateeb2014channel}
		A.~Alkhateeb, O.~El~Ayach, G.~Leus, and R.~W. Heath, ``Channel Estimation and Hybrid Precoding for Millimeter Wave Cellular Systems,'' {\em IEEE J. Sel. Topics Signal Process.}, vol.~8, no.~5, pp.~831--846, Oct.~2014.}
		
		
		
		
		
		\bibitem{cecomst}
		Y.~Liu, Z.~Tan, H.~Hu, L.~J. Cimini, and G.~Y. Li, ``Channel Estimation for OFDM,'' {\em IEEE Commun. Surveys Tuts.}, vol.~16, no.~4, pp.~1891--1908, 4th Quart., 2014.
		
		\bibitem{el2014spatially}
		O.~El~Ayach, S.~Rajagopal, S.~Abu-Surra, Z.~Pi, and R.~W. Heath, ``Spatially Sparse Precoding in Millimeter Wave MIMO Systems,'' \emph{IEEE Trans. Wireless Commun.}, vol.~13, no.~3, pp.~1499--1513, Mar.~2014.
		
		
		\bibitem{gao2021model}
		S.~Gao, X.~Cheng, L.~Fang, and L.~Yang, ``Model Enhanced Learning Based Detectors (Me-LeaD) for Wideband Multi-User 1-bit mmWave Communications,'' {\em IEEE Trans. Wireless Commun.}, vol.~20, no.~7, pp.~4646--4656, Jul.~2021.
  
		\bibitem{gaoHcodebook}
           S.~Gao, Y.~Dong, C.~Chen, and Y.~Jin, ``Hierarchical beam selection in mmWave multiuser MIMO systems with one-bit analog phase shifters, in {\em Proc. 8th IEEE Int. Conf. Wireless Commun. Signal Process. (WCSP)}, Yangzhou, China, Oct.~2016, pp.~1--5.
           
	
			\bibitem{ozdemir2007channel}
			M.~K.~Ozdemir and H.~Arslan, ``Channel estimation for wireless OFDM systems,'' \emph{ IEEE Commun. Surveys Tuts.}, vol.~9, no.~2, pp.~18--48, 2nd Quart., 2007.
   
           \bibitem{gao20estimation}
           S.~Gao, X.~Cheng, and L.~Yang, ``Estimating Doubly-Selective Channels for Hybrid mmWave Massive MIMO Systems: A Doubly-Sparse Approach," \emph{IEEE Trans. Wireless Commun.}, vol.~19, no.~9, pp.~5703--5715, Sep.~2020.
		
		\bibitem{brady2013beamspace}
		J.~Brady, N.~Behdad and A.~M.~Sayeed, ``Beamspace MIMO for Millimeter-Wave Communications: System Architecture, Modeling, Analysis, and Measurements," \emph{IEEE Trans. Ant. and Propag.}, vol.~61, no.~7, pp.~3814-3827, Jul.~2013.
		
		
		\bibitem{ye2017power}
		H.~Ye, G.~Y. Li, and B.-H. Juang, ``Power of Deep Learning for Channel Estimation and Signal Detection in OFDM Systems,'' \emph{IEEE Wireless Commun. Lett.}, vol.~7, no.~1, pp.~114--117, Feb.~2018.  
		
			{\color{black}
		\bibitem{dong2019deep}
		P.~Dong, H.~Zhang, G.~Y. Li, I.~S. Gaspar, and N.~NaderiAlizadeh, ``Deep CNN-Based Channel Estimation for mmWave Massive MIMO Systems,'' \emph{IEEE J. Sel. Topics Signal Process.}, vol.~13, no.~5, pp.~989--1000, Sep.~2019.}
		
		{\color{black}
		\bibitem{liao2019chanestnet}
		Y.~Liao, Y.~Hua, X.~Dai, H.~Yao, and X.~Yang, ``ChanEstNet: A Deep Learning Based Channel Estimation for High-Speed Scenarios,'' in \emph{Proc. IEEE Int. Conf. Commun. (ICC)}, Shanghai, China, May.~2019, pp.~1--6.}
		
		\bibitem{moon2020deep}
		S.~Moon, H.~Kim, and I.~Hwang, ``Deep learning-based channel estimation and tracking for millimeter-wave vehicular communications,'' \emph{J. Commun. Netw.}, vol.~22, no.~3, pp.~177--184, Jun.~2020.
		
		\bibitem{ma2020sparse}
		W.~Ma, C.~Qi, Z.~Zhang, and J.~Cheng, ``Sparse Channel Estimation and Hybrid Precoding Using Deep Learning for Millimeter Wave Massive MIMO,'' \emph{IEEE Trans. Commun.}, vol.~68, no.~5, pp.~2838--2849, May.~2020.
		
		\bibitem{wei2019ampbeamspace}
		Y.~Wei, M.~-M.~Zhao, M.~Zhao, M.~Lei and Q.~Yu, ``An AMP-Based Network With Deep Residual Learning for mmWave Beamspace Channel Estimation," \emph{IEEE Wireless Commun. Lett.}, vol.~8, no.~4, pp.~1289-1292, Aug.~2019.	
		
		\bibitem{liu2021beamspace}
		S.~Liu and X.~Huang, ``Sparsity-aware channel estimation for mmWave massive MIMO: A deep CNN-based approach," \emph{China Commun.}, vol.~18, no.~6, pp.~162-171, Jun.~2021.
		
		\bibitem{wei2021beamspace}
		X.~Wei, C.~Hu and L.~Dai, ``Deep Learning for Beamspace Channel Estimation in Millimeter-Wave Massive MIMO Systems,"  \emph{IEEE Trans. Commun.}, vol.~69, no.~1, pp.~182-193, Jan.~2021.
		
		
		\bibitem{gao2022beamspace}
		J.~Gao, C.~Zhong, G.~Y.~Li and Z.~Zhang, ``Deep Learning-Based Channel Estimation for Massive MIMO With Hybrid Transceivers," \emph{IEEE Trans. Wireless Commun.}, vol.~21, no.~7, pp.~5162-5174, Jul.~2022.
		
		
		\bibitem{he2023beamspace}
		H.~He, R.~Wang, W.~Jin, S.~Jin, C.~-K.~Wen and G.~Y.~Li, ``Beamspace Channel Estimation for Wideband Millimeter-Wave MIMO: A Model-Driven Unsupervised Learning Approach," \emph{IEEE Trans. Wireless Commun.}, vol.~22, no.~3, pp.~1808-1822, Mar.~2023.
		
		
		\bibitem{commmu2} Y.~Liu, G.~Liao and Z.~Yang, ``Range and angle estimation for MIMO-OFDM integrated radar and communication systems," in \emph{Proc. CIE Int. Conf. Radar}, Guangzhou, China, Oct.~2016, pp.~1--4.
		
		\bibitem{commu3} Z.~Cheng, B.~Liao, S.~Shi, Z.~He and J.~Li, ``Co-Design for Overlaid MIMO Radar and Downlink MISO Communication Systems via Cramér–Rao Bound Minimization," \emph{IEEE Trans. Signal Process.}, vol.~67, no.~24, pp.~6227--6240, Dec.~2019.
		
		\bibitem{sensingplus1} M.~Braun, C.~Sturm, A.~Niethammer and F.~K.~Jondral, ``Parametrization of joint OFDM-based radar and communication systems for vehicular applications,'' in \emph{Proc. IEEE 20th Int. Symp. Pers., Indoor Mobile Radio Commun.}, Tokyo, Japan, Sep.~2009, pp.~3020--3024.
		
		\bibitem{sensingplus2} Y.~Liu, G.~Liao, J.~Xu, Z.~Yang and Y.~Zhang, ``Adaptive OFDM Integrated Radar and Communications Waveform Design Based on Information Theory,'' \emph{IEEE Commun. Lett.}, vol.~21, no.~10, pp.~2174--2177, Oct.~2017.
		
		\bibitem{sensingplus3} D.~Li, M.~Zhan, H.~Liu, ``A Robust Translational Motion Compensation Method for ISAR Imaging Based on Keystone Transform and Fractional Fourier Transform Under Low SNR Environment,'' \emph{IEEE Trans. Aerosp. Electron. Syst.}, vol.~53, no.~3, pp.~2140--2156, Oct.~2017.
		
		\bibitem{sensing1} L.~Zhipeng, C.~Xingbo, W.~Xiaomo, S.~Xu and F.~Yuan, ``Communication analysis of integrated waveform based on LFM and MSK," in \emph{Proc. IET Int. Radar Conf.}, Hangzhou, China, Oct.~2015, pp.~1--5.
		
		\bibitem{sensing2} T.~Huang, N.~Shlezinger, X.~Xu, Y.~Liu and Y.~C.~Eldar,  ``MAJoRCom: A Dual-Function Radar Communication System Using Index Modulation," \emph{IEEE Trans. Signal Process.}, vol.~68, pp.~3423-3438, May.~2020.
		
		\bibitem{joint1} F.~Liu, C.~Masouros, A.~Li, H.~Sun and L.~Hanzo, ``MU-MIMO Communications With MIMO Radar: From Co-Existence to Joint Transmission,'' \emph{IEEE Trans. Wireless Commun.}, vol.~17, no.~4, pp.~2755--2770, Apr.~2018.
		
		\bibitem{joint2} F.~Liu, L.~Zhou, C.~Masouros, A.~Li, W.~Luo and A.~Petropulu, ``Toward Dual-functional Radar-Communication Systems: Optimal Waveform Design," \emph{IEEE Trans. Signal Process.}, vol.~66, no.~16, pp.~4264--4279, Aug.~2018.
		
		\bibitem{joint3} X.~Liu, T.~Huang, N.~Shlezinger, Y.~Liu, J.~Zhou and Y.~C.~Eldar, ``Joint Transmit Beamforming for Multiuser MIMO Communications and MIMO Radar," \emph{IEEE Trans. Signal Process.}, vol.~68, pp.~3929--3944, Jun.~2020.
		
		\bibitem{joint4} F.~Liu, C.~Masouros, A.~Li and T.~Ratnarajah, ``Robust MIMO Beamforming for Cellular and Radar Coexistence," \emph{IEEE Wireless Commun. Lett.}, vol.~6, no.~3, pp.~374--377, Jun.~2017.
		
			{\color{black}\bibitem{shijian1} S.~Gao, X.~Cheng and L.~Yang, ``Spatial Multiplexing With Limited RF Chains: Generalized Beamspace Modulation (GBM) for mmWave Massive MIMO,"  \emph{IEEE J. Sel. Areas Commun.}, vol.~37, no.~9, pp.~2029-2039, Sep.~2019.}
		
		\bibitem{shijian2} Y.~Fan, S.~Gao, X.~Cheng, L.~Yang, and N.~Wang, ``Wideband Generalized Beamspace Modulation (wGBM) for mmWave Massive MIMO Over Doubly-Selective Channels,'' \emph{IEEE Trans. Veh. Technol.}, vol.~70, no.~7, pp.~6869--6880, Jul.~2021.
		
		\bibitem{shijian3} S.~Gao, X.~Cheng and L.~Yang, ``Mutual Information Maximizing Wideband Multi-User (wMU) mmWave Massive MIMO," \emph{IEEE Trans. Commun.}, vol.~69, no.~5, pp.~3067--3078, May.~2021.
		
		\bibitem{shijian4} S.~Gao, J.~Li, X.~Cheng and L.~Yang, ``BER-Minimizing Precoded Wideband Generalized Beamspace Modulation for Hybrid mmWave Massive MIMO," \emph{IEEE Wireless Commun. Lett.}, vol.~11, no.~2, pp.~278--282, Feb.~2022.
		
		\bibitem{im} X.~Cheng, M.~Zhang, M.~Wen and L.~Yang, ``Index Modulation for 5G: Striving to Do More with Less,'' \emph{IEEE Wireless Commun.}, vol.~25, no.~2, pp.~126--132, Apr.~2018.
		
		\bibitem{AE} J.~M.~Mateos-Ramos \emph{et al.}, ``End-to-End Learning for Integrated Sensing and Communication,'' in \emph{Proc. IEEE Int. Conf. Commun. (ICC)}, Seoul, Korea, May.~2022, pp.~1942--1947.
		
		{\color{black}\bibitem{learningbeam} C.~Liu \emph{et al.}, ``Learning-Based Predictive Beamforming for Integrated Sensing and Communication in Vehicular Networks,'' \emph{IEEE J. Sel. Areas Commun.}, vol.~40, no.~8, pp.~2317--2334, Aug.~2022. }
		
		
		\bibitem{zeng2019accessing}
		Y.~Zeng, Q.~Wu, and R.~Zhang, ``Accessing From the Sky: A Tutorial on UAV Communications for 5G and Beyond,'' in {\em Proc. IEEE}, vol.~107, no.~12, pp.~2327--2375, Dec.~2019.
		
		\bibitem{yang2019beam}
		L.~Yang and W.~Zhang, ``Beam Tracking and Optimization for UAV Communications,'' {\em IEEE Trans. Wireless Commun.}, vol.~18, no.~11, pp.~5367--5379, Nov.~2019.
		
		\bibitem{kose2021beam}
		A.~Kose, H.~Lee, C.~H.~Foh, and M.~Dianati, ``Beam-Based Mobility Management in 5G Millimetre Wave V2X Communications: A Survey and Outlook,'' {\em IEEE Open J. Intell. Transp. Syst.}, vol.~2, pp.~347--363, Apr.~2021.
		
		{\color{black}\bibitem{wang2009beam}
		J.~Wang {\em et~al.}, ``Beam codebook based beamforming protocol for multi-Gbps millimeter-wave WPAN systems,'' {\em IEEE J. Sel. Areas Commun.}, vol.~27, no.~8, pp.~1390--1399, Oct.~2009.}
		
		\bibitem{tsang2011coding}
		Y.~M.~Tsang, A.~S.~Poon, and S.~Addepalli, ``Coding the Beams: Improving Beamforming Training in mmWave Communication System,'' in {\em Proc. IEEE Glob. Telecommun. Conf.}, Houston, TX, USA, Dec.~2011, pp.~1--6.
		
		\bibitem{zhang2019position}
		J.~Zhang, W.~Xu, H.~Gao, M.~Pan, Z.~Feng, and Z.~Han, ``Position-Attitude Prediction Based Beam Tracking for UAV mmWave Communications,'' in {\em Proc. IEEE Int. Conf. Commun. (ICC)}, Shanghai, China, May.~2019, pp.~1--7.

  		\bibitem{xu2021predictive}
		Y.~Xu, Y.~Guo, C.~Li, B.~Xia, and Z.~Chen, ``Predictive Beam Tracking with Cooperative Sensing for Vehicle-to-Infrastructure Communications,'' in {\em Proc. IEEE/CIC Int. Conf. Commun. China (ICCC)}, Xiamen, China, Jul.~2021, pp.~835--840.

		
		\bibitem{liu2020radar}
		F.~Liu, W.~Yuan, C.~Masouros, and J.~Yuan, ``Radar-Assisted Predictive Beamforming for Vehicular Links: Communication Served by Sensing,''  {\em IEEE Trans. Wireless Commun.}, vol.~19, no.~11, pp.~7704--7719, Nov.~2020.
		
		\bibitem{shaham2019fast}
		S.~Shaham, M.~Ding, M.~Kokshoorn, Z.~Lin, S.~Dang, and R.~Abbas, ``Fast Channel Estimation and Beam Tracking for Millimeter Wave Vehicular Communications,'' {\em IEEE Access}, vol.~7, pp.~141104--141118, 2019.
		
		\bibitem{liu2020tutorial}
		F.~Liu and C.~Masouros, ``A Tutorial on Joint Radar and Communication Transmission for Vehicular Networks—Part III: Predictive Beamforming Without State Models,'' {\em IEEE Commun. Lett.}, vol.~25, no.~2, pp.~332--336, Feb.~2021.
	
		
		\bibitem{mu2021integrated}
		J.~Mu, Y.~Gong, F.~Zhang, Y.~Cui, F.~Zheng, and X.~Jing, ``Integrated Sensing and Communication-Enabled Predictive Beamforming With Deep Learning in Vehicular Networks,'' {\em IEEE Commun. Lett.}, vol.~25, no.~10, pp.~3301--3304, Oct.~2021.	

		{\color{black}
		\bibitem{9777748}
		Q.~Zhang, X.~Zhang, and C.~Yang, ``Camera-Sensing-Assisted Fast mmWave Beam Tracking for Connected Automated Vehicles,'' {\em IEEE Internet Things J.}, vol.~9, no.~20, pp.~20630--20639, Oct.~2022.
	}
		
		\bibitem{dias2019position}
		M.~Dias, A.~Klautau, N.~Gonz{\'a}lez-Prelcic, and R.~W.~Heath, ``Position and LIDAR-Aided mmWave Beam Selection using Deep Learning,'' in {\em Proc. IEEE 20th Int. Workshop Signal Process. Adv. Wireless Commun. (SPAWC)}, Cannes, France, Jul.~2019, pp.~1--5.
		
		\bibitem{zheng2021deep}
		Y.~Zheng, S.~Chen, and R.~Zhao, ``A Deep Learning-Based mmWave Beam Selection Framework by Using LiDAR Data,'' in {\em 2021 33rd Chinese Control and Decision Conference (CCDC)}, Kunming, China, May.~2021, pp.~915--920.
		
		\bibitem{demirhan2022radar}
		U.~Demirhan and A.~Alkhateeb, ``Radar aided 6g beam prediction: Deep learning algorithms and real-world demonstration,'' in {\em Proc. IEEE Wireless 550 Commun. Netw. Conf. (WCNC)}, Austin, TX, USA, Apr.~2022, pp.~2655--2660.
		

		{\color{black}\bibitem{feifeisemantic}
		Y.~Yang, F.~Gao, X.~Tao, G.~Liu and C.~Pan, ``Environment Semantics Aided Wireless Communications: A Case Study of mmWave Beam Prediction and Blockage Prediction," {\em IEEE J. Sel. Areas Commun.}, vol.~41, no.~7, pp.~2025--2040, Jul.~2023.}	
				
  		\bibitem{miao2020lightweight}
		W.~Miao, C.~Luo, G.~Min, and Z.~Zhao, ``Lightweight 3-D Beamforming Design in 5G UAV Broadcasting Communications,'' \emph{IEEE Trans. Broadcast.}, vol.~66, no.~2, pp.~515--524, Jun.~2020.
  
  		\bibitem{MMFF}
  		H.~Zhang, S.~Gao, X.~Cheng, and L.~Yang, ``Integrated Sensing and Communications towards Proactive Beamforming in mmWave V2I via Multi-Modal Feature Fusion (MMFF),'' Oct.~2023, {\em arxiv:2310.02561}.
		
		
		\bibitem{gao2021fusionnet}
		F.~Gao, B.~Lin, C.~Bian, T.~Zhou, J.~Qian, and H.~Wang, ``FusionNet: Enhanced Beam Prediction for mmWave Communications Using Sub-6 GHz Channel and a Few Pilots,'' {\em IEEE Trans. Commun.}, vol.~69, no.~12, pp.~8488--8500, Dec.~2021.
		
				{\color{black}
		\bibitem{latenew1}
		S.~Pang, D.~Morris and H.~Radha, ``CLOCs: Camera-LiDAR Object Candidates Fusion for 3D Object Detection," in {\em IEEE Int. Conf. Intell. Rob. Syst. (IROS)}, Las Vegas, NV, USA, Oct.~2020, pp.~10386--10393.
		

		\bibitem{latenew3}
		G.~Melotti, C.~Premebida, N.~Goncalves, U.~Nunes and D.~Faria, ``Multimodal CNN Pedestrian Classification: A Study on Combining LIDAR and Camera Data," in {\em Proc. 14th Int. IEEE Conf. Intell. Transp. Syst. (ITSC)}, Maui, HI, USA, Nov.~2018, pp.~3138--3143.}
		
		\bibitem{hybrid1}
		X.~Chen, H.~Ma, J.~Wan, B.~Li, and T.~Xia, ``Multi-View 3D Object Detection Network for Autonomous Driving,'' in {\em Proc. IEEE Conf. Comput. Vis. Pattern Recognit. (CVPR)}, Honolulu, HI, USA, Jul.~2017, pp.~6526--6534.
		
		\bibitem{hybrid2}
		L.~Xie {\em et al.}, ``PI-RCNN: An Efficient Multi-Sensor 3D Object Detector with Point-Based Attentive Cont-Conv Fusion Module,'' in {\em Proc. AAAI Conf. Artif. Intell. (AAAI)}, New York, New York, USA, Feb.~2020, vol.~34, no.~7, pp.~12460--12467.


		\bibitem{hybridnew1}
		D.~Roy, Y.~Li, T.~Jian, P.~Tian, K.~R.~Chowdhury, and S.~Ioannidis, ``Multi-modality Sensing and Data Fusion for Multi-vehicle Detection,'' in {\em IEEE Trans. Multimedia}, vol.~25, pp.~2280-2295, Jun.~2023.
		
		\bibitem{hybrid3}
		C. R.~Qi, W.~Liu, C.~Wu, H.Su, and L.~J.~Guibas, L. J., ``Frustum PointNets for 3D Object Detection From RGB-D Data,'' in {\em Proc. IEEE Conf. Comput. Vis. Pattern Recognit. (CVPR)}, Salt Lake City, UT, USA, Jun.~2018, pp.~918--927.
		
		\bibitem{hybrid4}
		K.~Shin, Y.~P.~Kwon and M.~Tomizuka, ``RoarNet: A Robust 3D Object Detection based on Region Approximation Refinement,'' in {\em Proc. IEEE Intelligent Vehicles Symp. (IV)}, Paris, France, Jun.~2019, pp.~2510--2515.
		
		\bibitem{hybrid5}
		Z.~Wang, and K.~Jia, ``Frustum Convnet: Sliding Frustums to Aggregate Local Point-wise Features for Amodal 3D Object Detection,'' in {\em IEEE Int. Conf. Intell. Rob. Syst. (IROS)}, Macau, China, Nov.~2019, pp.~1742--1749.
		
		\bibitem{hybrid6}
		Z.~Yang, Y.~Sun, S.~Liu, X.~Shen, and J.~Jia, ``Ipod: Intensive Point-based Object Detector for Point Cloud,'' Dec.~2018, {\em arxiv:1812.05276}.
		
		\bibitem{raw4}
		H.~Li and F.~Nashashibi, ``Multi-vehicle Cooperative Perception and Augmented Reality for Driver Assistance: A Possibility to ‘See’ Through Front Vehicle," in {\em Proc. 14th Int. IEEE Conf. Intell. Transp. Syst. (ITSC)}, Washington, DC, USA, Oct.~2011, pp.~242--247.
		
		\bibitem{fakeautocast}
		S.~Kim {\em et al.}, ``Multivehicle Cooperative Driving Using Cooperative Perception: Design and Experimental Validation," {\em{IEEE} Trans. Intell. Transport. Syst.}, vol.~16, no.~2, pp.~663--680, Apr.~2015.
		
		\bibitem{raw3}
		Q.~Chen, S.~Tang, Q.~Yang and S.~Fu, ``Cooper: Cooperative Perception for Connected Autonomous Vehicles Based on 3D Point Clouds," in {\em Proc. IEEE 39th Int. Conf. Distrib. Comput. Syst. (ICDCS)}, Dallas, TX, USA, Jul.~2019, pp.~514--524.
		
		\bibitem{raw5}
		S.~Kim {\em et al.}, ``Multivehicle Cooperative Driving Using Cooperative Perception: Design and Experimental Validation," {\em{IEEE} Trans. Intell. Transport. Syst.}, vol.~16, no.~2, pp.~663--680, Apr.~2015.
		
		\bibitem{autocast}
		H.~Qiu, P.~Huang, N.~Asavisanu, X.~Liu, K.~Psounis, and R. Govindan, ``Autocast: Scalable infrastructure-less cooperative perception for distributed collaborative driving," Dec.~2021, {\em arxiv:2112.14947}.
		
		\bibitem{ann1voxel}
		Z.~Yin, and O.~Tuzel, ``VoxelNet: End-to-End Learning for Point Cloud Based 3D Object Detection," in {\em Proc. IEEE Conf. Comput. Vis. Pattern Recognit. (CVPR)}, Salt Lake City, UT, USA, Jun.~2018, pp.~4490--4499.
		
		\bibitem{ann2point}
		C.~R.~Qi, H.~Su, K.~Mo, and L.~J.~Guibas, ``PointNet: Deep Learning on Point Sets for 3D Classification and Segmentation," in {\em Proc. IEEE Conf. Comput. Vis. Pattern Recognit. (CVPR)}, Honolulu, HI, USA, Jul.~2017, pp.~652--660.
		
		\bibitem{ann3rcnn}
		R.~Girshick, ``Fast r-cnn," in {\em Proc. IEEE Int. Conf. Comput. Vis. (ICCV)}, Santiago, Chile, Apr.~2015, pp.~1440--1448.
		
		\bibitem{fcooper}
		Q.~Chen, X.~Ma, S.~Tang, J.~Guo, Q.~Yang, and S.~Fu, ``F-cooper: Feature based cooperative perception for autonomous vehicle edge computing system using 3D point clouds." in {\em Proc. 4th ACM/IEEE Symp.}, Washington, DC, USA, Nov.~2019, pp.~88--100.
		
		\bibitem{littlecooper}
		E.~Marvasti, A.~Raftari, A.~Marvasti, Y.~Fallah, R.~Guo, and H.~Lu, ``Feature Sharing and Integration for Cooperative Cognition and Perception with Volumetric Sensors," Nov.~2020, {\em arxiv:2011.08317}.
		
		\bibitem{coff}
		J.~Guo {\em et al.}, ``CoFF: Cooperative Spatial Feature Fusion for 3-D Object Detection on Autonomous Vehicles," {\em IEEE Internet Things J.}, vol.~8, no.~14, pp.~11078--11087, Jul.~2021.
		
		\bibitem{slimfcp}
		J.~Guo {\em et al.}, ``Slim-FCP: Lightweight-Feature-Based Cooperative Perception for Connected Automated Vehicles," {\em IEEE Internet Things J.}, vol.~9, no.~17, pp.~15630--15638, Sep.~2022.
		
		{\color{black}
		\bibitem{v2vnet}
		T.~Wang, S.~Manivasagam, M.~Liang, B.~Yang, W.~Zeng, and R.~Urtasun, ``V2VNet: Vehicle-to-Vehicle Communication for Joint Perception and Prediction," in {\em Proc. Eur. Conf. Comput. Vis. (ECCV)}, Glasgow, UK, Aug.~2020, pp.~605--621.}
		
		{\color{black}
		\bibitem{privacy}
		J.~Xiong, R.~Bi, M.~Zhao, J.~Guo, and Q.~Yang, ``Edge-Assisted Privacy-Preserving Raw Data Sharing Framework for Connected Autonomous Vehicles," {\em IEEE Wireless Commun.}, vol.~27, no.~3, pp.~24--30, Jun.~2020.}
		
		\bibitem{senet}
		J.~Hu, L.~Shen, G.~Sun, ``Squeeze-and-Excitation Networks," in {\em Proc. IEEE Conf. Comput. Vis. Pattern Recognit. (CVPR)}, Salt Lake City, UT, USA, Jun.~2018, pp.~7132--7141.
		
		\bibitem{semanticseg}
		Z.~Xiao, Z.~Mo, K.~Jiang, and D.~Yang, ``Multimedia Fusion at Semantic Level in Vehicle Cooperactive Perception," in {\em Proc. IEEE Int. Conf. Multimedia Expo Workshops}, San Diego, CA, USA, Jul.~2018, pp.~1--6.
		
		\bibitem{semanticob1}
		Z.~Y.~Rawashdeh and Z.~Wang, ``Collaborative Automated Driving: A Machine Learning-based Method to Enhance the Accuracy of Shared Information," in {\em Proc. Int. Conf. Intell. Transp. Syst.}, Maui, HI, USA, Nov.~2018, pp.~3961--3966.
		
		\bibitem{semanticob2}
		H.~Cho, Y.~-W.~Seo, B.~V.~K.~V.~Kumar and R.~R.~Rajkumar, ``A multi-sensor fusion system for moving object detection and tracking in urban driving environments," in {\em Proc. IEEE Int. Conf. Robot. Autom. (ICRA)}, Hong Kong, China, Sep.~2014, pp.~1836--1843.
		
		\bibitem{malicious1}
		B.~Hurl, R.~Cohen, K.~Czarnecki, and S.~Waslander, ``TruPercept: Trust Modelling for Autonomous Vehicle Cooperative Perception from Synthetic Data," in {\em Proc. IEEE Intell. Veh. Symp. (IV)}, Las Vegas, NV, USA, Oct.~2020, pp.~341--347.
  
            \bibitem{usecase1}
        W.~Zimmer {\em et al.}, ``Infradet3d: Multi-modal 3d object detection based on roadside infrastructure camera and lidar sensors,'' Apr.~2023, {\em arxiv:2305.00314}.
            \bibitem{usecase2}
        W.~Zimmer, C.~Creß, H. T.~Nguyen, and A. C.~Knoll, ``A9 Intersection Dataset: All You Need for Urban 3D Camera-LiDAR Roadside Perception,'' Jun.~2023, {\em arxiv:2306.09266}.
           \bibitem{usecase3}
        E.~Arnold, M.~Dianati, R.~de Temple and S.~Fallah, ``Cooperative Perception for 3D Object Detection in Driving Scenarios Using Infrastructure Sensors,'' {\em IEEE Trans. Intell. Transp. Syst.}, vol.~23, no.~3, pp.~1852--1864, Mar.~2022.
		\bibitem{2019evoke}
		T.~Stahlbuhk, B.~Shrader, and E.~Modiano, ``Learning algorithms for scheduling in wireless networks with unknown channel statistics,'' {\em  Ad Hoc Netw.}, vol.~85, pp.~131--144, Mar.~2019.
		
		\bibitem{hu2020deep}
		Q.~Hu, F.~Gao, H.~Zhang, S.~Jin, and G.~Y.~Li, ``Deep Learning for Channel Estimation: Interpretation, Performance, and Comparison,'' \emph{IEEE Trans. Wireless Commun.}, vol.~20, no.~4, pp.~2398--2412, Apr.~2020.
		
		\bibitem{yan2019optimal}
		J.~Yan, S.~Bi, Y.~J.~Zhang, and M.~Tao, ``Optimal Task Offloading and Resource Allocation in Mobile-Edge Computing With Inter-User Task Dependency,'' {\em IEEE Trans. Wireless Commun.}, vol.~19, no.~1, pp.~235--250, Jan.~2020.
		
		\bibitem{letaief2006dynamic}
		K.~Letaief and Y.~Zhang, ``Dynamic multiuser resource allocation and adaptation for wireless systems,'' {\em IEEE Wireless Commun.}, vol.~13, no.~4, pp.~38--47, Aug.~2006.
		
	\end{thebibliography}
\end{document}